\documentclass[11pt,tightenlines,eqsecnum,floats,aps,amsmath,amssymb,nofootinbib,prd,shownopacs,floatfix]{revtex4}
\usepackage{graphicx, wrapfig}
\usepackage{amssymb}
\usepackage{color}
\usepackage{mathrsfs}
\usepackage{subfigure}
\def\f{\frac}

\def\lp{l_{\rm Pl}}

\def\rhopl{\rho_{\rm Pl}}

\def\heff{\mathcal{H}_{\rm eff}}

\usepackage{enumerate}

\newcommand{\be}{\nopagebreak[3]\begin{equation}}
\newcommand{\ee}{\end{equation}}
\newcommand{\bfig}{\nopagebreak[3]\begin{figure}}
\newcommand{\efig}{\end{figure}}
\newcommand{\ba}{\nopagebreak[3]\begin{eqnarray}}
\newcommand{\ea}{\end{eqnarray}}

\newcommand{\bmult}{\nopagebreak[3]\begin{multline}}
\newcommand{\emult}{\end{multline}}
\newcommand{\fref}[1]{Fig.\,\ref{#1}}
\newcommand{\eref}[1]{eq.\,(\ref{#1})}

\begin{document}
\title{Numerical simulations of a loop quantum cosmos: robustness of the quantum bounce and the validity of effective dynamics}
\author{Peter Diener$^{1,2}$}
\email{diener@cct.lsu.edu}

\author{Brajesh Gupt$^2$}
\email{brajesh@phys.lsu.edu}

\author{Parampreet Singh$^2$}
\email{psingh@phys.lsu.edu}

\affiliation{
$^1$
Center for Computation and Technology, Louisiana State University, Baton Rouge, LA 70803, U.S.A.
}
\affiliation{
$^2$
Department of Physics and Astronomy, Louisiana State University, Baton Rouge, LA 70803, U.S.A.
}

\begin{abstract}
A key result of isotropic loop quantum cosmology is the existence of a quantum
bounce which occurs when the energy density of the matter field approaches a
universal maximum close to the Planck density. Though the bounce has been
exhibited in various matter models, due to 
severe computational challenges, some important questions have so far remained
unaddressed. These include the demonstration of the bounce for widely spread
states, its detailed properties for the states when matter
field probes regions close to the Planck volume and the reliability of the
continuum effective spacetime description in general. In this manuscript we 
rigorously answer these questions using the Chimera numerical scheme for 
the isotropic spatially flat model sourced
with a massless scalar field. We show that, as
expected from an exactly solvable model, the quantum bounce is a generic
feature of states even with a very wide spread, and for those which bounce 
much closer to the Planck volume. We perform a detailed analysis of the departures 
from the effective description and find some expected, and some surprising results. 
At a coarse level of description, the effective dynamics
can be regarded as a good approximation to the underlying quantum dynamics
unless the states correspond to small scalar field momenta, in which case they bounce
closer to the Planck volume or are very widely spread. Quantifying the amount of
discrepancy between the quantum and the effective dynamics, we find that the 
departure between them depends in a subtle and non-monotonic way on the field 
momentum and different fluctuations. Interestingly, the departures are generically 
found to be such that the effective dynamics overestimates the spacetime curvature, 
and underestimates the volume at the bounce. 

\end{abstract}

\maketitle

\section{Introduction}\label{sec:intro}
A reliable understanding of the physics of the very early universe and its initial conditions 
requires us to go beyond the realm of validity of Einstein's general relativity (GR). In the 
classical theory unless the matter is chosen to violate the weak energy condition, the 
backward evolution of a  spatially flat Friedmann-Robsertson-Walker (FRW) universe 
ends in a big bang singularity. The evolution of the FRW universe is hence past incomplete 
in GR. Singularity theorems of Penrose and Hawking show that the fate of other spacetimes 
in the presence of spatial curvature, anisotropies and inhomogeneities is similar. It has been 
long expected that this key limitation of GR -- its breakdown when the singularities are 
approached -- arises from assuming the validity of the classical notion of spacetime at all 
scales, and a quantum theory of spacetime and gravity would provide insights on the 
resolution of singularities. In the last decade, progress in quantization of cosmological 
models in loop quantum cosmology (LQC) has provided a glimpse of the new physics 
which may arise from the quantization of gravity using non-perturbative techniques in 
loop quantum gravity (LQG). For various matter models, a rigorous quantization has 
been performed, and a common result found in the study of all these models is the 
existence of a quantum bounce when the curvature of the spacetime becomes Planckian 
\cite{as1}. The backward evolution of an expanding branch of the universe in LQC 
undergoes a smooth non-singular bounce, as opposed to the classical theory where 
the evolution is singular \cite{aps1,aps2,aps3}.

The underlying quantum evolution equation in LQC is a discrete equation in the geometric 
representation. In the homogeneous and isotropic spacetime, the setting which we are 
interested in, the quantum Hamiltonian constraint is a difference equation with a uniform 
spacing in volume \cite{aps3,apsv,ap,bp,kv,szulc_open,warsaw_flat,warsaw_closed,mmo,aps4}.
This uniform spacing is fixed by the minimum eigenvalue of the area operator in LQG\footnote{Due to ambiguities in the quantum theory, one may be tempted to motivate 
a quantum Hamiltonian constraint with a different discretization property, such as uniform 
discretization in the scale factor. However, it turns out that all but uniform discretization in volume are 
ruled out by mathematical consistency \cite{cs1}. This is supported by factor
ordering 
issues \cite{mairi1} and the stability arguments of the difference equation 
\cite{brizuela_cartin_khanna,ps12}.}. The origin of the discretization lies in the non-local character 
of the gravitational field strength of the Ashtekar-Barbero connection expressed in terms of 
holonomies \cite{abl}.  The matter part of the Hamiltonian constraint is quantized using 
conventional Fock quantization, resulting in differential operators for the corresponding 
momenta. If the matter is chosen to be a massless scalar field $\phi$, it can act as an internal 
clock and unitary evolution can be studied using a Klein-Gordon like equation. The 
quantization procedure leads to some key differences at both kinematical as well as 
physical level with the Wheeler-DeWitt theory, including a universal maxima  for the 
expectation value of the energy density $\rho_{\rm{max}} \approx 0.409 \rho_{\rm{Pl}}$\cite{acs} 
and a zero probability for the occurrence of a big bang singularity \cite{craigh-singh13}. 
However, the quantum difference equation of LQC is very well approximated by the 
differential Wheeler-DeWitt equation when spacetime curvature becomes much smaller 
than the  Planckian value. For the massless scalar field model the curvature invariants 
can be completely captured by the energy density $\rho = p_\phi^2/2 V^2$, which implies, since the momentum $p_\phi$ is a constant of motion,
that the above agreement occurs when the physical volume $V$ is much larger than the Planck volume. 

It is interesting to note that for suitable semi-classical states, it is possible to derive an 
effective Hamiltonian for LQC using a geometrical formulation of quantum mechanics 
\cite{josh,vt,psvt}. The effective Hamiltonian provides a continuum description of the 
underlying quantum geometry in terms of the modified Friedmann and Raychaudhuri 
equations \cite{ps06}, and has served as a very valuable tool to gain many insights 
on the new physics in LQC \cite{as1}. A key result of the numerical simulations 
performed so far for the isotropic as well as anisotropic models in LQC, is that the effective 
dynamics obtained from the corresponding effective Hamiltonian is a very good 
approximation to the underlying quantum dynamics at all scales for initial states 
which are peaked at a large volume at late times \cite{aps1,aps2,aps3,apsv,kv,bp,ap,b1madrid1}. 
However, there are some important caveats. Firstly, it is  to be emphasized that the effective 
Hamiltonian is derived for a specific set of states which are semi-classical and sharply 
peaked at a classical trajectory at late times when the spacetime curvature is small 
compared to the Planckian curvature. Therefore, departures from the effective theory can 
be expected if one considers states which are widely spread. Secondly, an assumption 
underlying the derivation of the effective Hamiltonian in Refs. \cite{josh,vt,psvt} is that 
the volume is always greater than the Planck volume. This assumption is tied to the approximation of 
the eigenvalues of the inverse volume operator in the Hamiltonian constraint with $1/V$. 
It is remarkable that even at volumes as small as 10 times the Planck volume, the error in 
this approximation is less than 1\%. 
Note that in all the numerical investigations carried out so far, excellent 
agreement was found with the effective theory as, in these cases, the quantum 
bounce occurred at volumes far greater than the Planck volume.
As an example, in the study of spatially flat 
isotropic models with a massless scalar field \cite{aps3}, the value of the bounce volume was greater than 1000 times the Planck volume. Moreover, the initial states were always chosen to be
sharply peaked with relative fluctuations much smaller than unity. 

In previous works such as in \cite{aps1,aps2,aps3}, where the objective was to understand 
the fate of the classical singularity in the quantum theory for a universe which corresponds to a classical macroscopic universe at late 
times, the value of $p_\phi$ at which the initial semi-classical state was peaked was taken to be large. 
Our objective, in this article, is to extend the present understanding of 
the quantum bounce in LQC to more extreme initial conditions (including small $p_\phi$) 
for which the volume at the bounce is of the order of a few Planck volumes, and probe the 
validity of the effective dynamics for a wide variety of initial conditions.
Since the effective Hamiltonian approach is a very important tool to extract physics in LQC, 
it is pertinent to carefully understand the validity of the effective dynamics by probing regimes 
where we expect it to fail. This would be the case when the bounce happens close to the 
Planck volume and when the initial states are not be sharply peaked \cite{psvt}. Such an exercise 
will allow a rigorous test of the robustness of the quantum bounce and effective dynamics for far 
more general states than the ones considered so far. If the departures from the quantum theory 
are found, then this would open a new avenue to compute finer corrections to the modified 
Friedmann dynamics and understand the robustness of various phenomenological predictions extracted in LQC. 
This can be potentially very useful. In particular to gain a deeper understanding of the 
novel results found in the pre-inflationary epoch in LQC \cite{aan1} and by considering the perturbations of the effective Hamiltonian 
\cite{barrau_review}. 

Probing the effects of quantum discreteness at geometrical scales much smaller than 
those considered so far, and to check the robustness of effective theory for states which are 
not sharply peaked is computationally very challenging. To understand the associated 
difficulties, let us first note that the technical challenges for both types of states are 
essentially the same since a state probing quantum geometry at much smaller scales than 
those considered so far will be widely spread. We note that the scale at which a state bounces, 
which can be regarded as the minimum volume of quantum geometry probed by the initial state, 
directly depends on the value of the scalar field momentum $p_\phi$ on which the initial state is 
peaked. The effective dynamics predicts that the bounce occurs at volume 
$V_{\rm b}^{\rm(eff)} \approx 1.1\, p_\phi\, V_{\rm Pl}/\sqrt{G}\hbar$. If the field momentum is taken to be
$p_\phi=1000\sqrt{G}\hbar$, a typical value for which numerical simulations have been 
performed in LQC, then the universe will bounce at 
$V_{\rm b}^{\rm (eff)} \approx 1100\ V_{\rm Pl}$ according to the effective
theory. Indeed, it turns out that for sharply peaked states the quantum 
bounce occurs at approximately the same value \cite{aps1,aps2,aps3}. This bounce volume is larger 
than the underlying quantum discreteness in volume by approximately three orders of magnitude. 
Further, the  errors resulting from the approximation of inverse volume modifications considered in 
the effective Hamiltonian at this bounce volume are smaller than $10^{-10}$. To probe the quantum 
geometry at volumes, say one order greater than the Planck volume, we expect to have to consider much 
smaller values of $p_\phi$ than before. Such states are inevitably widely spread in 
volume. As an example, if an initial state is peaked at a very small value of $p_\phi$, say 
$p_\phi = 20 \sqrt{G} \hbar$ with $\Delta p_\phi = 1.59$ and peaked at $V = 68700\, V_{\rm Pl}$, 
then it has relative spread in volume as $\Delta V/V = 6.45$. 
One is thus dealing with a very quantum state.

The computational cost involved in simulations with states with large relative fluctuation in volume 
with the techniques used so far is extremely high compared to sharply peaked states.  A 
typical simulation for sharply peaked states, for example states which are 
peaked at 
$p_\phi=1500\,\sqrt{G}\hbar$ and with initial relative volume dispersion $\Delta V/V \approx 0.01$, 
requires $\sim 30,000$ grid points in the spatial direction. 
Such a simulation takes approximately $240\, s$ on a 2.4 GHz Sandybridge
computer with 16 cores. On the other hand, simulations with small 
$p_\phi$ such as $20\, \sqrt{G}\hbar$ with $\Delta V/V\approx 6$, 
requires the outer boundary in the numerical grid for volume to be placed at
$V_{\rm outer}\approx 2\times 10^{19}\, V_{\rm Pl}$. This would, in turn, 
require $1.25\times 10^{14}$ times more grid points in the spatial direction. 
Due to the stability requirements for the numerical simulation,\footnote{For 
the numerical simulations, the differential operators in the matter quantum 
Hamiltonian constraint need to be discretized. Unlike the discretization in 
volume, the numerical discretization in $\phi$ is not fixed by the underlying 
theory, but is constrained by the Courant-Friedrichs-Lewi (CFL) criteria for a
stable evolution of the difference equations \cite{cfl}. Due to the nature of
the equations, any increase in the location of the spatial boundary requires a
corresponding decrease in the time step for a stable evolution.} this amounts
to requiring a $1.25\times10^{14}$ times smaller time step for a stable
evolution. 
On a similar workstation, this simulation would take approximately 
$10^{23}\, {\rm years}$ if it is assumed that the simulation would fit
in memory.

To tackle this challenge, we recently developed a hybrid numerical
scheme, the Chimera scheme, by utilizing the fact that the LQC difference equation tends to the
Wheeler-DeWitt equation in the large colume regime~\cite{dgs1}. The Chimera scheme uses a 
hybrid spatial grid having two components: an inner grid where evolution is performed using the LQC 
difference equation, and an outer grid chosen at very small spacetime curvature where evolution is 
performed using a discretized version of the Wheeler-DeWitt equation. Since we 
are free to choose the underlying discretization on the outer grid, it leads 
to a significant computational cost reduction.
As an example, using the Chimera scheme for $p_\phi^*=20\,\sqrt{G}\hbar$ and 
$\Delta V/V \approx 6$, the numerical simulation can be performed in less than
10 minutes! 

In this work, we use the Chimera scheme to perform simulations for various values of
$p_\phi$ in the range between $20$ to $1500~\sqrt{G} \hbar$ for a large range of peakedness 
properties of the initial states. We perform numerical simulations with initial states constructed via 
three different methods originally proposed in Refs. \cite{aps2,aps3}.\footnote{Due to issues 
associated with the construction of the initial state, for the first method the smallest value of $p_\phi$ 
at which the initial state is peaked is chosen to be 500 $\sqrt{G} \hbar$.}   
The quantum bounce is shown to occur for all three kinds of states and for all initial data parameters. This is an expected 
result from the analytical studies on genericness of the bounce in Ref. \cite{acs}.\footnote{Here we 
note that the quantum constraint considered in the exactly solvable model in Ref. \cite{acs} was 
different from the one considered here.} The relative fluctuations across the bounce remain tightly 
constrained throughout the evolution, satisfying triangle inequalities found in the 
analysis of Ref. \cite{kp}. We obtain new and unexpected results when we 
analyze the departures of the effective theory from the underlying quantum 
evolution. The effective theory 
shows departures under certain conditions, and these departures have different features in the
different methods for constructing the initial state. These differences, which are summarized in 
table \ref{tab:summary}, become significantly less pronounced for sharply peaked states for large 
values of $p_\phi$. The general expectation that the differences between the
effective and the LQC theory grow as $p_\phi$ decreases and the fluctuations in volume increase, is found not to be true 
in general. We find that the results depend on the method chosen for the initial state construction 
and also on the way an initial state departs from a minimum uncertainty state with equal relative 
dispersions in the volume and the field momentum. For all the three methods, it is possible that the 
agreement between the effective theory and the quantum evolution improves if the relative fluctuation in 
volume is increased while keeping $p_\phi$ the same. On the 
other hand, if the relative fluctuation in volume is fixed, then a decrease in 
the value of $p_\phi$ at 
which the state is peaked always increases the departure between effective theory and quantum 
evolution for two of the three methods. For the third method, it depends on whether the initial state 
has absolute fluctuation in $p_\phi$ greater than a certain value.

Our results show that for an initial state peaked at a macroscopic universe at late times, 
the effective theory provides an excellent description of the quantum dynamics. It is only 
for the initial states which have large fluctuations that the large departures from the effective theory 
are found. Under certain conditions, there can be significant departures between the 
effective theory and the quantum evolution if the initial state is peaked at a small value of $p_\phi$.
As mentioned earlier, such states have wide spread even initially and they do not 
correspond to a universe which evolves to a classical macroscopic universe. It turns out that the 
quantitative differences can be fitted by simple polynomial functions when 
plotted against the spread in field momentum $\Delta p_\phi$.  Also, the departures in the
large $p_\phi$ regime, though small, show interesting behavior with respect to
the relative spread in $V$. A general result, found from the study of all the methods, 
is that the effective theory overestimates the value of spacetime curvature at the bounce and 
underestimates the bounce volume. Another general result is the non-monotonic behavior of 
the relative volume dispersion, as the dispersion in the field momentum varies for a fixed 
value of $p_\phi$. This result is in agreement with analytical studies performed with a different 
kind of initial states and a different quantum constraint \cite{montoya_corichi2}. The non-monotonicity in the volume 
dispersion is also reflected in the quantitative study of the differences between the effective and 
LQC trajectories, as the spread of the initial state is varied. 

This manuscript is organized as follows. For completeness, we provide a brief overview of 
the main results we need in studying the loop quantization of homogeneous and isotropic 
spacetime with a massless scalar field, in Sec. \ref{sec:lqc}. Our discussion is based on Ref. 
\cite{aps3}, which we refer the reader to for details. 
In Sec. \ref{sec:eff} we briefly discuss the effective theory and modified Friedmann equation 
for flat FRW model. In Sec. \ref{sec:initialdata} we describe the
three different methods of specifying the initial states, which were originally discussed in Ref. 
\cite{aps2}. Here we also discuss properties of uncertainty products and minimum uncertainty 
states.  In Sec. \ref{sec:chimera} we briefly discuss the Chimera scheme, which is used for 
the numerical simulations discussed in this paper. In Sec. \ref{sec:evolution} we show some 
representative results with one of the methods to provide a snap shot of the new findings. Here 
we show the contrast in results for sharply peaked initial states with a large value of $p_\phi$ at 
which the state is peaked, with a widely spread initial state chosen to be peaked at a small value of 
$p_\phi$. In the latter case, we show departures of the effective trajectory from the quantum 
expectation values of the volume observable (and their agreement within the variances) for 
different values of relative fluctuations.  We also demonstrate in this section, the way the relative 
fluctuations change and the way triangle inequalities between them are satisfied for different 
methods of construction of initial states. In Sec. \ref{sec:results}, we study the 
differences between the effective and the LQC trajectories quantitively in detail for different initial state  methods. This brings out several interesting features both in the 
small and large $p_\phi$ regime. In Sec. \ref{sec:disc} we provide a summary and discussion 
of the main results presented in the paper.

\section{Loop quantum cosmology: spatially flat, isotropic model}
\label{sec:lqc}
Loop quantization of the homogeneous and isotropic model is a constraint based quantization 
which closely follows the procedure in Loop Quantum Gravity (LQG) -- a candidate theory for 
canonical quantization of gravity based on Ashtekar variables: the SU(2) connection $A^i_a$ and 
the conjugate triad $E^a_i$. Due to the underlying symmetries of the homogeneous and isotropic 
spacetime, the Ashtekar-Barbero connection and the triad can be symmetry reduced to connection 
$c$ and triad $p$ \cite{abl}:
\be
A_a^i = c~\mathring{V}^{1/3}~ \mathring{\omega}_a^i, \quad  {\rm and} \quad E_i^a={\rm sgn}(p)\, p~\mathring{V}^{-2/3}\sqrt{q} ~\mathring{e}_i^a,
\ee
where $\mathring{e}_i^a$ and $\mathring{\omega}_a^i$ respectively denote the densitized triads 
and  co-triads compatible with the fiducial metric $\mathring{q}_{ab}$ on the spatial manifold which 
we consider to be a 3-torus.\footnote{Results of this analysis, are independent of the particular 
topology of the spatial manifold.}  In the above equation, $\mathring{V}$ denotes the fiducial 
volume: $\mathring{V} = \int_{\mathbb{T}^3} d^3 x \sqrt{|\mathring{q}|}$. The fiducial metric is 
related to the spatial metric $q_{ab}$ as: $q_{ab} = a^2 \mathring{q}_{ab}$, where $a$ denotes the 
scale factor in the spacetime line element:
\be
ds^2 = -N^2 dt^2 + a^2(t)\left(dx^2+dy^2+dz^2\right) ~.
\ee
In our analysis, we will choose the lapse $N = 1$. For this choice of lapse, the connection and its conjugate triad are related to the metric and its time derivative as\footnote{The following relation between $c$ and the time derivative of the scale factor is true only in the classical theory.}
\be
p=\mathring{V}^{1/3} a^2, \qquad c=\gamma \mathring{V}^{1/3}\dot{a},
\ee
where $\gamma \approx 0.2375$ is the Barbero-Immirizi parameter \cite{Meissner_gamma,Domagala_gamma}.

In the loop quantization of cosmological spacetimes, due to symmetry reduction, the only non-trivial constraint is the Hamiltonian constraint. Following LQG, the gravitational part of the classical Hamiltonian constraint is expressed in terms of the holonomies and fluxes of the symmetry reduced Ashtekar-Barbero connection $c$ and the conjugate triad $p$
\be
C_{\rm grav} = -\f{1}{\gamma^2} \int d^3x~\varepsilon_{ijk}~ \f{E^{ai}E^{bj}F^k_{ab}}{\sqrt{|\rm{det}(E)|}} = -\f{6}{\gamma^2}c^2~\sqrt{|p|},
\ee
where $F_{ab}^k$ denotes the field strength of the connection in terms of the holonomies over straight edges $\lambda \mathring{e}^a_k$
\be
h_k^{(\lambda)} = \cos({\lambda c/2}) \mathbb{I}+2\sin({\lambda c/2})\tau_k,
\ee
where $\tau_k = - i \sigma_k/2$, and $\sigma_i$ denote Pauli spin matrices. The field strength $F_{ab}^k$ is written in terms of the holonomy around a square loop, as in the classical gauge theory, as:
\be
F_{ab}^k = -2 \lim_{Ar_\Box\rightarrow 0} {\rm Tr}\left(\f{h_{\Box_{ij}}^{(\lambda)}-1}{\lambda^2\mathring{V}^{2/3}}\right)\tau^k \mathring{\omega}^i_a\mathring{\omega}^j_b,
\ee
where
\be
  h_{\Box_{ij}}^{(\lambda)} = h_i^{(\lambda)} h_j^{(\lambda)}\left(h_i^{(\lambda)}\right)^{-1} \left(h_j^{(\lambda)}\right)^{-1}.
\ee
In the quantum theory, due to the underlying quantum geometry, the area of the loop $ \lambda^2 
p$ has a non-zero minimum $\Delta=4\sqrt{3}\pi\gamma\lp^2$. Thus, 
$\lambda=\sqrt{\Delta}/|p|^{1/2}$. The dependence of the $\lambda$ on the triad $p$ leads 
to the volume representation as a  natural choice of the geometric representation in the 
quantum theory \cite{aps3,acs}. 
On the eigenkets $|v \rangle$ of the volume operator $\hat V = \widehat{|p|}^{3/2}$,
\be\label{V_eigen}
\hat V |v \rangle = \left(\f{8 \pi \gamma}{6}\right)^{3/2} \f{|v|}{K} \lp^3 |v \rangle , ~~~ \mathrm{where} ~~~ K = \f{2}{3\sqrt{3\sqrt{3}}},
\ee
 the action of the holonomy operators is of a simple translation
\be
\widehat{h^{(\lambda)}} |v\rangle = |v-1\rangle ~. 
\ee
The eigenkets $| v \rangle$ satisfy: $\langle v_1 | v_2 \rangle = \delta_{v_1,v_2}$, and form an orthonormal basis in the kinematical Hilbert space which is the space of the square integrable functions on the Bohr compactification of the real line. 

Here we are interested in a flat FRW model with the matter part of the Hamiltonian constraint 
provided by a massless scalar field $\phi$: $C_{\phi} = 8 \pi G |p|^{-3/2} p_\phi^2$. The physical 
states, $\Psi(v,\phi)$ are obtained by solving 
$(\widehat C_{\mathrm{grav}} + \widehat C_{\phi}) \Psi = 0$, which yields a Klein-Gordon type 
equation with $\phi$ playing the role of internal time \cite{aps3}:
\be
\label{eq:lqcevol}\f{\partial^2}{\partial \phi^2}\Psi(v,\,\phi)=-\widehat{\Theta}\Psi(v,\,\phi) ~.
\ee
Here $\Theta$ is the discrete spatial Laplacian defined as,
\be
\label{eq:lqctheta}\widehat{\Theta} :=-\f{1}{B(v)}\Bigg[C^+(v) \Psi(v+4, \phi) + C^o(v)\Psi(v,\,\phi)+ C^-(v)\Psi(v-4,\,\phi)\Bigg]
\ee
where the coefficients $C^\pm$ and $C^o$ are given as
\ba
  C^+(v) &=& \frac{3\pi K G}{8} |v+2|\left | |v+1|^{}-|v+3| \right |, \nonumber \\
  C^-(v) &=& C^+(v-4)=\frac{3\pi K G}{8} |v-2|\left | |v-3|-|v-1| \right |, \nonumber \\
  C^0(v) &=& -C^+(v)-C^-(v).
\ea
and $B(v) = \f{27}{8}K |v|\left(\arrowvert |v+1|^{1/3} - |v-1|^{1/3}\arrowvert\right)^3$ denotes the eigenvalue of the inverse volume operator,
\be
|p|^{-3/2} |v\rangle = \left(\frac{8 \pi \gamma \lp^2}{6}\right)^{-3/2} \, B(v) |v\rangle ~.
\ee 
The LQC evolution operator given in \eref{eq:lqctheta} is a difference operator with a uniform
discretization in $v$. As a result, the evolution equation in LQC turns out to be a difference
equation, as opposed to Wheeler-DeWitt theory, where the evolution equation is a 
partial differential equation.

Since we are considering non-fermionic matter, physics does not distinguish between the 
orientations of the triads.  The physical states are therefore chosen to be symmetric under 
the change of sign of the physical triads: $\Psi(v,\,\phi)=\Psi(-v,\,\phi)$. The inner product of 
the physical states can be obtained by using a group averaging procedure 
\cite{Ashtekar:1995zh, Marolf:1995cn}
\be
\langle \Psi_1|\Psi_2\rangle = \sum \overline{\Psi_1}(v,\phi) {B(v)}^{-1} {\Psi_2}(v,\phi).
\ee
To extract physical predictions, we need a set of Dirac observables. As the field momentum 
is a constant of motion in the classical theory, it turns out to be one of the Dirac observables. 
Another Dirac observable is the expectation value of $v$ at a fixed ``emergent time'' 
$\phi_o$, $v|_{\phi=\phi_o}$. The expectation value of these Dirac observables are 
computed for each value of the scalar field $\phi$ via
\ba
\label{eq:expectation}
\langle \Psi|\widehat{v}|_{\phi=\phi_o}|\Psi\rangle&=&||\Psi||^{-1} \sum_v B(v) |v||\Psi(v,~\phi_o)|^2 \nonumber \\
\langle \Psi|\widehat{p_\phi}|\Psi\rangle&=&||\Psi||^{-1} (-i\hbar)\sum_v B(v) \bar{\Psi}(v,~\phi)~\partial_\phi \Psi(v,~\phi).
\ea
Similarly, one can obtain the fluctuations of the expectation values. 
The expression for the energy density can be given as
\be
\rho=\frac{\langle p_\phi \rangle^2}{2 \langle V\rangle^2},
\ee
where $\langle p_\phi \rangle$ and $\langle V\rangle$ respectively are the 
expectation values of the field momentum and volume, obtained via 
\eref{eq:expectation}. We will use this definition of the energy density for the 
numerical simulations presented in this paper. Note that the physical 
states have support on the discrete lattices $\pm \epsilon + 4n$ where $\epsilon \in [0,4)$.  
These are preserved by the dynamical evolution, and hence there is a superselection in the 
physical Hilbert space. We will focus our analysis on the case of the lattice defined by 
$\epsilon = 0$ which includes the possibility of physical volume becoming zero in the evolution.

With the quantum evolution equations available, we can now consider a suitable initial state 
and study its evolution. By computing the expectation values of the Dirac observable, and 
comparing them with the classical trajectory, physical implications of the underlying quantum 
geometry can be studied. For the model under consideration, this task was first completed in 
Refs. \cite{aps1,aps2,aps3}, for initial states, with reasonably large $p_\phi$, which are 
sharply peaked on a classical trajectory at late times and at large volumes. The backward 
evolution of such states revealed the existence of a quantum bounce in the Planck regime, 
when the energy density of the scalar field reached a universal maximum $\rho_
{\mathrm{max}} \approx 0.409 \rho_{\mathrm{Pl}}$~\cite{acs}. It also turns out that the quantum 
evolution is very well approximated by a trajectory obtained from an effective Hamiltonian 
(see Sec.\ \ref{sec:eff}) at all scales, including the bounce for sharply peaked states.

As discussed in Sec.\ \ref{sec:intro}, our goal in this manuscript is to study the above quantum 
evolution in more detail to understand the way the quantum bounce and the agreement with the 
effective dynamical trajectory, obtained from the effective Hamiltonian, are affected by 
considering states which may not be sharply peaked and correspond to small values of 
$p_\phi$. For this purpose, and to facilitate a comparison with previous works, we will consider the 
form of the initial states identical to the ones in Refs. \cite{aps1,aps2,aps3} in Sec. 
\ref{sec:initialdata}.\ These initial states are constructed using an important property of the LQC 
quantum difference equation that in the large volume approximation, it is extremely well 
approximated by the Wheeler-DeWitt equation. In terms of $v$, the Wheeler-DeWitt equation 
can be written as \cite{aps3}
\be
\label{eq:wdw} \f{\partial^2}{\partial \phi^2} \underline{\Psi}(v,\,\phi)=-\widehat{\underline{\Theta}}\,\underline{\Psi}(v,\,\phi) := 12 \pi G v\f{\partial }{\partial v}\left(v\f{\partial }{\partial v}\right)\underline{\Psi}(v,\,\phi),
\ee
where $\underline{\widehat\Theta}$, unlike $\widehat\Theta$ in eq. (\ref{eq:lqctheta}), is a 
differential operator. It is important to note that the underlying geometry in the Wheeler-
DeWitt theory is continuous unlike the discrete quantum geometry in LQC. It is only in the 
regime where the spacetime curvature is small, that the discreteness of quantum geometry is 
well approximated by the smooth continuum description.

\section{Effective dynamics} \label{sec:eff}
Before we proceed to the construction of the initial states, we briefly discuss the
main features of the effective theory.
The effective spacetime description is derived using the geometric formulation
of quantum mechanics. Using the inner product in the Hilbert space, one can
define a Riemannian metric and the symplectic structure, and treat the
Hilbert space as a quantum phase space. In order to obtain the modifications to
the classical dynamical equations resulting from the quantum theory, one then
relates the quantum phase space variables to their classical counterparts \cite{schilling_aa}. 
In LQC, this has been achieved using two independent approaches under different
sets of assumptions. The first is the embedding approach \cite{josh,vt} which has been
extensively used for phenomenological applications and analysis of
cosmological perturbations in LQC, and which is also used in this manuscript. The second 
is the truncation approach \cite{mb_rev_eff} which relates the 
expectation values of the basic conjugate variables in the quantum phase space
to those in the classical phase space using an order by order approximation. 

The embedding approach is based on utilizing a fiber bundle structure in the
quantum phase space whose base space is the classical phase space. Using the
expectation values of the basic quantum phase space variables, which remain
constant along the fibers of the bundle, one seeks  a projection of the
quantum phase space into the classical phase space. In particular, with a
judicious choice of states, one aims to construct approximately horizontal
sections which are preserved under evolution. If such sections can be found, then
a faithful embedding of quantum phase in the classical phase space is
achieved and corrections to the classical dynamical equations can be obtained. In
LQC, this approach has been used for Gaussian states to compute modifications
to the classical Friedmann equation in the spatially flat model with a
massless scalar field \cite{vt}, and for dust and radiation matter fields albeit for an
older form of quantum constraint \cite{josh}. These results have been extended
to a general equation of state of matter recently \cite{psvt}. The validity of
the modified Friedmann dynamics in the effective theory has so far been
verified with the evolution obtained from the quantum Hamiltonian constraint
in LQC, using sharply peaked states, for isotropic spacetimes in the presence and
absence of spatial curvature for massless scalar field \cite{aps3,apsv,kv}, in
the presence of a cosmological constant \cite{bp,ap} and inflationary potential
\cite{aps4}, as well as in the case of vacuum Bianchi-I spacetime \cite{b1madrid2}.
Recently, these  tests have also been conducted in the case of a cyclic
potential \cite{cyclic}.

In the truncation approach, which in principle is applicable for a wider
variety of states, there are infinitely many  moments which are coupled by
non-linear differential equations. By taking a truncation to a lower order of
moments, one can compute corrections  (valid to that small order) to the
classical dynamical equations. In this approach, it is important to check the
self-consistency of the truncated equations and to also ensure that ignored
higher moments remain insignificant throughout the evolution. Otherwise the
truncated equations may not reliably capture the full quantum evolution from the quantum 
Hamiltonian constraint. In the LQC literature, the phenomenological applications of 
the truncation approach are not as widely studied as the embedding approach and few 
tests have been performed to determine how well the truncation approach agrees with the 
full quantum evolution. There have been numerical investigations of the backreaction 
effects due to higher order moments on Gaussian initial states in this approach to 
the effective theory, see Ref. \cite{bojowald_backrec}.
In the following, we will focus our discussion 
to the effective dynamics as derived using the embedding approach.

In the embedding approach, for the the case of the massless scalar field in
the spatially flat isotropic model the effective Hamiltonian constraint up to
the terms which depend on the fluctuations is given as:
\be
\label{eq:heff}\heff = -\f{3V}{8\pi G \gamma^2}\f{\sin^2\left(\lambda \beta\right)}{\lambda^2}  + {\cal H}_\phi \approx 0, ~~~ {\rm where} ~~~ {\cal H}_\phi =\f{p_\phi^2}{2 V} ~.
\ee
Here $\beta$ is the conjugate variable to $V$, satisfying $\{\beta,V\} = 4 \pi G\gamma$. 
In the derivation of the above constraint,  the initial state is
peaked at small spacetime curvature at a large volume. The constraint is
derived under the approximation that volume remains larger than the Planck volume\footnote{As explained in Sec.\ \ref{sec:intro}, 
the volume does not need to be exceedingly large than the Planck volume for the 
approximation to be valid.}
and the relative fluctuations of the initial state remain small compared to
unity throughout the evolution \cite{psvt}.

Using Hamilton's equation, the time derivative of volume can be easily computed:
\be
\dot V = - 4 \pi G \gamma \f{\partial}{\partial \beta} \heff = \f{3}{2 \gamma \lambda} \, \sin(2 \lambda \beta) V, \\
\ee 
where $\lambda$ is the same value as in the quantum theory: $\lambda = 2
(\sqrt{3} \pi \gamma)^{1/2} \lp$. Using the effective Hamiltonian
constraint given in eq.\ (\ref{eq:heff}), and the expression for energy
density $\rho = p_\phi^2/(2\,V^2)$ we obtain the following modified Friedmann
equation
\be
  H^2=\left(\f{\dot V}{V}\right)^2 = \f{8\pi
G}{3}\rho\left(1-\f{\rho}{\rho_{\rm max}^{\rm(eff)}}\right),
\ee
where $H$ is the Hubble rate and
$\rho_{\rm max}^{\rm(eff)}$ is the
upper bound on the energy density given as
\be\label{rhomax_eff}
  \rho_{\rm max}^{\rm(eff)} = \f{3}{8\pi G \gamma^2\lambda^2}\approx 0.409\,\rhopl.
\ee
The modifications to the classical Friedmann equation, that are independent of the choice of state, 
present a very good approximation to the underlying quantum geometry, and the resulting 
trajectories agree very well with the full quantum evolution as long as the volume is larger than the 
Planck volume \cite{psvt}. It has also been argued that the potential corrections due to the state 
dependent quantum fluctuations in the non-compact model are negligible, 
in the limit of taking the volume of the fiducial cell to infinity \cite{rovelli_wilson}.

Hamilton's equation of motion for the matter field yields the following 
equation, which is equivalent to the conservation of the energy momentum tensor,
\be
\dot \rho + 3 H (\rho + P) = 0.
\ee
Here, $\rho$ is the energy density and $P=-\partial {\mathcal H}_{\phi}/\partial V$ is the pressure of the 
matter field. Using the modified Friedmann equation and the expression for the energy
density of the massless scalar field, it is straightforward to find that the
volume at the bounce is determined by the values of $p_\phi$ as follows:
\be\label{bounceeff}
V_{\rm b}^{\rm (eff)} = 1.1051 \, p_\phi\, V_{\rm Pl}/\sqrt{G}\hbar~.
\ee
Thus, the effective dynamics predicts that the bounce volume should decrease as
$p_\phi$ is decreased. This insight plays an important role in performing
numerical simulations in this paper. Later we will use the above value of 
$V_{\rm b}^{\rm (eff)}$ to define a quantity, $\delta$ which quantifies the departure 
between the effective theory and the quantum evolution in LQC as follows:
\be
\label{eq:delta}\delta= \f{({V_{\rm b}}  -{V_{\rm b}}^{(\rm eff)})}{{V_{\rm b}}^{(\rm eff)}}.
\ee

\section{Initial data}\label{sec:initialdata}
In the previous section we discussed the loop quantization of spatially flat isotropic FRW
spacetimes. We also discussed that in the large volume limit the LQC evolution equation
can be approximated by the Wheeler-DeWitt evolution equation. In this section we will utilize 
this property of the LQC evolution equation to construct the initial state for the numerical
evolution. In all the simulations, the initial conditions are provided so that the initial energy
density is very small and the initial state is peaked at a classical trajectory.
Therefore the initial state can be described by a Wheeler-DeWitt state, as the
Wheeler-DeWitt equation is the continuum limit of the LQC difference equation. A general 
solution of the Wheeler-DeWitt equation (\ref{eq:wdw}) is 
\be
\label{eq:wdwsol}\underline{\Psi}(v,\,\phi) = \int_{-\infty}^{+\infty} \Psi(\omega) \underline{e}_\omega(v)e^{i\omega(\phi-\phi_o)}\,d\omega,
\ee
where $\Psi(\omega)$ is the wave profile in $\omega$ space, and $e_{\omega}(v)$ are the eigenfunction of the ${\underline{\widehat{\Theta}}}$ operator (\ref{eq:wdw}):
\be\label{wdw_eigen}
\underline{e}_{\omega}(v)=\f{1}{2\pi}e^{i\,\omega\,\ln(v)/\sqrt{12\pi G}} ~.
\ee
Given a form of the wave profile, one can evaluate the above integral 
(numerically or analytically where possible) to obtain the form of the solution 
$\underline{\Psi}(v,\,\phi)$ at a given value of the `emergent time'
$\phi$ and construct an initial state for the evolution in LQC.

 In this paper, we
construct the initial data in form of a Gaussian wavepacket.
Such a class of initial state is characterized by three main parameters:
(i) the volume at which the initial state is peaked $v^*$
(ii) the scalar field momentum $p_\phi$ and
(iii) the spread of the state $\sigma$ or $\sigma_v$.
The value of $v^*$ is kept large enough to make sure the classicality of
the initial trajectory. Here we will consider three different types of initial data,
based on Ref. \cite{aps2}. The first
is a Gaussian in $v$, and the second and third types of initial states are Gaussian in
$\omega$, characterized by $\Psi(\omega)$ as follows
\be\label{psi_profile}
\Psi(\omega) = e^{-\left(\omega-\omega^*\right)^2/2\sigma^2},
\ee
where $\omega=p_\phi/\hbar$, $\omega^*$ is the value of $\omega$ at which the initial state
is peaked and $\sigma$ represents the spread of the Gaussian waveform. The value of 
$\omega^*(=p_\phi^*/\hbar)$ governs the value of the bounce volume.
In the effective theory the bounce volume can be exactly predicted to be
$v_{\rm b} \approx 0.32\, p_\phi^*/\sqrt{G}\hbar$ \cite{aps3}. Thus, larger $p_\phi^*$ corresponds to larger 
volume at the bounce. The relative spread in volume $\Delta v/v$ is inversely
proportional to the spread in  the field momentum $\sigma$.
Hence, by changing $\sigma$ in the initial data, we have a control
over the relative spread in volume, of the initial state.

We now discuss the three types of initial states considered in this paper.
\subsection{Method-1: Gaussian in volume (v)}
In this initial data construction method, we choose the
wavefunction to be a Gaussian in $v$, peaked at large volume $v^*$. Such a 
state at a given value of the scalar field $\phi=\phi_o$ can be written as
\be
\underline{\Psi}(v)|_{\phi=\phi_o} = e^{-\f{(v-v^*)^2}{2\sigma_v^2}}\ e^{ib^*(v-v^*)}.
\ee
where $b$ is the variable conjugate to $v$ and $\sigma_v$ is the spread of the Gaussian 
waveform. Here, $b$ is related to $\beta$ in \eref{eq:heff} as 
$\beta = 6\,b\,K\sqrt{3/16\pi G\hbar \gamma}$. Using the classical Hamiltonian constraint, 
the $\phi$ derivative of the above wavefunction can be shown to be
\be
\label{derivsm1}\f{d}{d\phi} \underline{\Psi} (v, \omega)|_{\phi_o} = \sqrt{12 \pi G} \Bigg[\f{v(v-v^*)}{\sigma_v^2}+ib^* v\Bigg] \underline{\Psi}(v)|_{\phi=\phi_o},
\ee
where $b^*$  is the initial value of $b$, given by
\be
b^* =\pm \f{\omega^*}{\sqrt{12\pi G}v^*}.
\ee
Substituting the value of $b^*$ in \eref{derivsm1} we obtain the ``time'' derivative
of the initial state and hence have the complete set of initial conditions
needed in order to study the evolution.
Note that the above relation between $b^*$ and $\omega^*$ holds true only in the
classical limit, i.e.\
when the LQC difference equation can be safely approximated by the Wheeler-DeWitt equation.
To satisfy this condition one has to make sure to choose $v^*$ large enough.

\subsection{Method-2: Wheeler-DeWitt initial state}
We consider a solution of the Wheeler-DeWitt equation given by the integral 
\eref{eq:wdwsol} as the second initial data method with the profile given by eq.
(\ref{psi_profile}). Using the eigenfunctions for the Wheeler-DeWitt evolution operator 
(\ref{wdw_eigen}), the integral in \eref{eq:wdwsol} yields the following form of the 
wave-function
\be
\underline{\Psi}(\phi, v) = \f{\sigma}{\sqrt{2 \pi}} \exp\Bigg(-\f{1}{2} \left(\f{\ln\left(v/v^*\right)}{\sqrt{12\pi G}}-\phi\right) \left( \sigma^2\f{\ln\left(v/v^*\right)}{\sqrt{12\pi G}} -\sigma^2\phi - 2 i\omega^* \right)\Bigg)~~.
\ee
The initial state given above is a continuous function of $v$, but in the numerical simulations,
since the spatial grid in LQC is discrete with uniform discreteness $\Delta v=4$, we compute
the states only at the discrete points on which an LQC state has support on. In order to provide all the necessary data for evolution,
one also needs the value of the time derivative of state
$\partial_\phi \Psi|_{\phi=\phi_o}$, which is given by
\be
\partial_\phi \underline{\Psi}(\phi, v) = \left(\sigma^2\f{\ln(v/v*)}{\sqrt{12\pi G}} - \sigma^2\phi - i \omega_o\right) \Psi(\phi, v).
\ee

\subsection{Method-3: Rotated Wheeler-DeWitt initial state}
In this initial data construction method we consider a variation of method-2 described above. The initial state is obtained by rotating an initial Wheeler-DeWitt state by an
$\omega$ dependent phase factor by multiplying  the eigenfunctions of the Wheeler-DeWitt
operator with $e^{-i\alpha}$, where $\alpha$ is given by \cite{aps2}
\be
\alpha = k\, (\ln(|k|)-1)
\ee
with $k=-\omega/\sqrt{12\pi G}$. This phase is introduced to match the eigenfunctions of the 
LQC evolution operator with those of the Wheeler-DeWitt operator in the large volume 
regime. The initial state is then obtained by numerically evaluating the following integral
\be
\underline{\Psi}(v, \phi) = \int dk\, \Psi(k) e^{-i\alpha} \underline{e}_{k}(v) e^{i\omega(\phi-\phi_o)}
\ee
with $\Psi(k)$ chosen to be a Gaussian peaked on $k^*=-p_\phi^*/\sqrt{12\pi G}$ with a spread $\sigma_k=\sigma/\sqrt{12\pi G}$:
\be
\Psi(k)=e^{-(k-k^*)/2\sigma_k^2}.
\ee
Further, the time derivative of the initial state can be
numerically evaluated as follows
\be
\partial_\phi\underline{\Psi}(v, \phi) = \int dk\, i\omega\, \Psi(k) e^{-i\alpha} \underline{e}_{k}(v) e^{i\omega(\phi-\phi_o)}.
\ee

\subsection{Uncertainty products and relative dispersions for method 1, 2 and 3 initial states }

The state dependent properties of the evolution is captured by the
product of the uncertainties in $V$ and
$p_\phi$ (which are related through the uncertainty relation between $\phi$ and $p_\phi$).
An important feature of the choice of initial state is
the symmetry of the dispersion of the state across the bounce. It turns
out (also noted previously in Refs. \cite{aps2,aps3}) that the relative dispersion in $V$ may or may
not be symmetric on the two sides of the bounce depending on the type of
initial state and the dispersion in $p_\phi$.

In the Wheeler-DeWitt theory the time parameter `$\phi$' is related to the volume  $V$ via the following relation\footnote{This result is straightforward to derive using the classical evolution equations. The Wheeler-DeWitt states are peaked on this classical trajectory (see Refs. \cite{aps2,aps3,dgs1} for details).}
\be
\phi = \f{1}{\sqrt{12\pi G}} \ln\left(\f{V}{V_o}\right) + \phi_o.
\ee
Using this relation we can obtain a relation between the
dispersion in $\phi$ and the relative dispersion in
volume $\Delta V/V$ as follows
\be
\label{dphiwdw}\Delta \phi = \f{1}{\sqrt{12\pi G}} \f{\Delta V}{V}.
\ee
The product of the uncertainties of the matter sector  is
given as
\be
\Delta\phi\Delta p_\phi \ge \f{\hbar}{2}.
\ee
Since, the evolution equations are de-parameterized with respect to the
scalar field $\phi$ which plays the role of emergent time, we can write
the above uncertainty product as the product of the
relative dispersion in volume $\Delta V/V$ and the dispersion in the field momentum
$\Delta p_\phi$ as follows
\be
\f{\Delta V}{V}\Delta p_\phi \ge \sqrt{3\pi G} \hbar.
\ee
Thus, for a minimum uncertainty initial state the above inequality turns to the following equality
\be\label{min_uncertainty_product}
\bigg(\f{\Delta V}{V}\Delta p_\phi\bigg)_{\rm min}=\sqrt{3\pi G}\hbar.
\ee

The initial state for method-2 is a minimum uncertainty state by construction \cite{aps2},
and the above equality holds for all values of $\Delta p_\phi$ and $\Delta V/V$.
Let us now consider a particular choice $\widetilde{\Delta p_\phi}$ and $\widetilde{\f{\Delta V}{V}}$ such that the relative
dispersions in volume and the field momentum are equal. Together with eq.(\ref{min_uncertainty_product}), this yields the following relation for method-2
\be
\f{\widetilde{\Delta p_\phi}}{{p_\phi}}=\widetilde{\f{\Delta V}{V}}=\left(\f{\sqrt{3 \pi G}\hbar}{p_\phi}\right)^{1/2}
\label{eq:uncertm2}
\ee
so that
\be\label{tildedeltapphi}
 \widetilde{\Delta p_\phi} = (3\pi p_\phi^2 G\hbar^2)^{1/4}, ~~~ \mathrm{and} ~~~ \widetilde{\f{\Delta V}{V}}=\left(\f{3\pi} {p_\phi^2} G\hbar^2\right)^{1/4}. 
\ee
It is straightforward to verify that
for $\widetilde{\Delta p_\phi} $ and $\widetilde{\f{\Delta V}{V}}$ the uncertainty product remains minimum during the evolution.
For method-1 and method-3 initial data on the other hand,
if we choose $\Delta p_{\phi}=\widetilde{\Delta p_\phi}$ the resulting $\Delta V/V$ will not be equal to $\widetilde{\Delta V}/V$, i.e. the state is not a
minimal uncertainty state.
Further, the numerical simulations
show that the following turns out to be true for method-1 and method-3:
\begin{itemize}
	\item For method-1: The input parameter in this initial data method is $\sigma_v$ which
	  directly determines $\Delta V/V$. Moreover, $\Delta p_\phi$ is a non-monotonic function of $\f{\Delta V}{V}$
	  and becomes minimum at $\widetilde{\f{\Delta V}{V}}=\left(\f{3\pi} {p_\phi^2} G\hbar^2\right)^{1/4}$.
	
	\item For method-3: The input parameter for this method is $\sigma$ which directly determines
	$\Delta p_\phi$ and $\f{\Delta V}{V}$ is non-monotonic function of $\Delta p_\phi$. It turns out that
        at $\widetilde{\Delta p_\phi} = (3\pi p_\phi^2 G\hbar^2)^{1/4}$, $\f{\Delta V}{V}$ is minimized.
\end{itemize}

The non-monotonic nature of the relative volume dispersion for a
Gaussian state was earlier found and studied in Ref. \cite{montoya_corichi2,montoya_corichi1}, where the state considered was different from the initial states discussed in this work. However, the value of $\Delta p_\phi$ which minimizes
the relative volume dispersion turns out to be the  same as we find in our analysis.
We will discuss later that $\widetilde{\Delta p_\phi}$ and  $\widetilde{\f{\Delta V}{V}}$ play
important roles in understanding the behavior of the deviation of the effective trajectory from the LQC
trajectory. It turns out that these values are directly related to the non-monotonic behavior of the
quantity that measures the deviation of effective theory from LQC.

\section{Implementation of the Chimera scheme}\label{sec:chimera}
Our goal in this analysis is to consider initial states with a wide variety of values of $p_\phi^*$ and dispersions, including those with small $p_\phi^*$, and those with a large dispersion in volume. Evolution with such states present severe  computational challenges. They both require the boundary of the
spatial grid to be sufficiently large in order to contain the state
within the computational domain during the evolution. Due to the Courant-Friedrichs-Lewy (CFL) stability
condition, for a stable evolution, a larger spatial grid requires a smaller time step \cite{cfl}.
Thus, if the state is widely spread, the time steps in the evolution have to be
small, which leads to a larger computation time. Note that the discretization in the quantum difference equation (\ref{eq:lqcevol}), denoted by  $\Delta v$, is 4 times the Planck volume and is fixed by the underlying quantum geometry. 
Since the discreteness in the spatial grid can not be changed, the computational cost 
grows quadratically with the location of the outer boundary. Therefore, it is difficult to study the numerical evolution of very widely spread states
with the numerical approaches used in Refs. \cite{aps1,aps2,aps3}.
To circumvent this problem we use a hybrid numerical scheme, named Chimera, which was introduced by the authors recently \cite{dgs1}. It is an explicit time integration scheme based on the idea of a hybrid spatial grid which 
takes advantage of the fact that the LQC difference equation can be
approximated by the Wheeler-DeWitt equation at large volume. Therefore, we can divide the entire
domain into two parts: the inner grid on which we solve the LQC difference equation
and an outer grid where we solve the Wheeler-DeWitt equation. On the outer grid we take
advantage of the fact that the Wheeler-DeWitt equation has a continuum limit allowing
us to transform to numerically better behaved coordinates, so that the time
step no longer depends on the outer boundary location. Below we summarize the way this scheme is implemented. A more elaborate
discussion of the scheme including several robustness tests has been provided
in our previous article, which is focused on the details of the numerical
properties of the scheme \cite{dgs1}.

Let us denote the boundary of the inner and the outer grids as $v_{\mathrm{interface}}$. This boundary is chosen at a volume large enough, where the LQC difference equation is extremely well approximated by the Wheeler-DeWitt equation.  On the inner grid, where the LQC difference equations are evolved,
 we use $v$ as our variable with discreteness\footnote{In the discussion in this section, $\Delta$ should not be confused with the symbol used for dispersion else where in this manuscript.} $\Delta v = 4$. On the outer grid we use
$x = \ln v$ as our variable.
This scheme now allows us to place the outer boundary at very large
values of $v_{\mathrm{outer}} = \exp (x_{\mathrm{outer}})$ without (i) having to
use extremely large number of grid points, and (ii) having to use an extremely small time step in $\phi$. This leads to a very large reduction in computational cost and makes the numerical evolution of states with small $p_\phi^*$, and those with 
large dispersion in volume practically feasible.

In the Chimera scheme, it is possible to use different methods to solve the
Wheeler-DeWitt equation, which is a partial differential equation, on the outer grid.
In our previous article \cite{dgs1}, we introduced two separate
implementations of the Chimera scheme:
(i) a Finite Difference (FD) implementation where we use standard centered
second order finite differencing on the outer grid, and
(ii) a Discontinuous Galerkin (DG) implementation where we use the DG method to solve
the Wheeler-DeWitt equation on the outer grid.
We found that the DG implementation of the Chimera scheme is an order of
magnitude more efficient than the FD implementation, since the higher accuracy
provided by DG requires less grid points.
Using the Chimera scheme we can perform the simulation of a widely spread state
with $p_\phi^*=20\sqrt{G}\hbar$, which would have taken $~10^{27}\, {\rm years}$,
in $1118\, {\rm seconds}$ using the FD implementation with $253,953$ grid points on the outer grid, and in $144\,{\rm seconds}$ using the DG implementation with $8,125$ grid points on the
outer grid, while, in both cases, choosing $7,500$ grid points on the inner grid. A numerical
evolution with twice the resolution takes $9148\,{\rm seconds}$ using the FD implementation and $472\,{\rm seconds}$ using the DG implementation.
(For a detailed discussion of
the efficiency of the scheme and its performance, see Ref. \cite{dgs1}). Since the DG implementation is more efficient than the FD implementation we use it
exclusively for the simulations presented here.

\section{Evolution of initial states}\label{sec:evolution}
With the Chimera scheme at our disposal, we can now extend
the numerical simulations to initial data parameter regimes, which were not 
accessible in previous numerical investigations. We have performed a large number of 
simulations using initial states from the 3 different initial data methods 
described in Sec. IV.  In this section our goal is 
to discuss a sample of these results, obtained from evolution of initial states peaked at large and 
small values of $p_\phi^*$ and different values of dispersion in volume, as a demonstration of the 
implementation of the Chimera scheme and to provide a glimpse  of  some of the results. Detailed 
discussions of the simulations performed with the different methods and various results is provided in the 
next section.  
\begin{figure}[tbh!]
\includegraphics[width=0.75\textwidth]{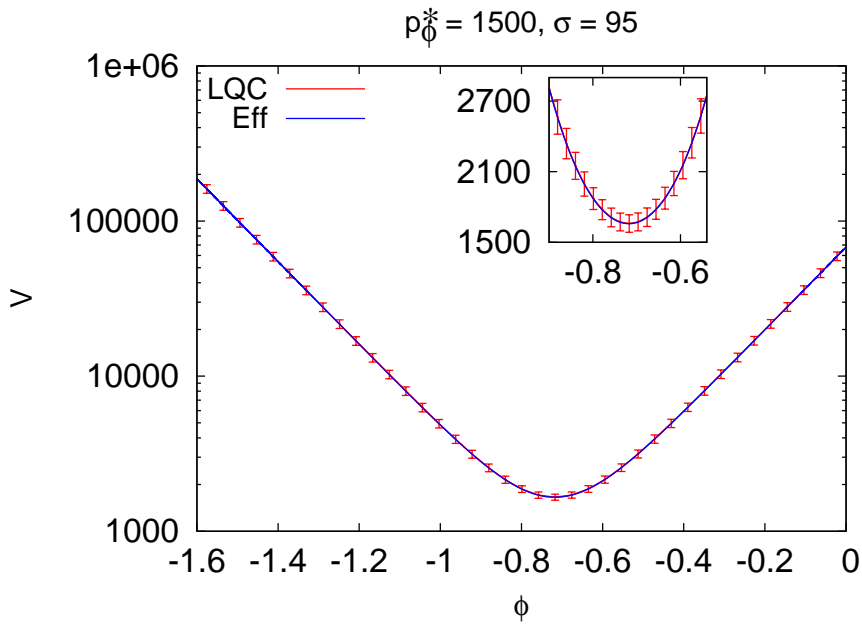}
\caption{The plot shows the evolution of the expectation value and dispersion of the
volume for an initial state peaked at a large value of $p_\phi$: 
$p_\phi^*=1500\sqrt{G}\hbar$ and $\sigma=95$ for method-3 initial data. The
dispersion is depicted with error bars.
The bounce happens far away from the Planck volume and the effective trajectory 
is an excellent approximation to the LQC trajectory.}
\label{fig:largepphi}
\end{figure}
In the following subsections,
we first compare the expectation values of volume in LQC and the effective theory, for the case of 
initial states constructed with method-3. In the subsequent part of this section, we discuss the 
results corresponding to some values of $p_\phi^*$  for relative volume dispersion, and the 
constraints on the fluctuations in pre and post bounce phases for all the methods.  
As discussed before,
the initial state is characterized by three main parameters: the field momentum at which it is 
peaked $p_\phi^*$, the width of the state in the field momentum $\sigma$ and the volume $v^*$ 
at which the initial state is peaked. Here we will consider a variety of values of $p_\phi^*$ and 
$\sigma$ while keeping the value of the volume to be large. A large value of $v^*$ is taken 
to ensure that the initial state is peaked at a classical trajectory. The value of $p_\phi^*$ and 
$\sigma$ on the other hand govern whether the state is sharply peaked. The larger the value of 
$\sigma$, the more sharply peaked the state is in $v$\footnote{In the following the results are 
plotted and discussed in terms of the physical volume $V$, where as for the evolution we use the 
variable $v$ which is related to $V$ via eq. (\ref{V_eigen}).}.
\begin{figure}[tbh!]
  \subfigure[]
  {
     \includegraphics[width=0.75\textwidth]{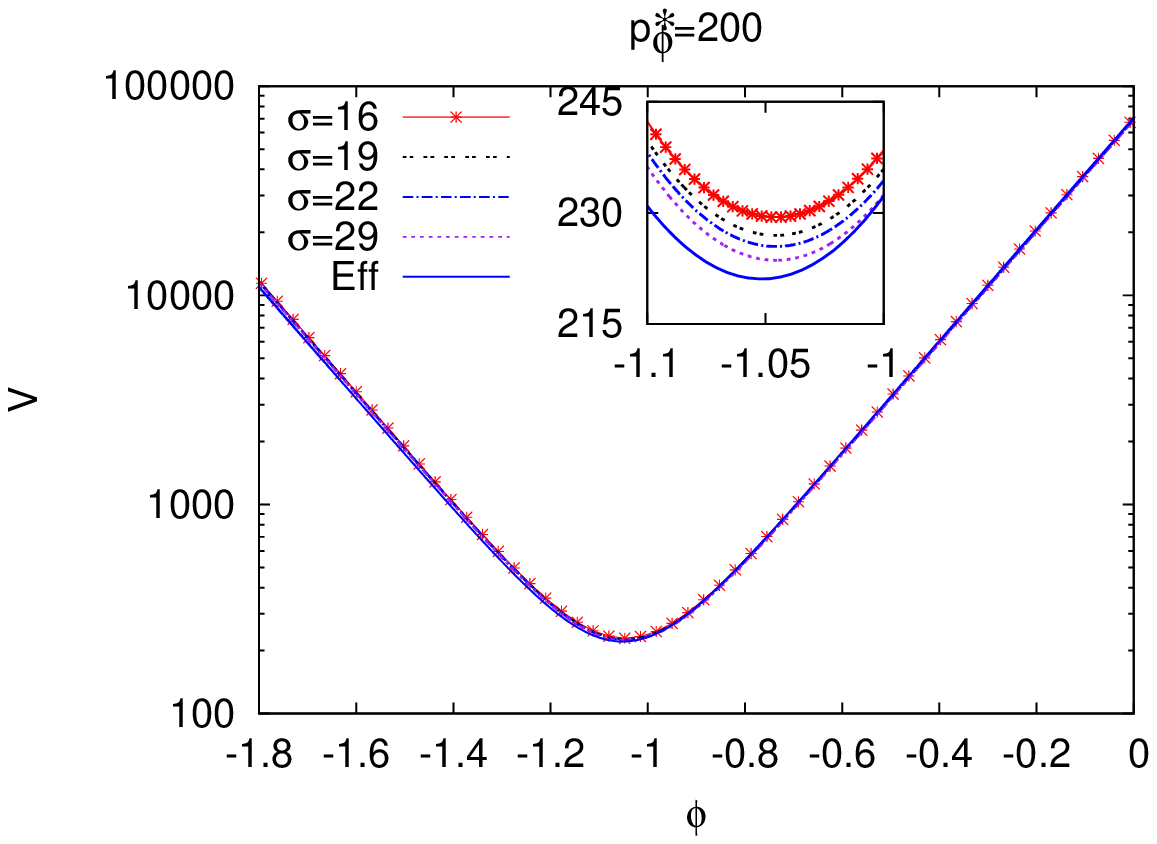}
  }
  \subfigure[]
  {
     \includegraphics[width=0.75\textwidth]{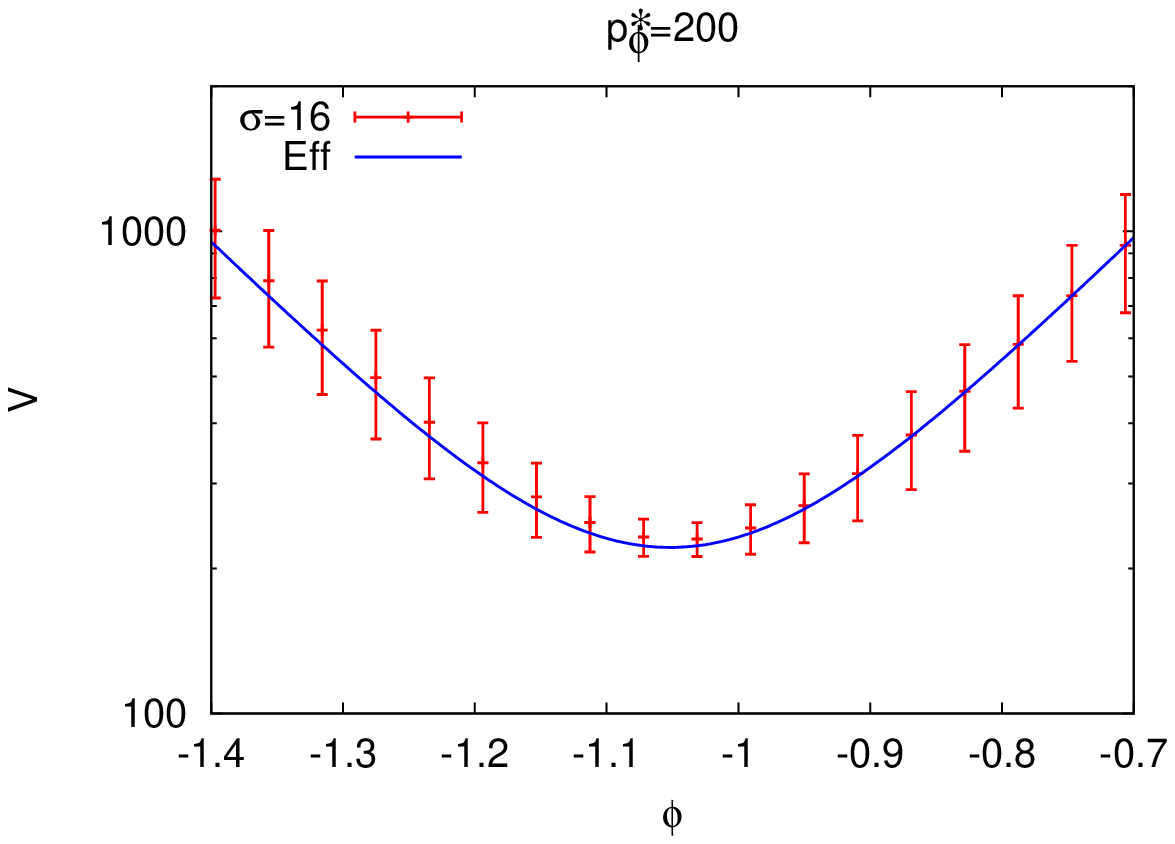}
  }
  \caption{The expectation value of the volume $V$ and the associated dispersion  $\Delta V$ (shown as errorbars), for 
  $p_\phi^*=200\sqrt{G}\hbar$ for different values of $\sigma$ is plotted for method-3 initial data in panel (a).
  We find  that the bounce happens much closer to the Planck 
  volume than in \fref{fig:largepphi} (for $p_\phi^*=1500\sqrt{G}\hbar$)
  and the effective trajectory shows small deviations from the LQC
  trajectory (larger for smaller values of $\sigma$).
  In panel (b), we see (for $\sigma=16$ only) that although there
  are differences between the LQC and effective trajectories, the effective theory lies 
  within the dispersions. These trajectories correspond to $\Delta p_\phi<\widetilde{\Delta p_\phi}$,
  for which $\sigma<\widetilde{\sigma}=\sqrt{2}\widetilde{\Delta p_\phi}=35.04$ for 
  $p_\phi^*=200\sqrt{G}\hbar$. In this regime, increasing $\Delta V/V$ (decreasing $\sigma$) 
  results in larger deviation from the effective theory,
  as discussed further in Table-\ref{tab:summary}.}
  \label{fig:expectation1}
\end{figure}
\begin{figure}[tbh!]
  \subfigure[]
  {
     \includegraphics[width=0.75\textwidth]{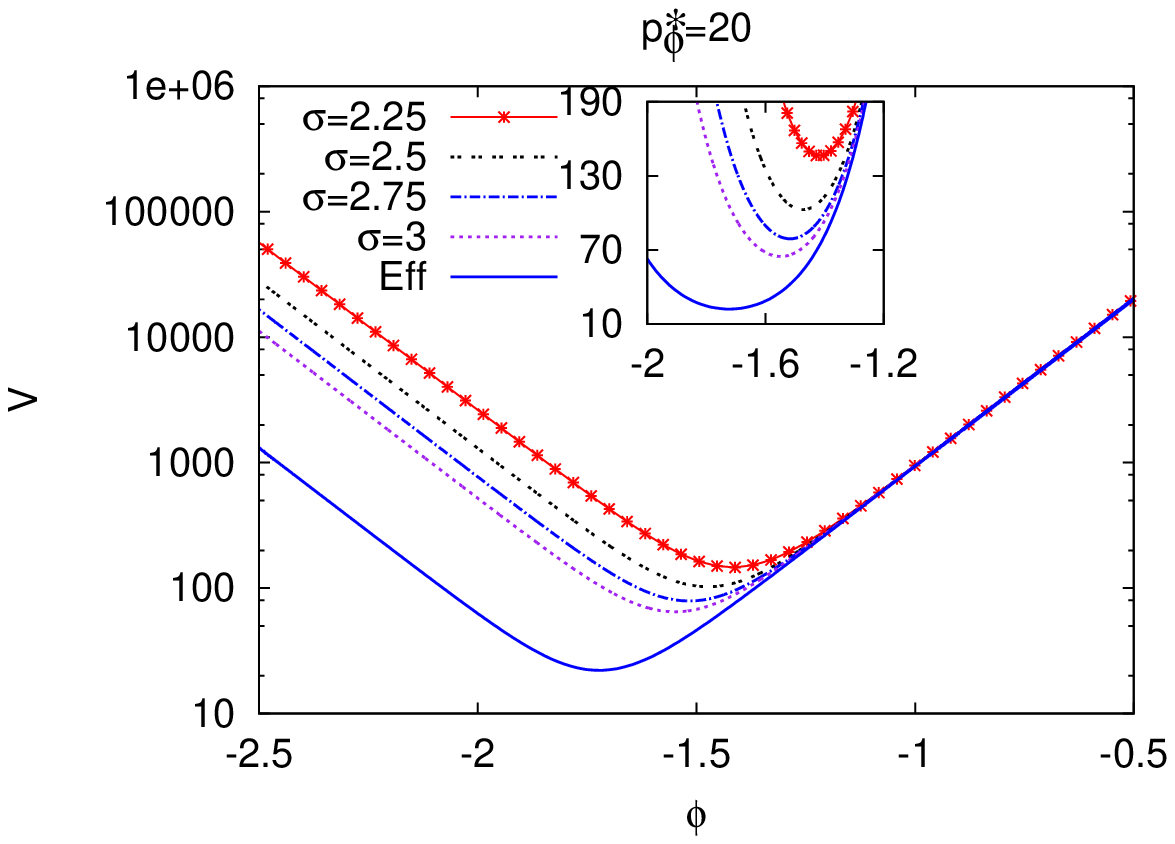}
  }
  \subfigure[]
  {
     \includegraphics[width=0.75\textwidth]{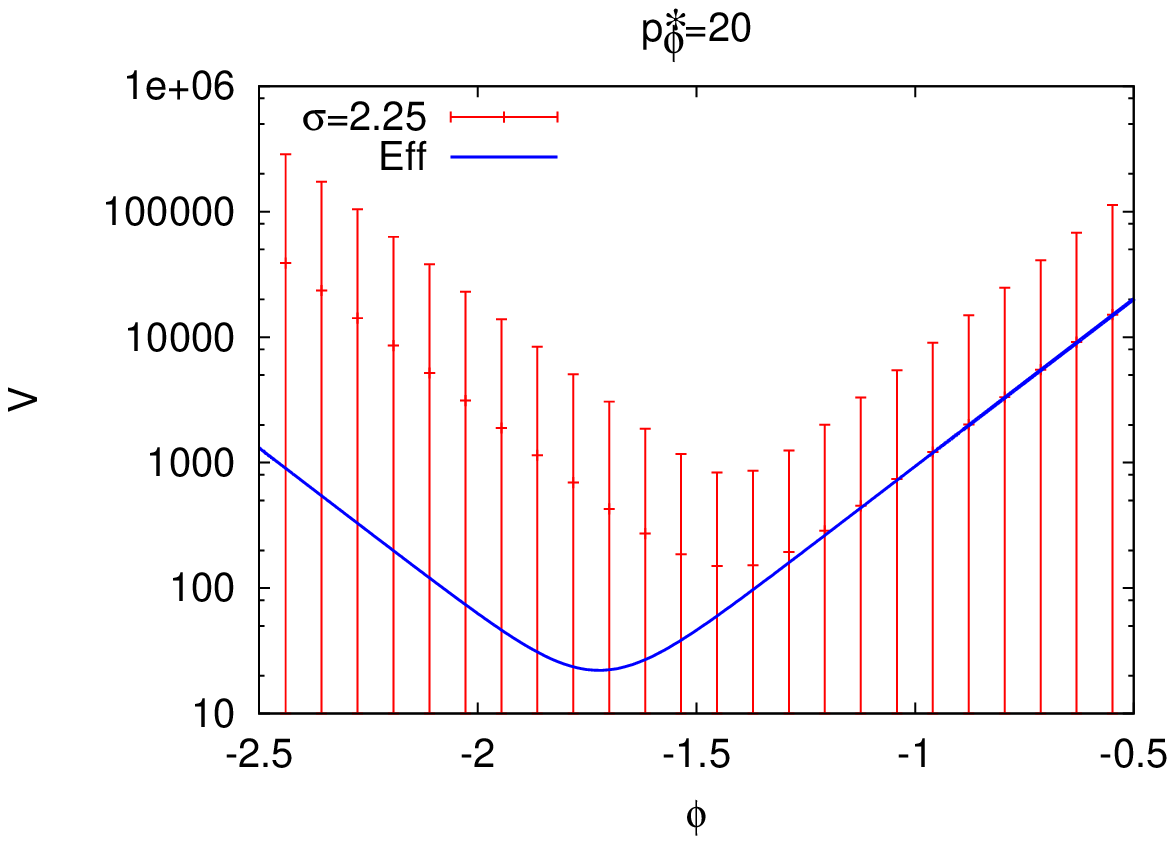}
  }
  \caption{The expectation value of the volume $V$ and the associated dispersion  $\Delta V$ (shown as errorbars), for $p_\phi^*=20\sqrt{G}\hbar$ for different values of $\sigma$ is plotted for method-3 initial data in panel (a).
  We find that the bounce happens much closer to the
  Planck volume than in \fref{fig:largepphi} (for $p_\phi^*=1500\sqrt{G}\hbar$)
  and there are significant differences between the effective trajectory and the LQC
  trajectories for various field dispersions (larger for smaller values of $\sigma$). 
  Panel (b) shows the plot of the expectation value of volume, associated volume dispersion,
  and the corresponding effective trajectory for $\sigma=2.25$. It is evident that
  although there are differences between these trajectories, the effective
  theory lies within $\Delta V$.  These trajectories correspond to 
  $\Delta p_\phi<\widetilde{\Delta p_\phi}$, for which
  $\sigma<\widetilde{\sigma}=\sqrt{2}\widetilde{\Delta p_\phi}=11.08$ 
  for $p_\phi^*=20\sqrt{G}\hbar$.
  As discussed in Table-\ref{tab:summary}, increasing $\Delta V/V$ (decreasing $\sigma$) 
  results in larger deviation from the effective theory, in this regime.}
  \label{fig:expectation2}
\end{figure}

\subsection{Expectation value of $V$}
\label{subsec:expv}
\fref{fig:largepphi} shows the evolution of the expectation value and dispersion of the
volume observable for the initial state constructed with method-3 peaked at 
$p_\phi^*=1500\sqrt{G}\hbar$ and with $\sigma=95$. We make a comparison with the trajectory 
obtained from the effective spacetime description (blue solid curve). It can be seen that the
expectation value of the volume is in excellent agreement with the effective 
evolution throughout the evolution including the bounce point at volume $V_{\rm b} \approx 1660\,\lp^3$,
which according to eq.(\ref{bounceeff}) occurs at approximately 
1658 $\lp^3$ in the effective theory. In this particular simulation, the
properties of the state are compatible with the assumptions made in the
derivation of the effective theory -- the state is sharply peaked and the volume never gets close to the Planck volume during evolution. Therefore, the agreement of the LQC evolution
with the effective dynamical trajectory is expected.

We now discuss the cases when we decrease the value of $p_\phi^*$. We choose two representative values $p_\phi^*=200\sqrt{G}\hbar$ and $p_\phi^*=20\sqrt{G}\hbar$.
In these simulations, the spread $\sigma$ for each of these
values is kept in the range of $8-15\%$ of $p_\phi^*$.
We find that 
a smaller value of $p_\phi^*$ leads to a smaller bounce volume, and
as the spread in $p_\phi^*$ decreases, the relative spread in volume of the
state increases, leading to deviation from sharp peakedness.
\fref{fig:expectation1} (corresponding to $p_\phi^*=200\sqrt{G}\hbar$) and \fref{fig:expectation2}
(corresponding to $p_\phi^*=20\sqrt{G}\hbar$) show the evolution of the expectation
value and dispersion of the volume observable, and the respective effective trajectories. For these particular cases, as the value of
$p_\phi^*$ is decreased, the deviation between the effective and LQC trajectory increases. We will later see that this is not a generic feature independent of the initial state construction method.
The differences between the effective and the LQC trajectory
are in particular quite prominent for $p_\phi^*=20\sqrt{G}\hbar$, where the value of the
spread $\sigma$ in our simulations ranges between $2.25$ and $3.0$. The volume at the bounce for
this $p_\phi^*$, in the effective theory, is $V_{\rm b}^{(\mathrm{eff})} \approx  22\, \lp^3$, whereas
the bounce volume for LQC trajectories are larger.
We make two important observations based on these simulations:
(i) there are dispersion dependent deviations in the LQC trajectory from
the effective trajectory, and
(ii) the deviations seem to grow as the spread $\sigma$ is decreased for a
fixed $p_\phi^*$.
In the next section, we will discuss the way these results are affected with different choices of initial state construction. For a quick comparison see Table~\ref{tab:summary}. 
Note that the  LQC trajectories shown in \fref{fig:expectation1} and \ref{fig:expectation2},
correspond to a fixed $p_\phi^*$ with different values of $\Delta p_\phi$ such that
$\Delta p_\phi<\widetilde{\Delta p_\phi}$, defined via \eref{tildedeltapphi}. In this regime, as the value of $\sigma$
decreases, $\Delta p_\phi$ decreases which in turn causes the relative volume dispersion
to increase and leads to a larger deviation from the effective
trajectory. 

In Table \ \ref{tab:bouncevol}, we provide the values of the bounce volume for
different values of the field momentum $p_\phi^*$ and spread $\sigma$. It is clear
that the bounce volume depends more strongly on the spread of the state
for small values of $p_\phi^*$.
\begin{table}[tbh!]
\caption{The dependence of the value of the bounce volume for
different values of $p_\phi^*$ on the spread $\sigma$. The bounce volume $V_b$ in the quantum theory is in the Planck units.}
  \begin{tabular}{c c c c c c c c}
   \hline
   \multicolumn{2}{c}{$p_\phi^*=20\sqrt{G}\hbar$} & \multicolumn{2}{c}{$p_\phi^*=80\sqrt{G}\hbar$} & \multicolumn{2}{c}{$p_\phi^*=140\sqrt{G}\hbar$} & \multicolumn{2}{c}{$p_\phi^*=200\sqrt{G}\hbar$} \\
   $\sigma$ & $V_b$ & $\sigma$ & $V_b$ & $\sigma$ & $V_b$ & $\sigma$ & $V_b$ \\
   \hline
   \hline
    2.25  & 146.22 &    5.0  & 129.07 &  10.0   & 170.12 &  16.0   & 229.41 \\
    2.5   & 102.66 &    6.0  & 115.03 &  13.0   & 163.69 &  19.0   & 226.97 \\
    2.75  & 79.03 &    7.0  &  107.31 &  15.0   & 161.43 &  22.0   & 225.47 \\
    3.9   & 42.25 &    9.0  &  99.46 &  17.0    & 159.94 &  26.0   & 224.22 \\
   \hline
  \end{tabular}
\label{tab:bouncevol}
\end{table}

\begin{figure}[tbh!]
  \subfigure[]
  {
    \includegraphics[width=0.77\textwidth]{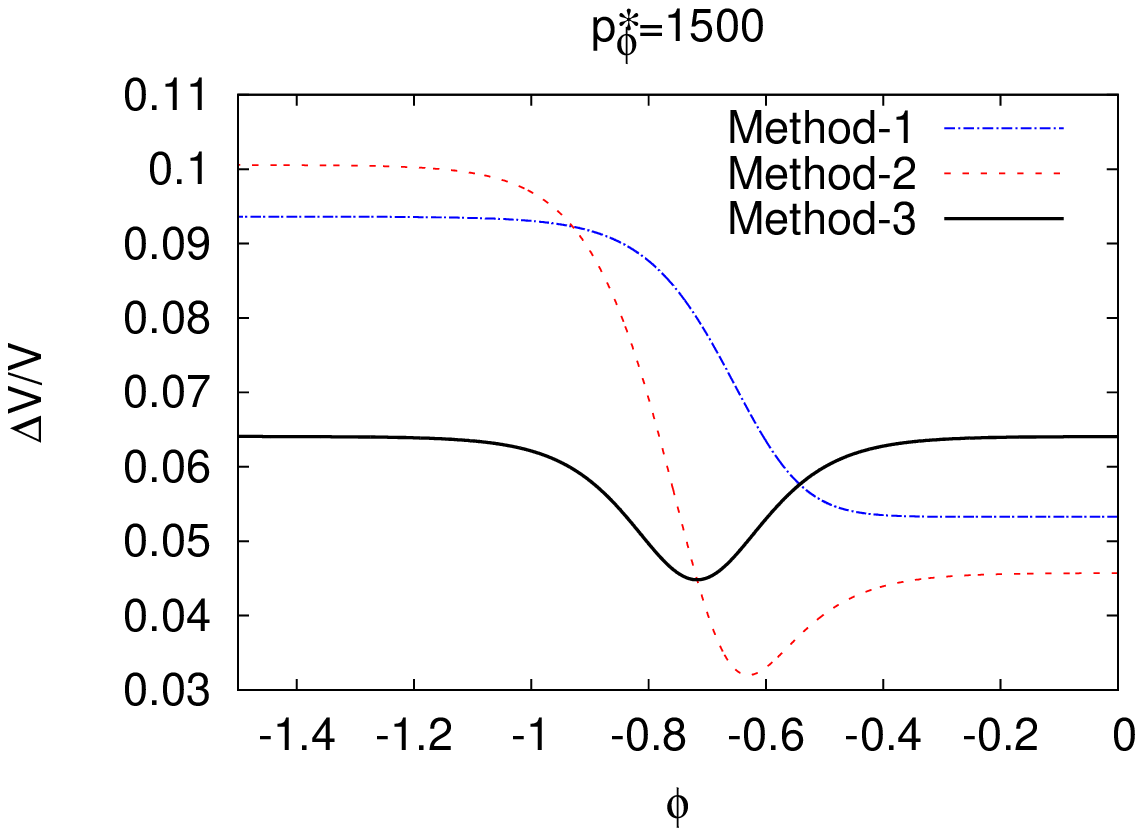}
    \label{fig:dispersion1}
  }
  \subfigure[]
  {
    \includegraphics[width=0.77\textwidth]{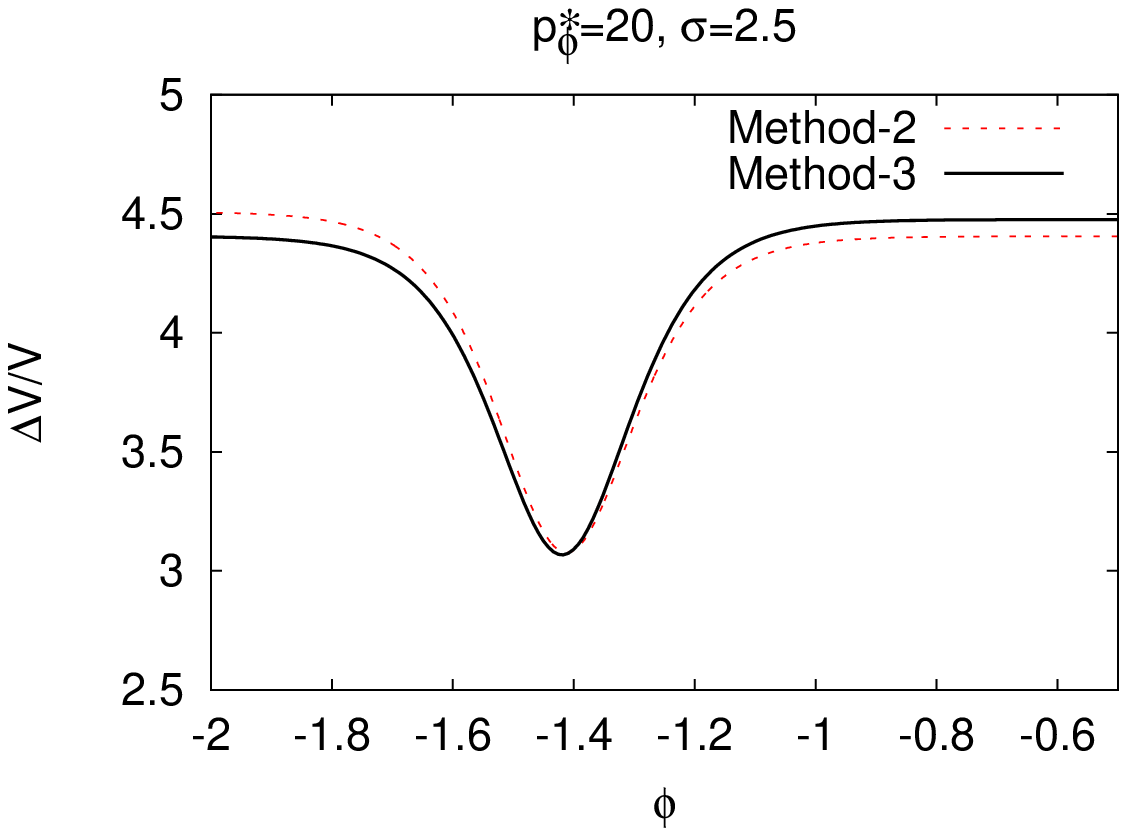}
    \label{fig:dispersion2}
  }
\caption{Evolution of the relative dispersion in volume $\Delta V/V$. Panel (a)
shows the evolution for $p_\phi^*=1500\sqrt{G}\hbar$ for initial data method-1 ($\sigma_v=1000$),
method-2 and method-3 (both with $\sigma=95$). Panel (b) shows the evolution of $\Delta V/V$
for $p_\phi^*=20\sqrt{G}\hbar$ $\sigma=2.5$ for initial data method-2 and 3.
}
\label{fig:dispersion}
\end{figure}

\subsection{Relative dispersion $\f{\Delta V}{V}$}\label{sec:relativedisp}
In this subsection we study the relation between the relative volume dispersion
and the relative dispersion in the field momentum, on the two sides of the bounce.
We will again consider the evolution of states with a wide variety of
initial conditions. These give rise to a wide range of initial values of the
relative dispersion.
The study of the evolution of the dispersion
of the states are important for several reasons, the most important one being the
issue of semi-classicality in the low curvature regime on the two sides of the
bounce \cite{Corichi:2007am,kp,montoya_corichi2}. These issues can also be analytically 
understood using triangle inequalities derived in Ref. \cite{kp}. A comparison of our results with the constraints from triangle inequalities are discussed in the next subsection.

\begin{figure}[tbh!]
   \subfigure[]
   {
    \includegraphics[width=0.77\textwidth]{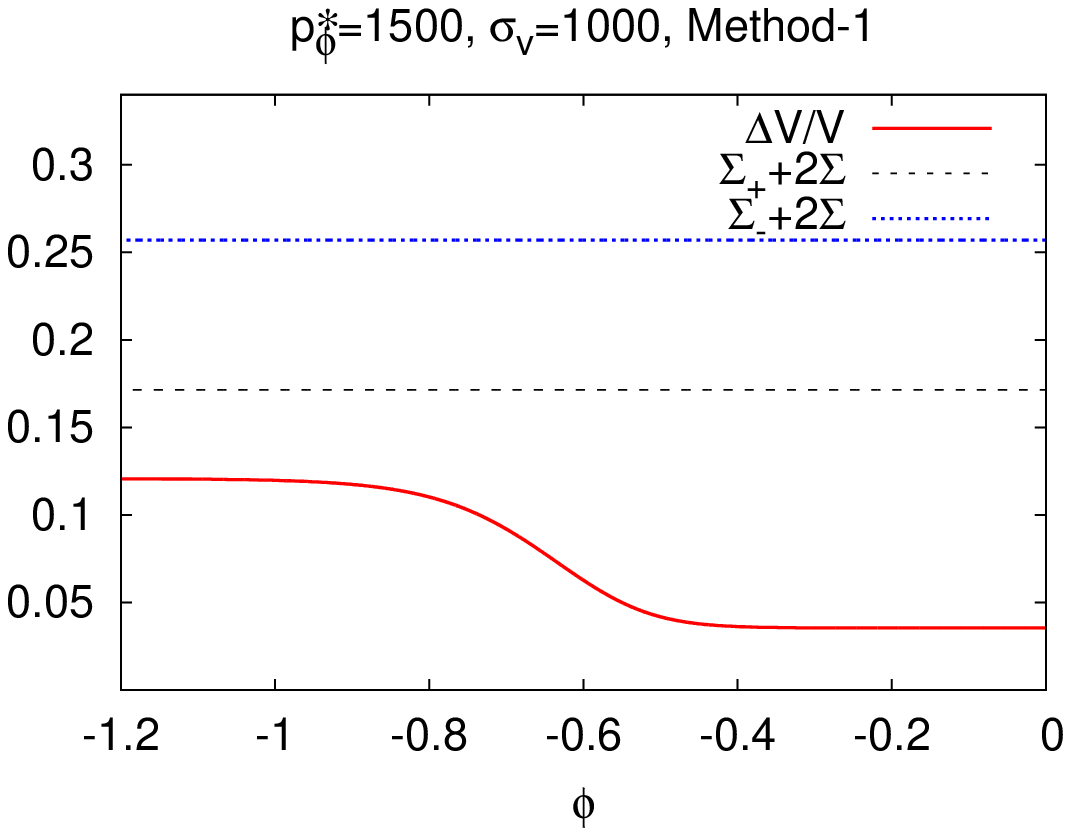}
    }
   \subfigure[]
   {
    \includegraphics[width=0.77\textwidth]{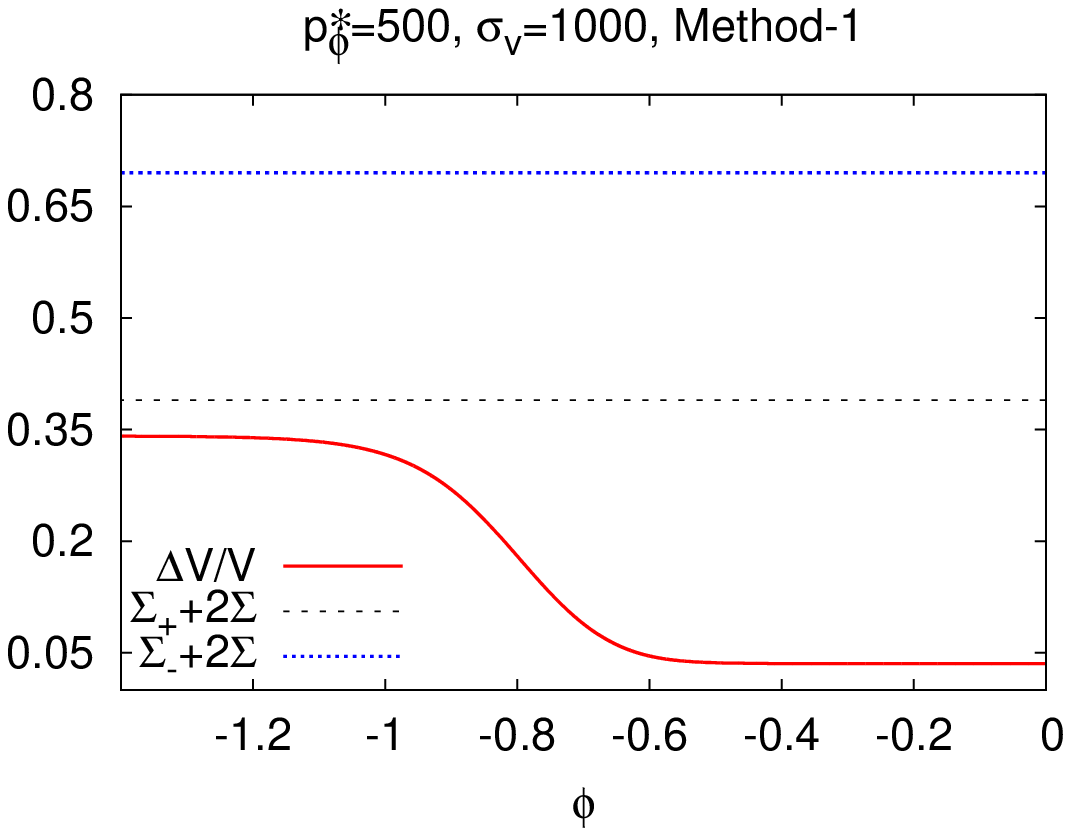}
    }
    \caption{Triangle inequalities for method-1 initial data for
    $p_\phi^*=1500\sqrt{G}\hbar$ (panel (a)) and $p_\phi^*=500\sqrt{G}\hbar$
    (panel (b)) and $\sigma_v=1000$ are shown to be valid  
     throughout the evolution.}
    \label{fig:trianglew1500m1}
\end{figure}

Let us first discuss the evolution of $\Delta V/V$ in LQC. \fref{fig:dispersion1}
shows the evolution of the relative volume dispersion for a state which is initially sharply
peaked on a classical trajectory. The value of the field
momentum is $p_\phi^*=1500\sqrt{G}\hbar$ for all initial data methods and
$\sigma_v=1000$ for method-1 while $\sigma=95$ for method-2 and 3. From
the figure it is clear that far
from bounce in the low curvature regime, $\Delta V/V$ tends to a constant value
in the asymptotic limit. In method-2 and method-3, as the state is evolved backwards, and the Planckian
curvature is reached, the relative volume dispersion decreases. As the evolution is continued backwards, it achieves a minimum value at some point and then starts
increasing again. For method-1 initial data, $\Delta V/V$ monotonically increases when evolved backwards. On the other side of the bounce, as the low curvature regime
is reached once again, the relative volume dispersion tends to a constant
value. The asymptotic
constant values on the two sides of the bounce, are equal if the initial state is
chosen according to method-3. For method-1
and method-2 initial data, they are different, in general. However, it turns out that for method-2, the
asymptotic values of the relative volume dispersion can be brought closer to
each other by
choosing appropriate values of the relative dispersion in the field momentum.

\begin{figure}[tbh!]
  \subfigure[]
  {
    \includegraphics[width=0.77\textwidth]{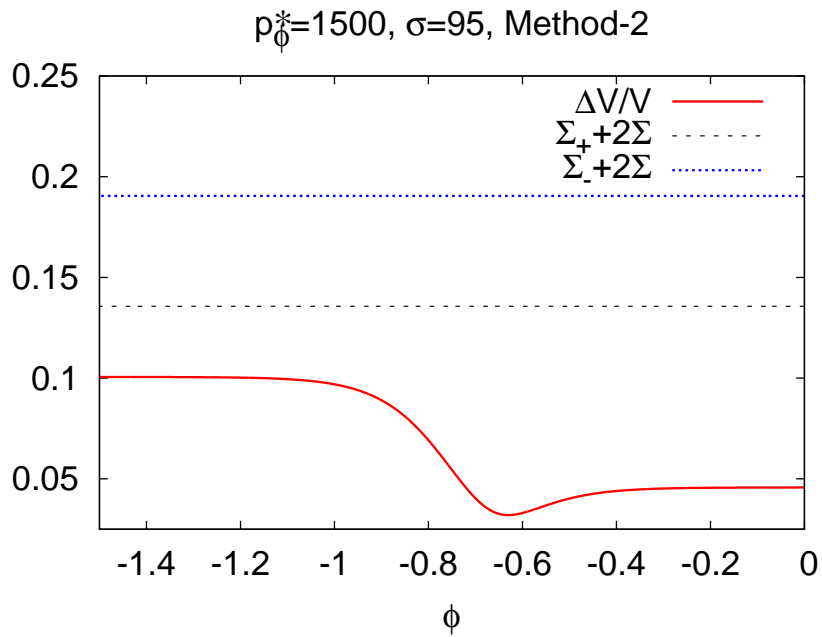}
  }
  \subfigure[]
  {
    \includegraphics[width=0.75\textwidth]{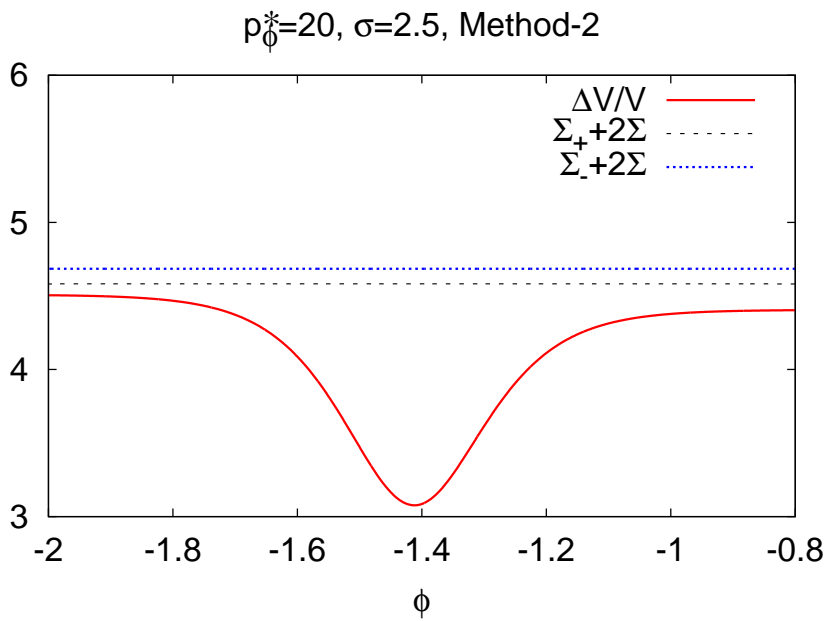}
  }
  \caption{This figure shows the evolution of the relative volume dispersion and the
  validity of the triangle inequality for $p_\phi^*=1500\sqrt{G}\hbar$ (panel (a)) and $p_\phi^*=20\sqrt{G}\hbar$ (panel (b)) and $\sigma=95$
  for method-2 initial data. We find that the triangle inequalities are valid throughout the evolution for both large and small
  $p_\phi^*$.}
  \label{fig:trianglew1500m2}
\end{figure}
\fref{fig:dispersion2} shows the evolution of the relative volume dispersion $\Delta V/V$ for
$p_\phi^*=20\sqrt{G}\hbar$ and $\sigma=2.5$ for method-2 and 3\footnote{For method-1, there are difficulties associated with a reliable 
construction of initial states for small values of $p_\phi^*$. These difficulties seem to be 
independent of the numerical procedure used for the evolution. Due to this reason, in our analysis 
small values of field momentum are not discussed for method-1, and we restrict ourselves to 
$p_\phi^* \geq 500$.\label{fn1}}.
As can be seen from the figure, the initial value
of the fractional dispersion in volume is as large as $4.3$. Due to such a large
dispersion in volume, the state is not sharply peaked.
However, the evolution across the bounce shows that the relative volume dispersion
again takes a constant value in the asymptotic limit on both sides of the bounce.
It is important to note that in both cases, i.e.\ as shown in \fref{fig:dispersion1}
and \fref{fig:dispersion2}, $\Delta V/V$ approaches a constant value in the
asymptotic limit, far from the bounce. Also, the value of the relative dispersion near
the bounce is smaller than these asymptotic values for method-2 and 3 initial data. 

\subsection{Triangle inequalities of the dispersion}
Let us now turn to the discussion of constraints on the difference of the relative volume dispersion on the two sides of the bounce.
A large disparity between the dispersion on the two sides may imply the loss of
semi-classicality as the universe evolves from one side of the bounce to the other.
Analytical investigations in this direction \cite{Corichi:2007am,kp,montoya_corichi2}, however,
establish that the state of the universe remains semi-classical if one starts
with a semi-classical state on one side of the bounce and evolves it across. 
According to
Ref. \cite{kp} the asymptotic values of the relative dispersions,
far from the bounce obey triangle inequalities given as follows
\be
\label{eq:triangleineq} \Sigma_- \leq \Sigma_+ + 2\,\Sigma \quad {\rm and} \quad \Sigma_+ \leq \Sigma_- + 2\,\Sigma,
\ee
where $\Sigma_+$ and $\Sigma_-$ are the asymptotic values of the relative volume
dispersion in the expanding and contracting branches respectively and
$\Sigma = \Delta p_\phi/p_\phi$ is the relative dispersion in the field momentum.
The triangle inequalities given in \eref{eq:triangleineq}, in a sense, place bounds on the
asymmetry in the relative volume dispersion on the two sides of the bounce
\be
  |\Sigma_--\Sigma_+| \leq 2\,\Sigma.
\ee
\begin{figure}[tbh!]
  \subfigure[]
  {
    \includegraphics[width=0.77\textwidth]{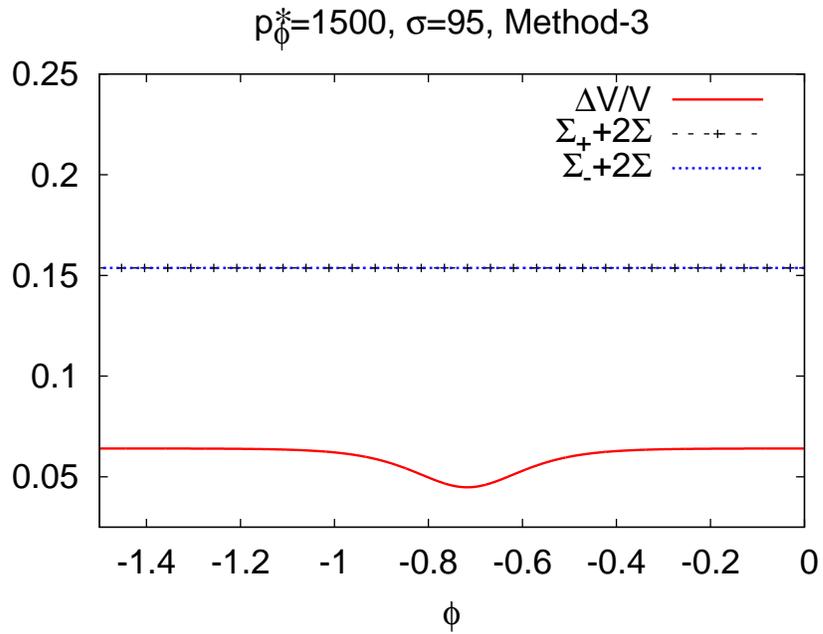}
  }
  \subfigure[]
  {
    \includegraphics[width=0.75\textwidth]{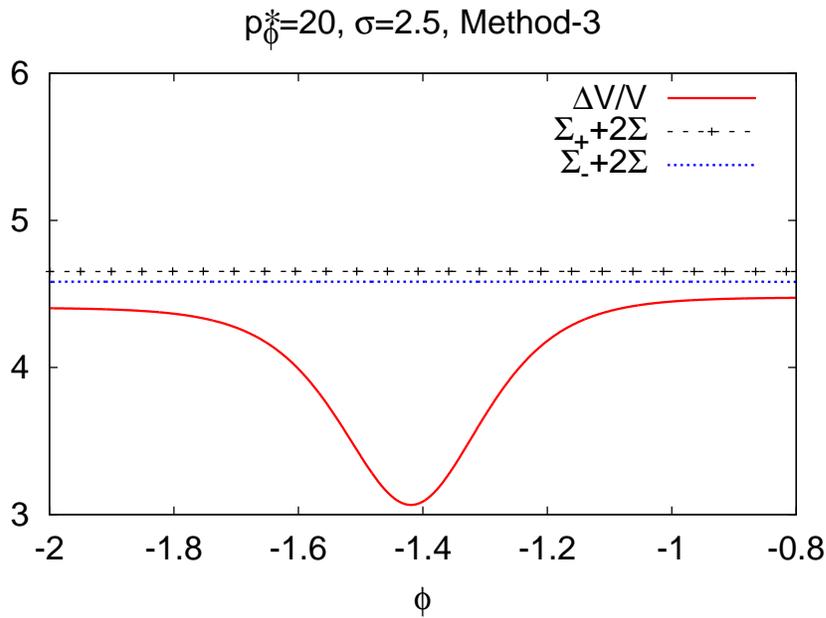}
  }
  \caption{This figure shows the evolution of the relative volume dispersion and the
  validity of the triangle inequality for $p_\phi^*=1500\sqrt{G}\hbar$ and $\sigma=95$
  for method-3 initial data in panel (a) and for $p_\phi^*=20\sqrt{G}\hbar$
  and $\sigma=2.5$. As in the case of method-2, we see that the triangle 
  inequalities are valid throughout the evolution for both large and small
  $p_\phi^*$.}
  \label{fig:trianglew1500m3}
\end{figure}

\begin{figure}[tbh!]
  \subfigure[]
  {
    \includegraphics[width=0.75\textwidth]{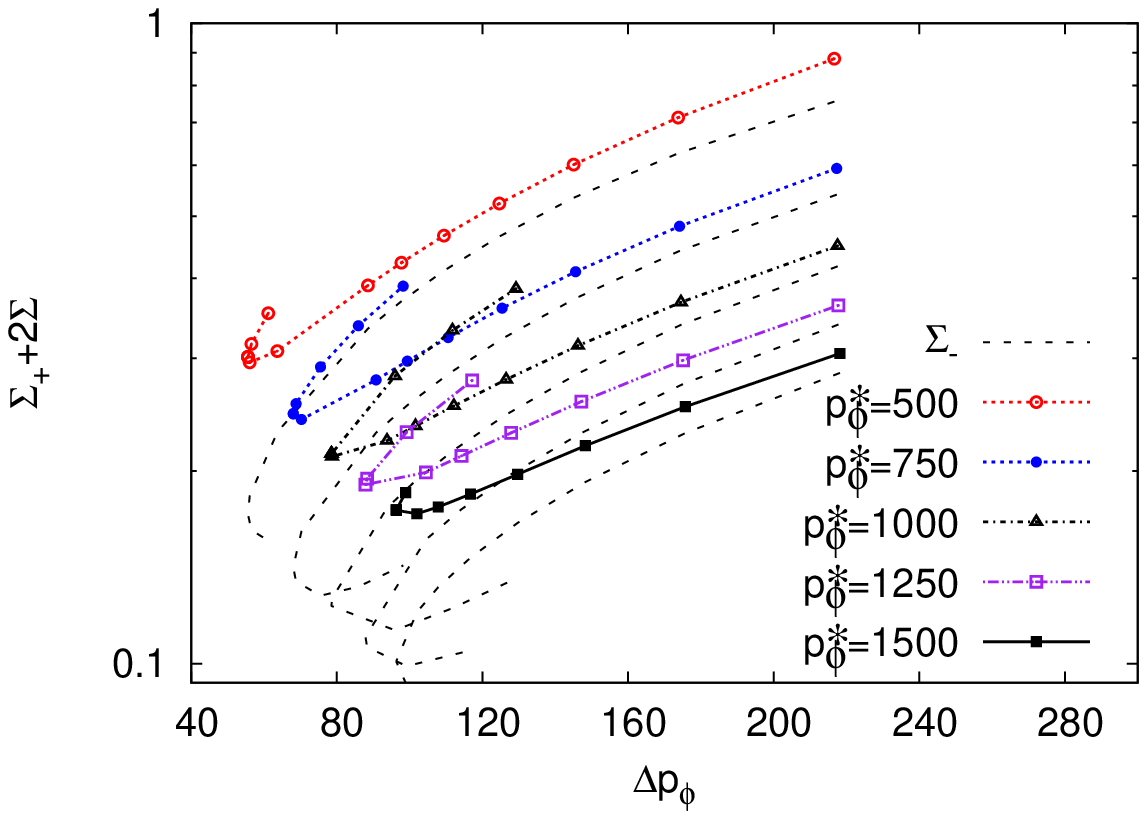}
    \label{fig:triangle1m1}
  }
  \subfigure[]
  {
    \includegraphics[width=0.75\textwidth]{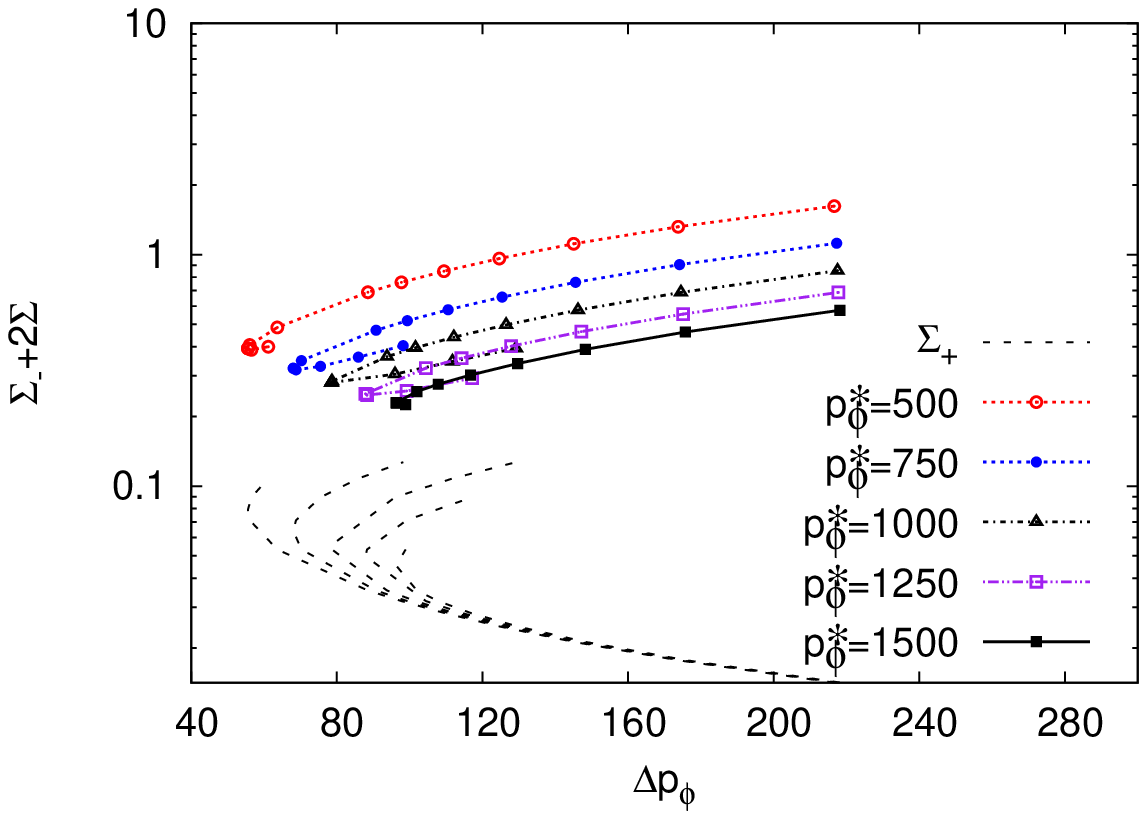}
    \label{fig:triangle2m1}
  }
  \caption{The validity of the triangle inequalities for method-1 initial data for varying 
  $\Delta p_\phi$: panel (a) and (b)
  show the variation of $\Sigma_++2\Sigma$ and $\Sigma_-+2\Sigma$, respectively,
  demonstrating the validity of the first and the second inequality in \eref{eq:triangleineq}. 
  The dashed curves in the upper panel shows the value of $\Sigma_-$ for different values 
  of $p_\phi^*$. The upper most dashed curve correspond to $p_\phi^*=500\,\sqrt{G}\hbar$ 
  and as $p_\phi^*$ increases the corresponding dashed curve for $\Sigma_-$ shifts downwards.
  Similarly, in the lower panel, the dashed curves correspond to $\Sigma_+$
  for different values of $p_\phi^*$. The left most curve corresponds to $p_\phi^*=500\,\sqrt{G}\hbar$
  and the right most to $p_\phi^*=1500\,\sqrt{G}\hbar$. }
  \label{fig:trianglem1}
\end{figure}
\begin{figure}[tbh!]
  \subfigure[]
  {
    \includegraphics[width=0.77\textwidth]{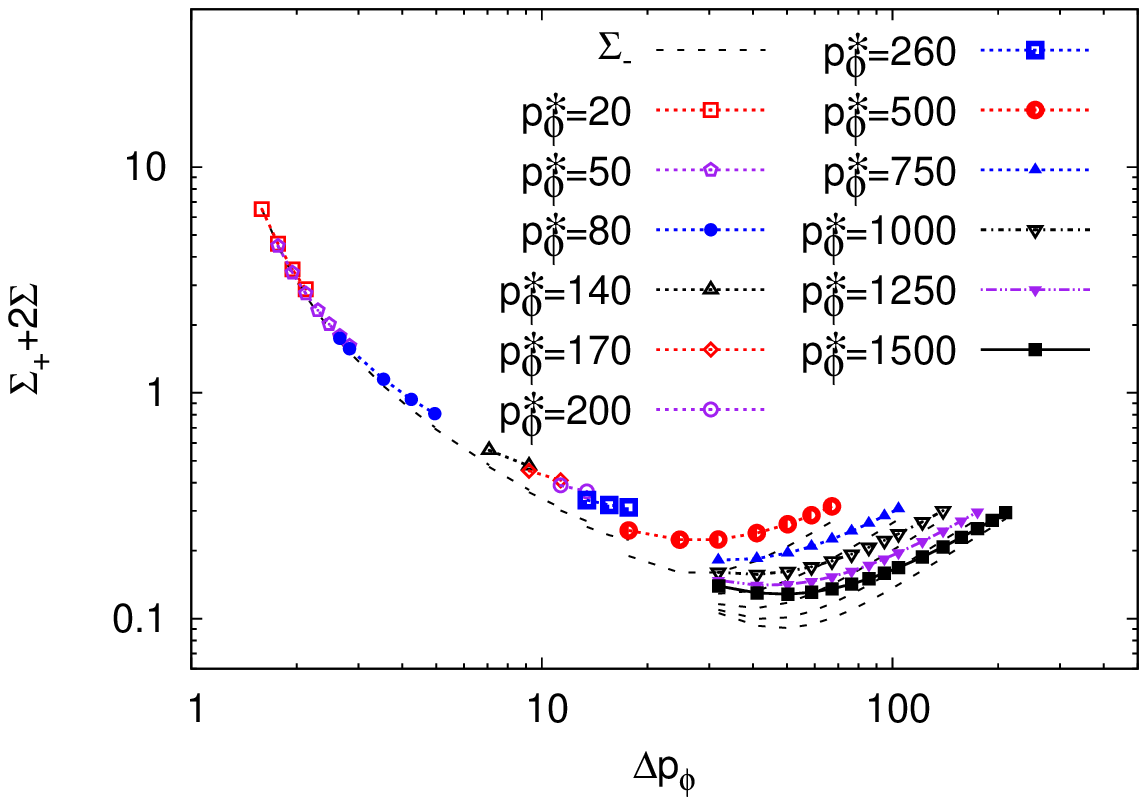}
    \label{fig:triangle1m2}
  }
  \subfigure[]
  {
    \includegraphics[width=0.77\textwidth]{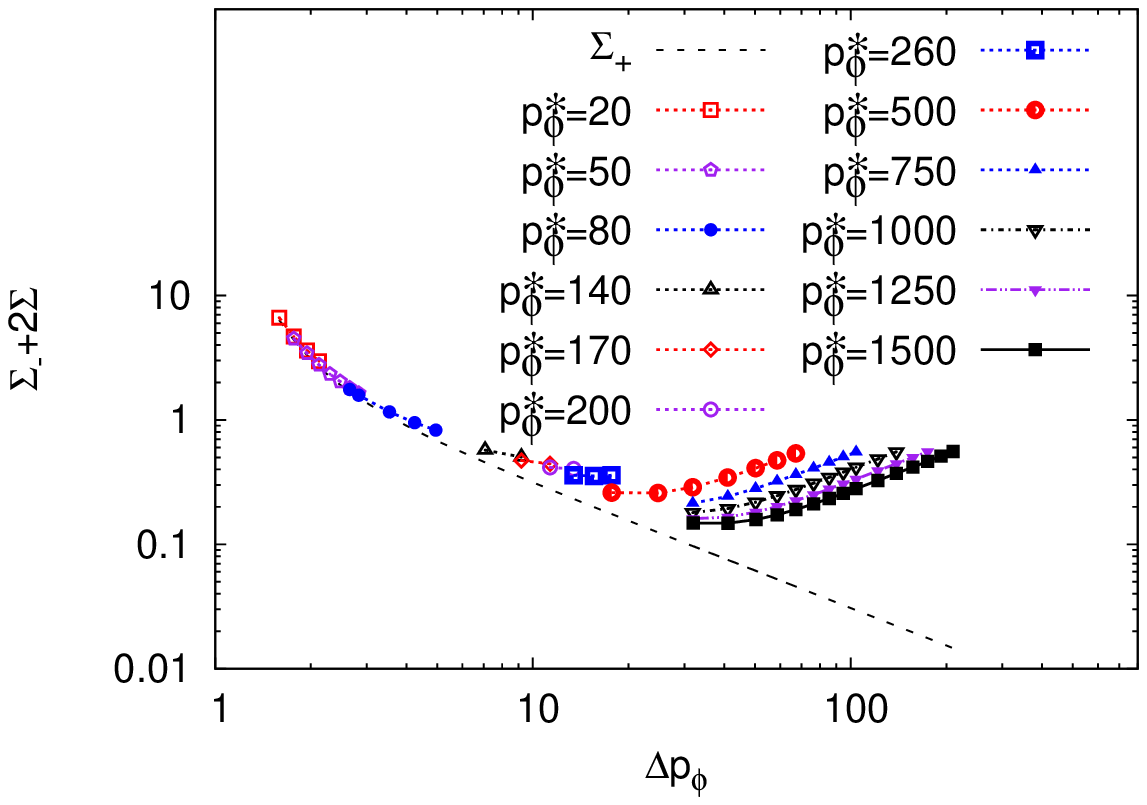}
    \label{fig:triangle2m2}
  }
  \caption{This figure shows the validity of the triangle inequalities for initial
  data of type method-2 for varying $\Delta p_\phi$. Panel (a) and (b) show the
  variation of $\Sigma_++2\Sigma$ and $\Sigma_-+2\Sigma$, respectively,
  demonstrating the validity of the first and the second inequality in
  \eref{eq:triangleineq}. It is interesting to note that the inequalities are 
  saturated in the low $\Delta p_\phi$ region.}
  \label{fig:trianglem2}
\end{figure}
\begin{figure}[tbh!]
  \subfigure[]
  {
    \includegraphics[width=0.75\textwidth]{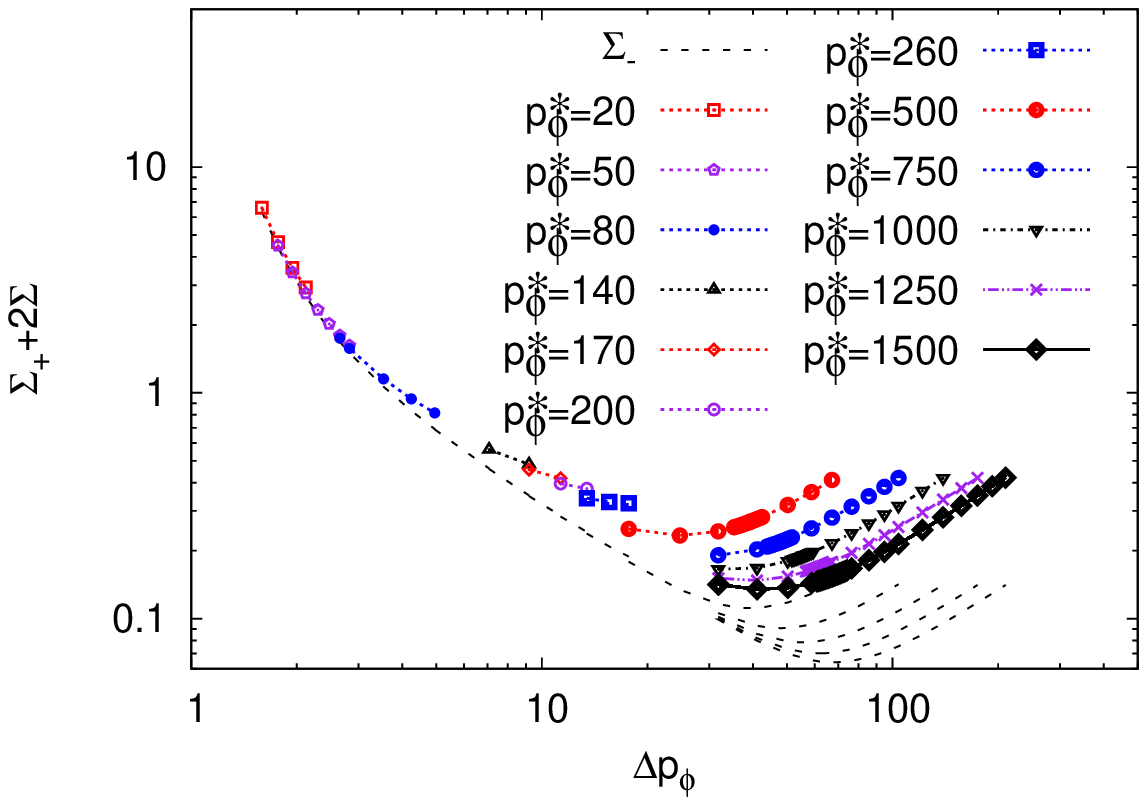}
    \label{fig:triangle1m3}
  }
  \subfigure[]
  {
    \includegraphics[width=0.75\textwidth]{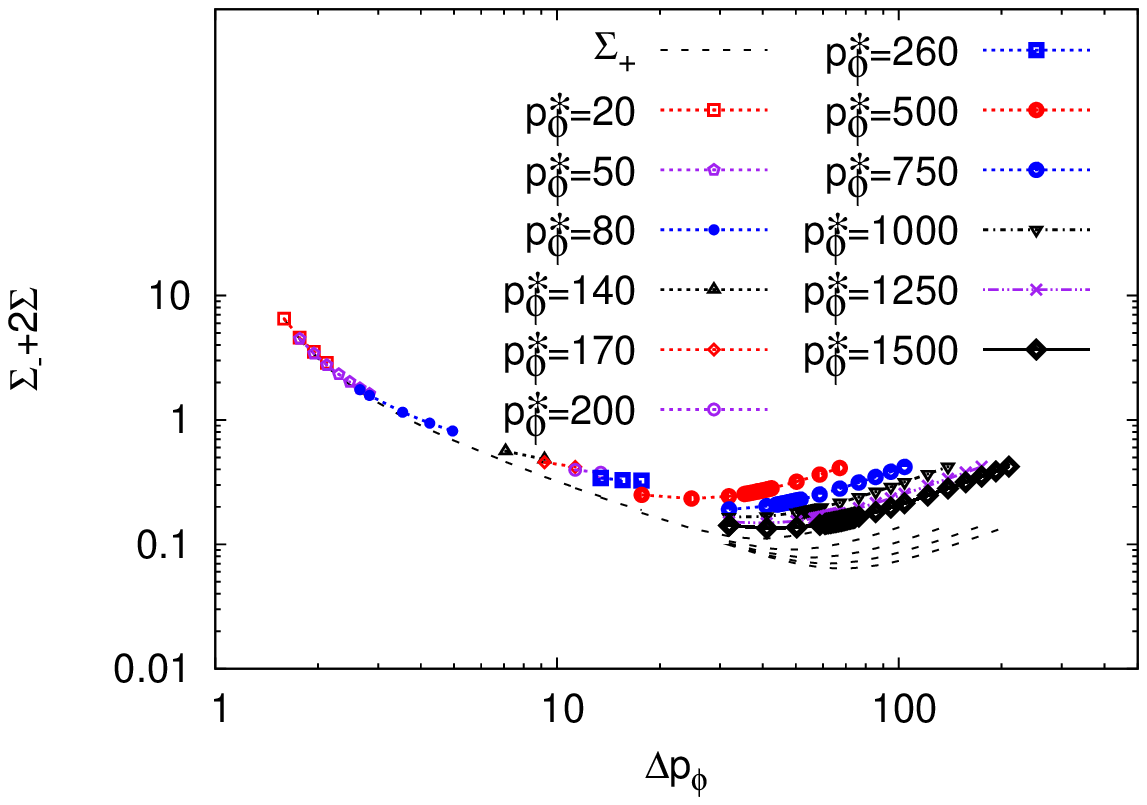}
    \label{fig:triangle2m3}
  }
  \caption{The plots in this figure demonstrate the validity of the triangle inequalities for initial
  data corresponding to method-3 for varying $\Delta p_\phi$ is shown.  Panel (a) and (b) show the
  variation of $\Sigma_++2\Sigma$ and $\Sigma_-+2\Sigma$, respectively,
  demonstrating the validity of the first and the second inequality in
  \eref{eq:triangleineq}.
 As in the case of method-2, the inequalities are 
  saturated in the low $\Delta p_\phi$ region.}
  \label{fig:trianglem3}
\end{figure}

\fref{fig:trianglew1500m1}, \fref{fig:trianglew1500m2} and \fref{fig:trianglew1500m3}
show the evolution of the relative volume dispersion for method-1, 2
and 3 respectively, with large and small values of the field momentum.
For method-2 and 3 we show the simulations for $p_\phi^*$ as small as
$20\sqrt{G}\hbar$, whereas for method-1 we restrict ourselves to $p_\phi^* \geq 500$ (see footnote \ref{fn1}). The dotted and dashed curves show
$\Sigma_\pm + 2\,\Sigma$. The figures clearly show that on both sides
of the bounce, the triangle inequalities are satisfied i.e.\
$\Sigma_\pm < \Sigma_\mp + 2\,\Sigma$. It is also important to note that these
inequalities holds true for all three initial data methods.
\fref{fig:trianglem1} shows the asymptotic values of the volume dispersion,
where $\Sigma_\pm$ is compared with $\Sigma_\mp+2\,\Sigma$, for method-1 initial data.
Similarly, \fref{fig:trianglem2} and \fref{fig:trianglem3} show the
asymptotic values of the relative volume dispersion, where $\Sigma_\pm$ is compared with
$\Sigma_\mp+2\,\Sigma$, for method-2 and method-3 initial data.
The dashed curves in the upper panel of these figures show $\Sigma_-$ and
in the lower panel $\Sigma_+$ for various values of $p_\phi^*$.

\section{Results}\label{sec:results}

In this section, we present results based on a large number of simulations by
considering states peaked at various values of $p_\phi^*$ in the range 
$20\sqrt{G}\hbar<p_\phi^*<1500\sqrt{G}\hbar$ for method-2 and 3 initial data,
and for $p_\phi^*\geq500\sqrt{G}\hbar$ for method-1. For each value of $p_\phi^*$
we vary $\Delta p_\phi$ (the 
spread of the Gaussian) in the range $5\%\, {\rm to}\, 15\%$ of the mean 
value of the field momentum $p_\phi^*$. The value of $p_\phi^*$
determines the bounce volume while $\Delta p_\phi$ dictates whether the state is sharply
peaked initially. For each of the simulations we compute
the expectation values of various quantities and their relative
dispersions both at the bounce and in the asymptotic limits far from the bounce.
This way we are able to make a detailed quantitative investigation of the spread
dependent properties of the initial state during evolution. The analysis performed here enables us 
to explore in a rigorous way the deviation of the LQC trajectory from the effective trajectory in a 
quantitative fashion. The study of the spread dependent corrections provides us valuable insights into which regime the effective theory is reliable.

As in \eref{eq:delta}, in order to quantify the deviation of LQC
trajectory from the effective dynamical trajectory, we define the quantity
\[
\delta= \f{\left ({V_{\rm b}}  -{V_{\rm b}}^{(\rm eff)}\right )}{{V_{\rm b}}^{(\rm eff)}},
\]
where ${V_{\rm b}}$ denotes the expectation value of the bounce volume in the LQC trajectory
and ${V_{\rm b}}^{(\rm eff)}$ is the bounce volume in the effective trajectory calculated using eq.
(\ref{bounceeff}) by setting $p_\phi = p_\phi^*$ (the field momentum at which the initial state is peaked). The quantity 
$\delta$ denotes the fractional difference in the bounce volumes in LQC and effective theory
with respect to the bounce volume in the effective theory.

\begin{table}[tbh!]
\caption{Summary of the main results.}
  \begin{tabular}{c c c c }
   \hline
   Initial data & Fixed $\f{\Delta V}{V}$ increasing $p_\phi^*$ & Fixed $p_\phi^*$ increasing $\f{\Delta V}{V}$ & Fixed $\Delta p_\phi$  increasing $p_\phi^*$\\
    \hline     \hline
Method-1 & $\delta$ decreases  & $\delta$ decreases & $\delta$ decreases if $\f{\Delta V}{V} < \widetilde{\f{\Delta V}{V}}$ \\
	&	&	& $\delta$ increases if $\f{\Delta V}{V} > \widetilde{\f{\Delta V}{V}}$ \\
\hline
Method-2 & $\delta$ decreases & $\delta$ increases if $\Delta p_\phi < \widetilde{\Delta p_\phi}$ & $\delta$ decreases \\
		&				& $\delta$ decreases if $\Delta p_\phi > \widetilde{\Delta p_\phi}$ & \\
		\hline
Method-3 & $\delta$ increases if $\Delta p_\phi < \widetilde{\Delta p_\phi}$ &  $\delta$ increases if $\Delta p_\phi < \widetilde{\Delta p_\phi}$ &  \\
			& $\delta$ decreases if $\Delta p_\phi > \widetilde{\Delta p_\phi}$ &  $\delta$ decreases if $\Delta p_\phi > \widetilde{\Delta p_\phi}$  & $\delta$ is independent of $p_\phi$ \\
   \hline
  \end{tabular}
\label{tab:summary}
\end{table}

Before we discuss the results from each of the methods, we summarize the key
results regarding the dependence of the quantity $\delta$ on the initial data 
parameters in Table \ref{tab:summary}. 
It is evident from the table that $\delta$ depends on several different factors 
including $p_\phi^*$, $\Delta p_\phi$ and $\Delta V/V$ and the initial data 
method. For example, if one fixes the relative volume dispersion and increases
the field momentum $p_\phi^*$, then $\delta$ decreases for method-1 and method-2 for all $\Delta p_\phi$, 
whereas for method-3, $\delta$ increases if $\Delta p_\phi<\widetilde{\Delta p_\phi}$.
Similarly, for a fixed value of the field momentum, $\delta$ decreases with increasing $\Delta V/V$ 
for all $\Delta p_\phi$ for method-1, whereas for method-2 and 3, the behavior is different for the 
two regimes $\Delta p_\phi<\widetilde{\Delta p_\phi}$ and $\Delta p_\phi>\widetilde{\Delta p_\phi}$.
Note that, for method-1, the parameters of the initial data are different compared to method-2 
and 3. The initial data in method-1 is a Gaussian which is characterized by 
$\sigma_v(=\sqrt{2}\Delta V)$, $v_*$ and $p_\phi^*$, and the value of $\Delta p_\phi$ is then 
derived from these. On the other hand, for method-2 and method-3 the parameters of the initial 
data are $\sigma(=\sqrt{2}\Delta p_\phi)$, $v_*$ and $p_\phi^*$. In the following, we consider the 
results for the three different initial data methods in separate subsections.

\subsection{Method-1: Gaussian state}
\begin{enumerate}
\item As discussed previously in Sec. \ref{sec:initialdata}, for method-1 
	 initial data, $\Delta V/V$
         appears as an input parameter and the value of the dispersion in the
         field momentum is computed from it. It turns out that $\Delta p_\phi$
         depends non-monotonically on $\Delta V/V$ as can be seen from
         \fref{fig:reldispm1a} which shows the initial field dispersion $\Delta p_\phi$
         plotted against the initial relative volume dispersion far from the bounce in the 
         expanding branch. The turn around in the plots correspond to the values of
         dispersion for which $\f{\Delta V}{V}=\widetilde{\f{\Delta V}{V}}=
         \left(\f{3\pi} {{p_\phi^*}^2} G\hbar^2\right)^{1/4}$.
         \fref{fig:reldispm1b} depicts the results from the same simulations in a slightly different way
         where $\f{\Delta V}{V}\sqrt{p_\phi^*}$ is plotted against $\Delta p_\phi$.
         This plot shows that irrespective of the value of
         $p_\phi^*$, the minimum of $\Delta p_\phi$ occurs when
         $\f{\Delta V}{V}\sqrt{p_\phi}=(3\pi G \hbar^2)^{1/4}$ (we use
	 $G=\hbar=1$ in the plots).
         At the minimum, the value of the relative volume
         dispersion is $\f{\Delta V}{V}=\widetilde{\f{\Delta V}{V}}$. 	
         Due to this non-monotonic dependence of $\Delta p_\phi$ on
	 $\Delta V/V$ any quantity varying monotonically with respect
	 to $\Delta V/V$ will vary non-monotonically with respect to
	 $\Delta p_\phi$.

\begin{figure}[tbh!]
  \subfigure[]
  {
     \includegraphics[width=0.75\textwidth]{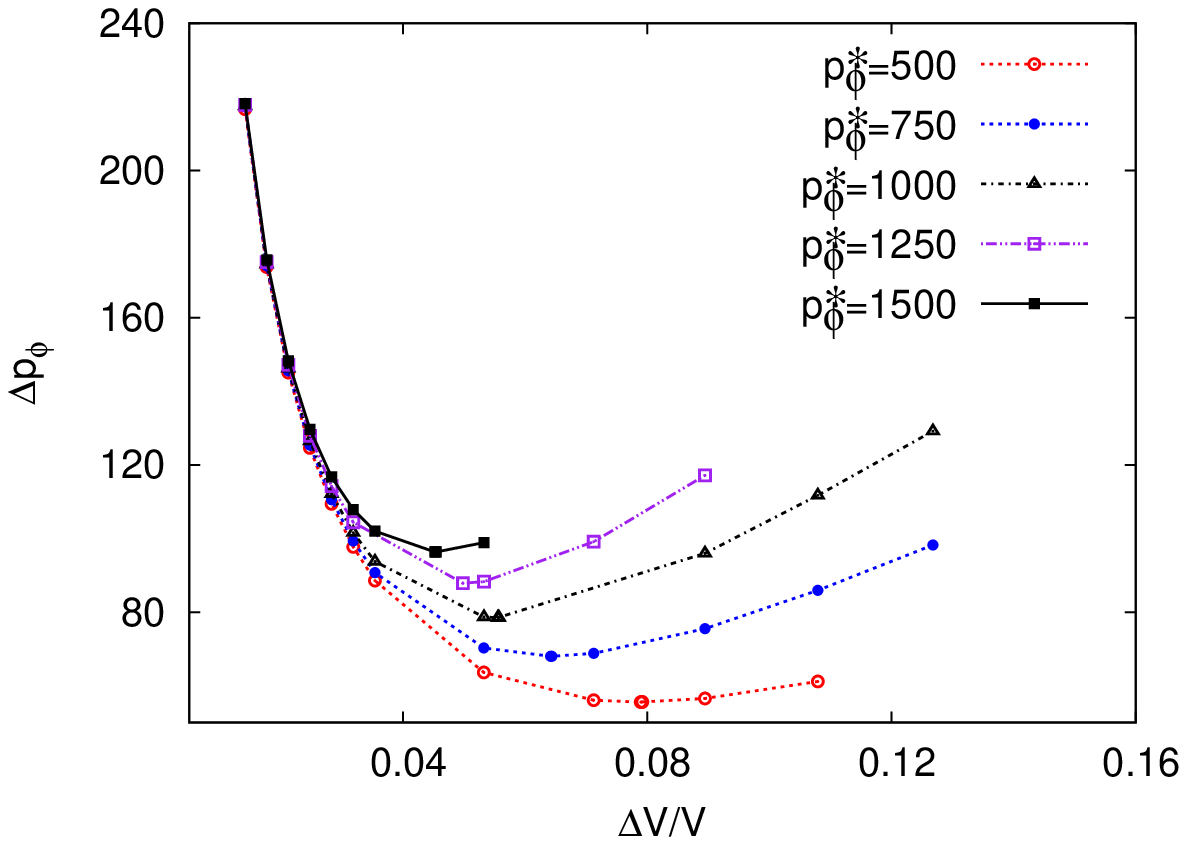}
     \label{fig:reldispm1a}
  }
  \subfigure[]
  {
	\includegraphics[width=0.75\textwidth]{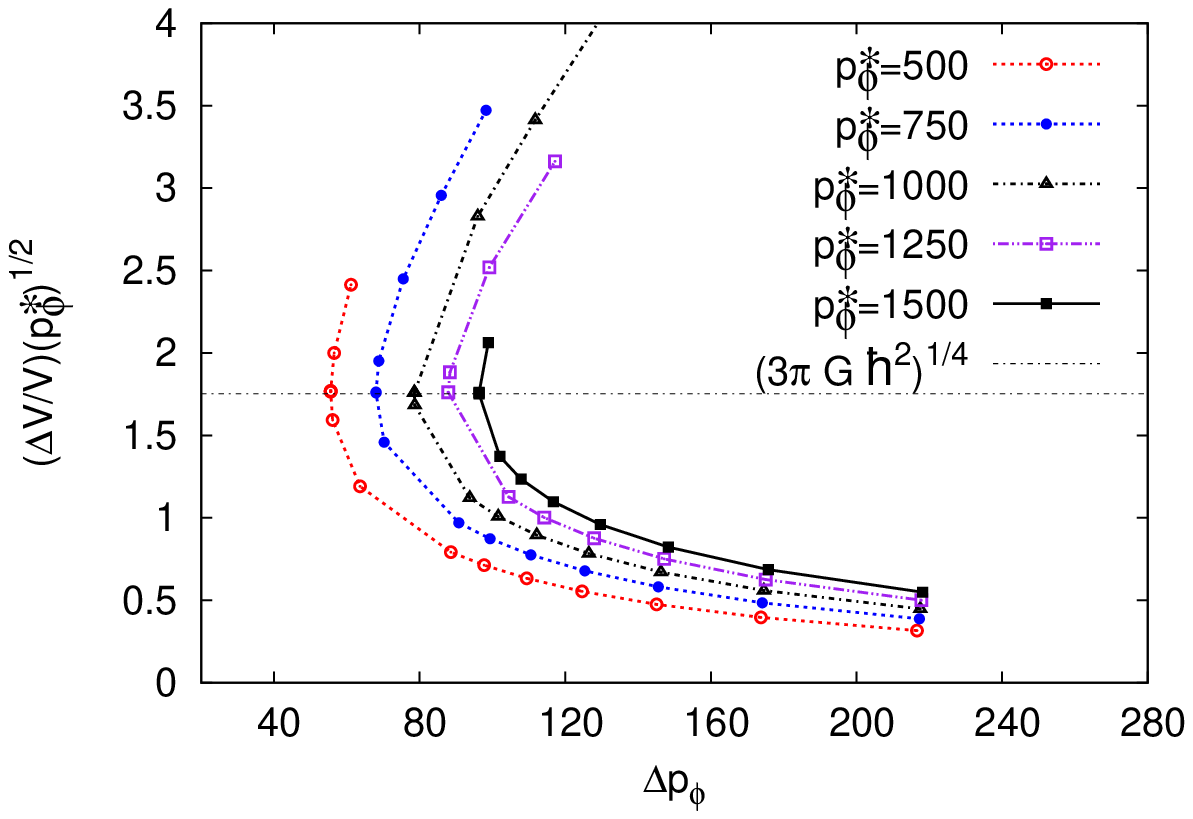}
     \label{fig:reldispm1b}
  }
\caption{Panel (a) shows the behavior of $\Delta p_\phi$ plotted against 
         $\Delta V/V$ for method-1 initial data. It is clear that $\Delta p_\phi$ 
         shows non-monotonic behavior with $\Delta V/V$. In panel (b), 
         $\f{\Delta V}{V}\sqrt{p_\phi^*}$ is plotted against $\Delta p_\phi$, 
         in order to make it clear that irrespective of the value of $p_\phi^*$, the minimum of
         $\Delta p_\phi$ occurs when $\f{\Delta V}{V}\sqrt{p_\phi^*}=(3\pi G \hbar^2)^{1/4}$.
         }
\label{fig:reldispm1}
\end{figure}

\item \fref{fig:vbouncem1} shows the variation of $\delta$ for different
         values of $p_\phi^*$ as function of the initial relative dispersion in volume
         $\Delta V/V$ for method-1 initial data. As $\Delta V/V$ increases for a fixed
         value of $p_\phi^*$, $\delta$ decreases monotonically. 
	 This behavior is summarized in the row for method-1 and the column
	 with fixed $p_\phi^*$ in Table \ref{tab:summary}. This presents 
         a counter example to the expectation that a wide-spread state should show
         larger differences between the effective and the LQC trajectory than
	 a narrow state.
         It is also evident from the figure that for a fixed value of
         $\Delta V/V$, a lower value of $p_\phi^*$ results in a larger
         value of $\delta$ (summarized in the row for method-1 and column for
	 fixed $\Delta V/V$ in Table \ref{tab:summary}). The variation of $\delta$ with respect to
         $\Delta p_\phi$ is shown in \fref{fig:vbouncem3delpphim1}. Here we see 
         a clear non-monotonic behaviour. The turnaround
         occurs exactly when $\f{\Delta V}{V}=\widetilde{\f{\Delta V}{V}}$.
         The upper branch, where $\delta$ decreases with increasing $p_\phi^*$
         for fixed $\Delta p_\phi$, corresponds to
         $\f{\Delta V}{V}<\widetilde{\f{\Delta V}{V}}$, while the lower branch,
         where $\delta$ increases with increasing $p_\phi^*$ for fixed
         $\Delta p_\phi$, corresponds to
         $\f{\Delta V}{V}>\widetilde{\f{\Delta V}{V}}$ as summarized in the
	 row for method-1 and the column for fixed $\Delta p_\phi$ in
         Table \ref{tab:summary}.

\begin{figure}[tbh!]
    \includegraphics[width=0.75\textwidth]{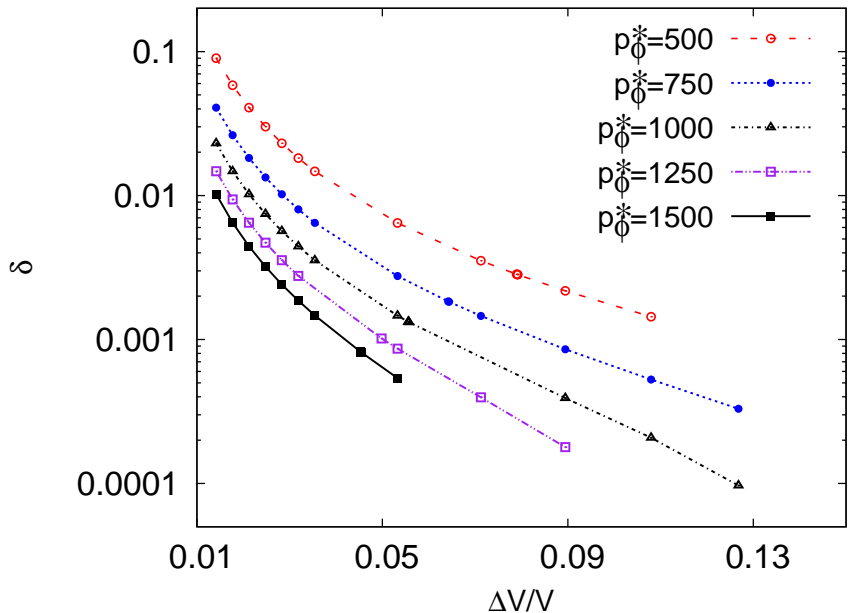}
  \caption{The variation of the quantity $\delta$ is plotted against the relative
  dispersion in volume $\Delta V/V$ for method-1 initial data.
  For all data points, the state is initially peaked at $v^*=20000$. The
  different lines correspond to different values of the field momentum
  $p_\phi^*$ in units of $\sqrt{G}\hbar$.}
\label{fig:vbouncem1}
\end{figure}
\begin{figure}[tbh!]
    \includegraphics[width=0.75\textwidth]{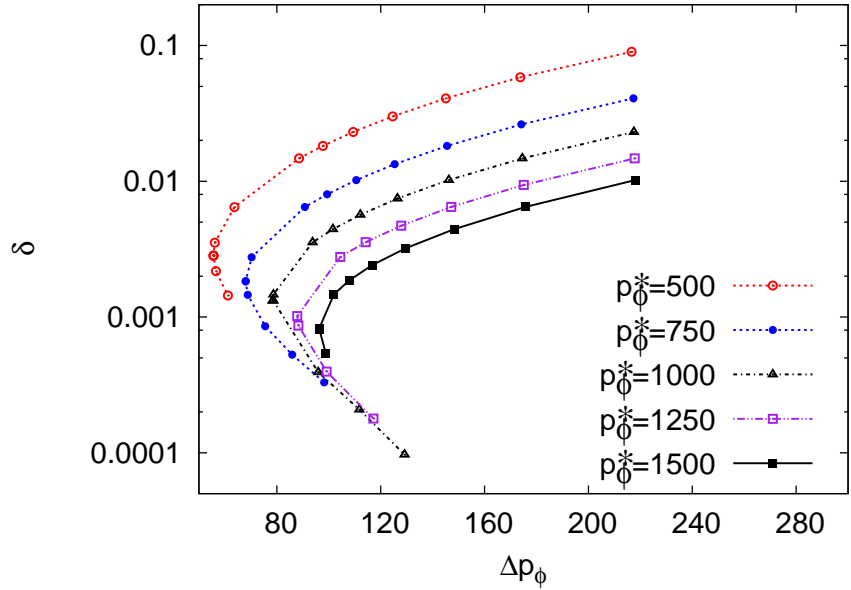}
\caption{This plot shows the behavior of the quantity $\delta$ with respect 
             to the field momentum $\Delta p_\phi$ for method-1 initial data
	     for several different values of $p_\phi^*$. The value of 
	     $v^*$ is the same as in \fref{fig:vbouncem1}.}
\label{fig:vbouncem3delpphim1}
\end{figure}

\item As discussed before, the scalar field $\phi$ plays the role of internal time in our analysis. 
      Its absolute value does not have a physical meaning as it can be
      shifted  without affecting the physics. However,
      a ``time'' difference is a useful quantity to study.  
      We consider here the difference between the values of the scalar field
      at the bounce in the effective and the LQC trajectory, i.e. 
      $\left (\phi_{\rm b}-\phi_{\rm b}^{(\rm eff)}\right )$, and plot it against $\Delta p_\phi$
      in \fref{fig:phim1}. It is evident that the LQC trajectory
      bounces earlier than the effective one and the difference in bounce time increases with
      increasing $\Delta p_\phi$ if $\f{\Delta V}{V}<\widetilde{\f{\Delta V}{V}}$ and 
      decreases with increasing $\Delta p_\phi$ if $\f{\Delta V}{V}>\widetilde{\f{\Delta V}{V}}$.
      Moreover, for a fixed $\Delta p_\phi$ this difference decreases
      with increasing $p_\phi^*$ if $\f{\Delta V}{V}<\widetilde{\f{\Delta V}{V}}$ and increases with
      $p_\phi^*$ if $\f{\Delta V}{V}>\widetilde{\f{\Delta V}{V}}$.
\begin{figure}[tbh!]
    \includegraphics[width=0.75\textwidth]{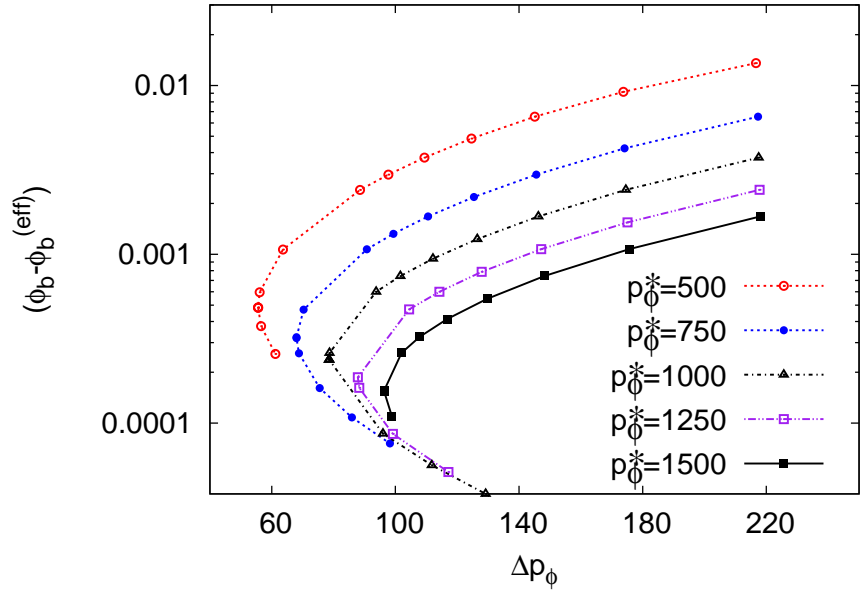}
\caption{This plot shows the variation of the difference between the bounce
         ``time'' in LQC and the effective theory 
	 $\left (\phi_{\rm b}-\phi_{\rm b}^{(\rm eff)}\right )$ plotted 
	 against the dispersion in the field momentum $\Delta p_\phi$ for 
	 method-1 initial data. }
\label{fig:phim1}
\end{figure}

\item The energy density at the bounce $\rho_{\rm b}$ also depends on the relative volume dispersion of
         the initial state $\Delta V/V$. As shown in \fref{fig:rhovm1}, for a fixed value of $p_\phi^*$,
         $\rho_{\rm b}$ decreases with a decreasing spread in the volume. On the other hand,
         if the dispersion in volume is fixed, then $\rho_{\rm b}$ takes smaller value as 
         $p_\phi^*$ decreases. It is to be noted that the energy density never exceeds
         the upper bound on $\rho_{\rm b}$, i.e.\ $\rho_{\rm b}\leq \rho_{\rm max}\approx0.409\rho_{\rm Pl}$ derived from the exactly solvable model in Ref.\cite{acs} and which is also predicted by the effective theory \eref{rhomax_eff}.
         The deviation of $\rho_{\rm b}$ from $\rho_{\rm max}$ is also a measure of the deviation
         of the effective theory from LQC. That is, the larger the difference between
         $\rho_{\rm b}$ and $\rho_{\rm max}$, the larger is the departure
         between the LQC and effective trajectory and the larger is $\delta$.
         \fref{fig:rhopphim1} shows the variation of $\rho_{\rm b}$ with respect to
         the dispersion in the field momentum $\Delta p_\phi$. The inset in this figure
	 shows the non-monotonic variation of $\rho_{\rm b}$ with $\Delta p_\phi$.
         It can also be seen that the energy density always remains below the 
	 critical value for all values of the field momentum.
\begin{figure}[tbh!]
\subfigure[]
{
\includegraphics[width=0.75\textwidth]{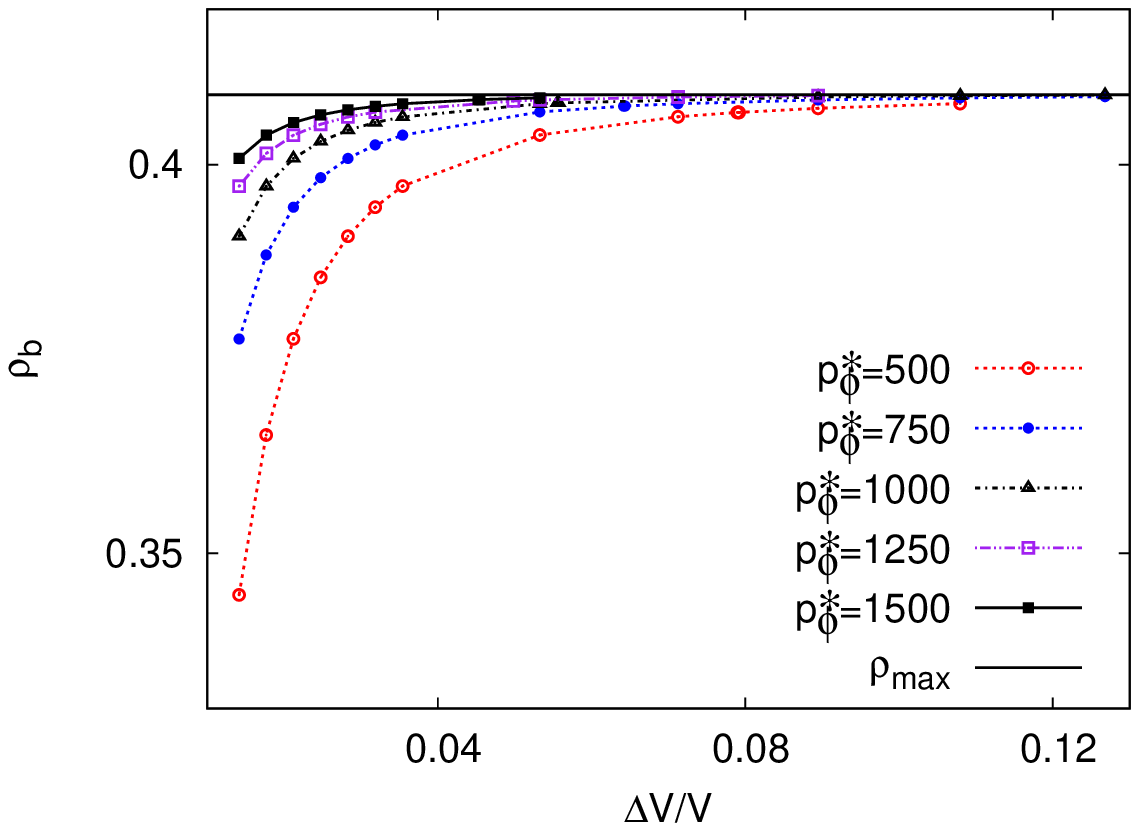}
\label{fig:rhovm1}
}
\subfigure[]
{
\includegraphics[width=0.75\textwidth]{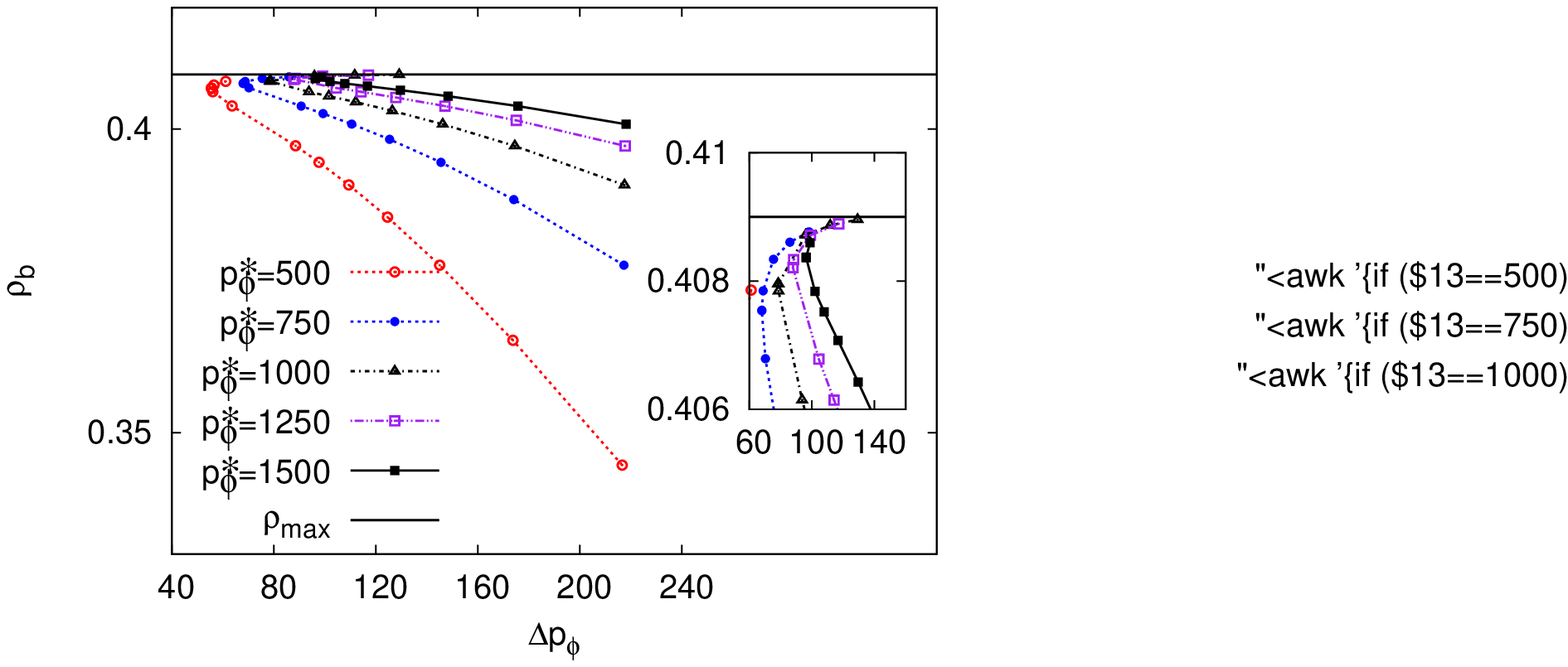}
\label{fig:rhopphim1}
}
\caption{The energy density at the bounce $\rho_{\rm b}$ is plotted with
  respect to the initial relative volume dispersion $\Delta V/V$ in panel (a)
  and with respect to the spread in the field momentum $\Delta p_\phi$
  in panel (b) for method-1
  initial data. It is evident from panel (a) that $\rho_{\rm b}$ decreases as
  $\Delta V/V$ decreases for a given value of $p_\phi$. For fixed
  $\Delta V/V$, as the value of the field momentum decreases $\rho_{\rm b}$
  also decreases. In panel (b) the non-monotonic variation of $\rho_{\rm b}$
  with respect to $\Delta p_\phi$ is apparent. From these figures
  it is also evident that the energy density at the bounce is always less than $\rho_{\rm max}$
  i.e.\ $\rho_{\rm b}<\rho_{\rm max}=0.409\,\rho_{\rm Pl}$ obtained from the exactly solvable model in LQC \cite{acs} and also predicted by the effective theory (see eq.(\ref{rhomax_eff})).}
\label{fig:rhom1}
\end{figure}

\end{enumerate}

\subsection{Method-2: Wheeler-DeWitt initial state}\label{sec:method2}
\begin{enumerate}
\item \fref{fig:reldispm2} shows the variation of the initial relative volume dispersion $\Delta V/V$
  with varying initial spread in the field momentum $\Delta p_\phi$, far from bounce in the 
  expanding branch. For the entire range of $\Delta p_\phi$,
  $\Delta V/V$ decreases monotonically with increasing $\Delta p_\phi$.
  This implies that if a quantity varies monotonically with respect to
  $\Delta p_\phi$ it should do so with respect to $\Delta V/V$ as well.
  Note that a method-2 initial state is a minimal uncertainty state in 
  $(\phi,\,p_\phi)$, and the
  variation of relative volume dispersion with respect to the dispersion in
  field momentum is monotonic. This is not the case for method-1 and 3 initial
  state, which are not minimal uncertainty states in $(\phi,\,p_\phi)$.
\begin{figure}[tbh!]
     \includegraphics[width=0.75\textwidth]{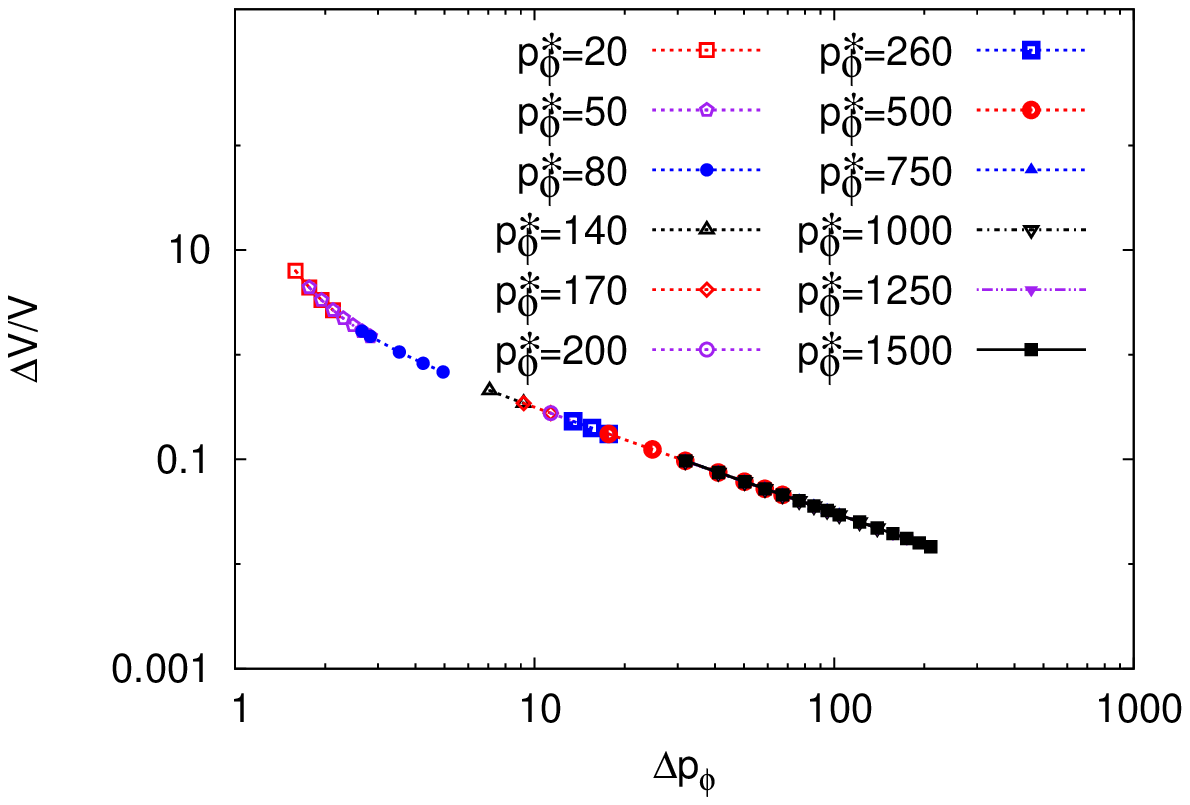}
\caption{Plot of the variation in $\Delta V/V$ with respect to $\Delta p_\phi$
  for method-2 initial data for different values of $p_\phi^*$. It is clear that the relative volume dispersion 
  varies with $\Delta p_\phi$ in a monotonic fashion. The values of the field momentum
  are given in the units of $\sqrt{G}\hbar$.}
\label{fig:reldispm2}
\end{figure}

\item \fref{fig:vbouncem3delpphim2} shows the variation of $\delta$ with
  respect to $\Delta p_\phi$ for method-2 initial data, in a log-log plot. It
  is evident from the figure that in the low $\Delta p_\phi$ regime, $\delta$
  is large, and as $\Delta p_\phi$ is increased, $\delta$ initially decreases monotonically.
  Then, around $\Delta p_\phi\approx 10$, there is a small interval of
  almost linear behavior. As  $\Delta p_\phi$ is increased further, the curves for
  different values of $p_\phi^*$ start to deviate from each other, reach a
  minimum value (at a $p_\phi^*$ dependent value of $\Delta p_\phi$) and then
  start to increase again. In this way, the curve for $\delta$ shows non-monotonicity 
  with respect to the dispersion in the field momentum in the large $\Delta p_\phi$ 
  regime. It is noteworthy that the minimum of $\delta$ occurs at 
  $\Delta p_\phi=\widetilde{\Delta p_\phi}=(3\pi {p_\phi^*}^2 G\hbar^2)^{1/4}$. 
  This figure also shows that for a fixed value of $\Delta p_\phi$, the value of 
  $\delta$ decreases with increasing $p_\phi^*$. This behavior is summarized in 
  the row for method-2 and column for fixed $\Delta p_\phi$ in Table 
  \ref{tab:summary}.
\begin{figure}[tbh!]
\includegraphics[width=0.75\textwidth]{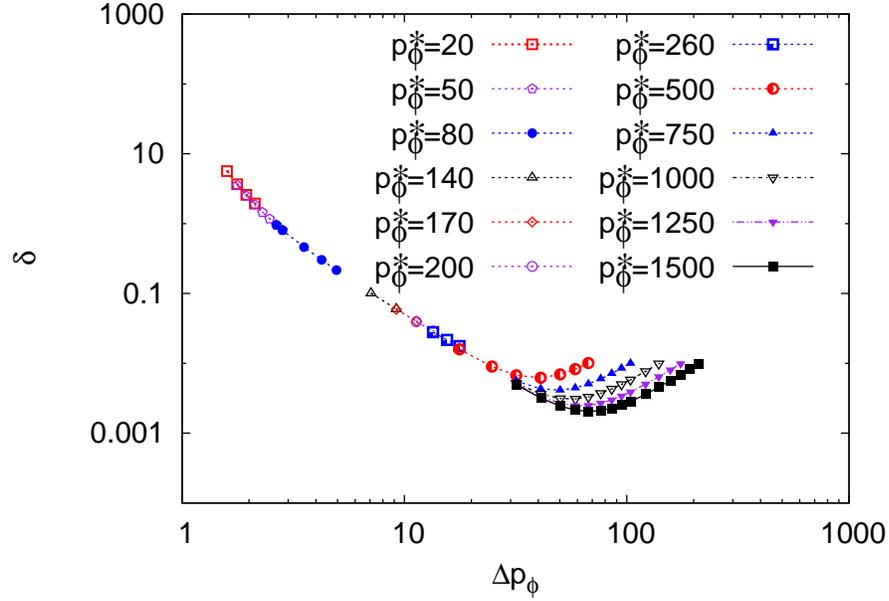}
\caption{The figure shows the variation of $\delta$ with respect to the spread in the field
  momentum $\Delta p_\phi$ for method-2 initial data for different values of
  $p_\phi^*$. It is
  clear that, in the large $\Delta p_\phi$ regime, $\delta$ behaves
  non-monotonically and attains a minimum at some $\Delta p_\phi$ for a
  given value of the field momentum $p_\phi^*$.}
\label{fig:vbouncem3delpphim2}
\end{figure}

\begin{figure}[tbh!]
  \subfigure[]
  {
    \includegraphics[width=0.73\textwidth]{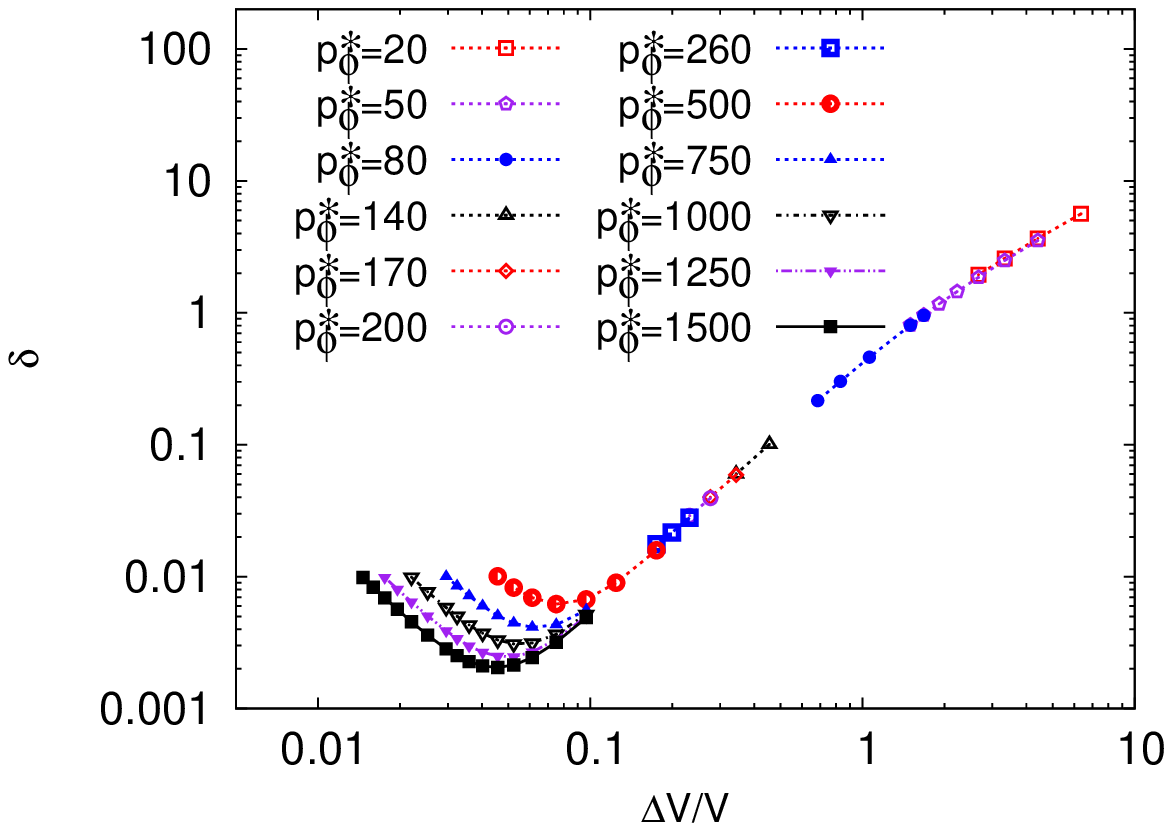}
    \label{fig:vbouncem2a}
  }
  \subfigure[]  {
    \includegraphics[width=0.73\textwidth]{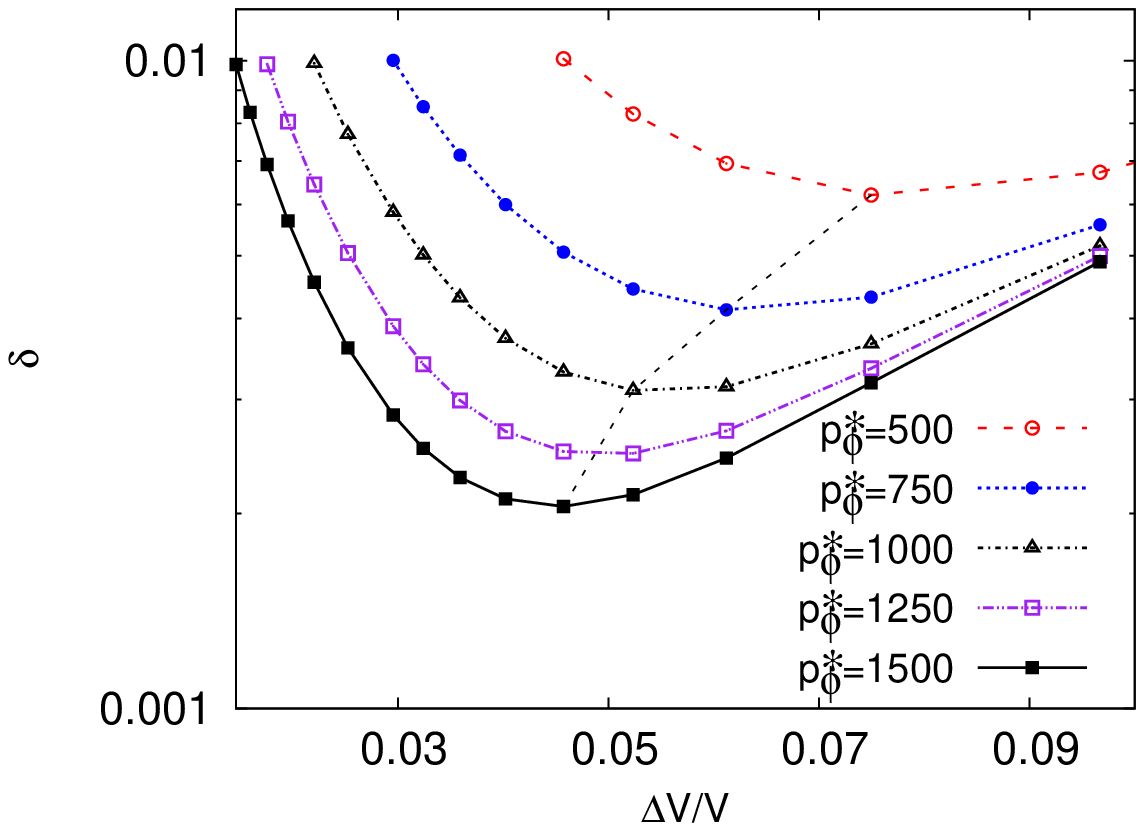}
    \label{fig:vbouncem2b}
  }\caption{The quantity $\delta$ plotted with respect
  to the relative dispersion in volume $\Delta V/V$ for method-2 initial data. It is clear from \fref{fig:vbouncem2a}
  that $\delta$ increases with increasing $\Delta V/V$ in the large $\Delta V/V$ regime. 
  \fref{fig:vbouncem2b} shows the same plot zoomed into the small
  $\Delta V/V$ regime, where the non-monotonicity of $\delta$ is apparent. 
  The dashed curve in \fref{fig:vbouncem2b} intersects the curves with various 
  values of $p_\phi^*$ at the points where the relative fluctuation in $p_\phi$ and $V$ are equal, i.e.
  $\f{\Delta p_\phi}{p_\phi}=\f{\Delta V}{V}=\left(\f{\sqrt{3\pi G}\hbar}{p_\phi^*}\right)^{1/2}$
  (see \eref{eq:uncertm2}).}
\label{fig:vbouncem2}
\end{figure}

\item  \fref{fig:vbouncem2a} shows the behavior of $\delta$ with varying
  relative dispersion in volume $\Delta V/V$ for different values of 
  $p_\phi^*$ and $\Delta p_\phi$. Each curve corresponds to different values 
  of $p_\phi^*$ with the data points corresponding to different 
  $\Delta p_\phi$. We find that in the large $\Delta V/V$ 
  regime $\delta$ is large. As the value of $\Delta V/V$ is decreased, $\delta$
  decreases until it reaches a minimum value at 
  $\Delta V/V = \left({3\pi} G\hbar^2/{p_\phi^2} \right)^{1/4}$ after
  which it starts to increase again. 
  \fref{fig:vbouncem2b} shows the behavior of $\delta$ zoomed in near the
  location of the minima to highlight the non-monotonicity.
  The dashed curve
  joining the separate $p_\phi^*$ curves correspond to the points where
  $\delta$ is minimum and
  $\Delta p_\phi=\widetilde{\Delta p_\phi}=(3\pi {p_\phi^*}^2 G\hbar^2)^{1/4}$.
  To the left of the dashed curve $\Delta p_\phi>\widetilde{\Delta p_\phi}$ and
  $\delta$ decreases with increasing $\Delta V/V$, while to the right of the dashed
  curve $\Delta p_\phi<\widetilde{\Delta p_\phi}$ and $\delta$ increases with 
  increasing $\Delta V/V$. This behavior is summarized in the row for method-2
  and the column for fixed $p_\phi^*$ in Table \ref{tab:summary}.
  It can also be seen that $\delta$ decreases with increasing $p_\phi^*$ for
  a fixed value of $\Delta V/V$ as summarized in the row for method-2 and the
  column for fixed $\Delta V/V$ in Table \ref{tab:summary}.

\item \fref{fig:powerlawphim2} shows the variation of the difference in the
  bounce time between the LQC and effective trajectories
  $\left ({\phi_{\rm b}}-{\phi_{\rm b}}^{(\rm eff)}\right )$ with varying
  $\Delta p_\phi$ in a log-log plot. It is clear that in the low
  $\Delta p_\phi$ regime, the curve is linear, which implies a power-law
  relation between $\left ({\phi_{\rm b}}-{\phi_{\rm b}}^{(\rm eff)}\right )$ and
  $\Delta p_\phi$
  \be
    \left ({\phi_{\rm b}}-{\phi_{\rm b}}^{(\rm eff)}\right ) \propto
    \left(\Delta p_\phi\right)^\nu,
  \label{eq:powerlawphim2}
  \ee
  where $\nu$ is an exponent. A numerical fit to the curve in the linear
  region gives the value of the exponent to be $\nu\approx -2.017\pm0.008$. 
  As we will see in the next subsection, this power law also holds true for method-3
  initial data, with the same exponent to three significant figures. In the large
  $\Delta p_\phi$ region there is non-monotonic behavior and
  $\left ({\phi_{\rm b}}-{\phi_{\rm b}}^{(\rm eff)}\right )$ reaches a minimum at
  $\Delta p_\phi=\widetilde{\Delta p_\phi}=(3\pi {p_\phi^*}^2 G\hbar^2)^{1/4}$.
  \begin{figure}[tbh!]
  \includegraphics[width=0.75\textwidth]{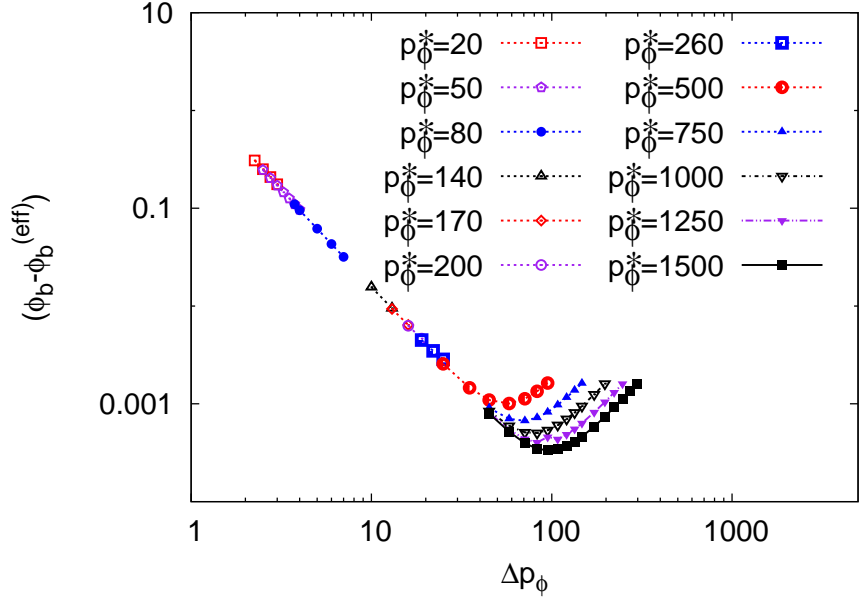}
  \caption{The variation of the difference in the time of bounce in the LQC
    evolution and the effective dynamical evolution as a function of
    $\Delta p_\phi$ for different values of $p_\phi^*$ for method-2 initial
    data is shown. Clearly, in the small $\Delta p_\phi$ regime, the curve is
    linear, implying a power law relation between
    $\left ({\phi_{\rm b}}-{\phi_{\rm b}}^{(\rm eff)}\right )$ and $\Delta p_\phi$
    given in \eref{eq:powerlawphim2}. The slope of the curves is $\nu=-2.017\pm0.008$. }
  \label{fig:powerlawphim2}
\end{figure}

\item \fref{fig:rhovm2} and \fref{fig:rhopphim2} show the variation in the
  energy density at the bounce with respect to the relative volume dispersion
  $\Delta V/V$ and the field dispersion $\Delta p_\phi$, respectively.
  It is evident from the figure, that the energy density at the bounce is
  very close to the maximum allowed energy density
  $\rho_{\rm max}=0.409\,\rho_{\rm Pl}$ for sharply peaked states
  (corresponding to small $\Delta V/V$). It is, however, noteworthy that the
  energy density at the bounce always remains smaller than $\rho_{\rm max}$.
  A similar dependence of the energy density at the bounce is discussed in
  \cite{montoya_corichi2} when the wavefunction is taken to be a squeezed
  state. There, it was found that increasing squeezing leads to an increasing
  $\Delta V/V$, which decreases $\rho_{\rm b}$.
\begin{figure}[tbh!]
\subfigure[]
{
\includegraphics[width=0.8\textwidth]{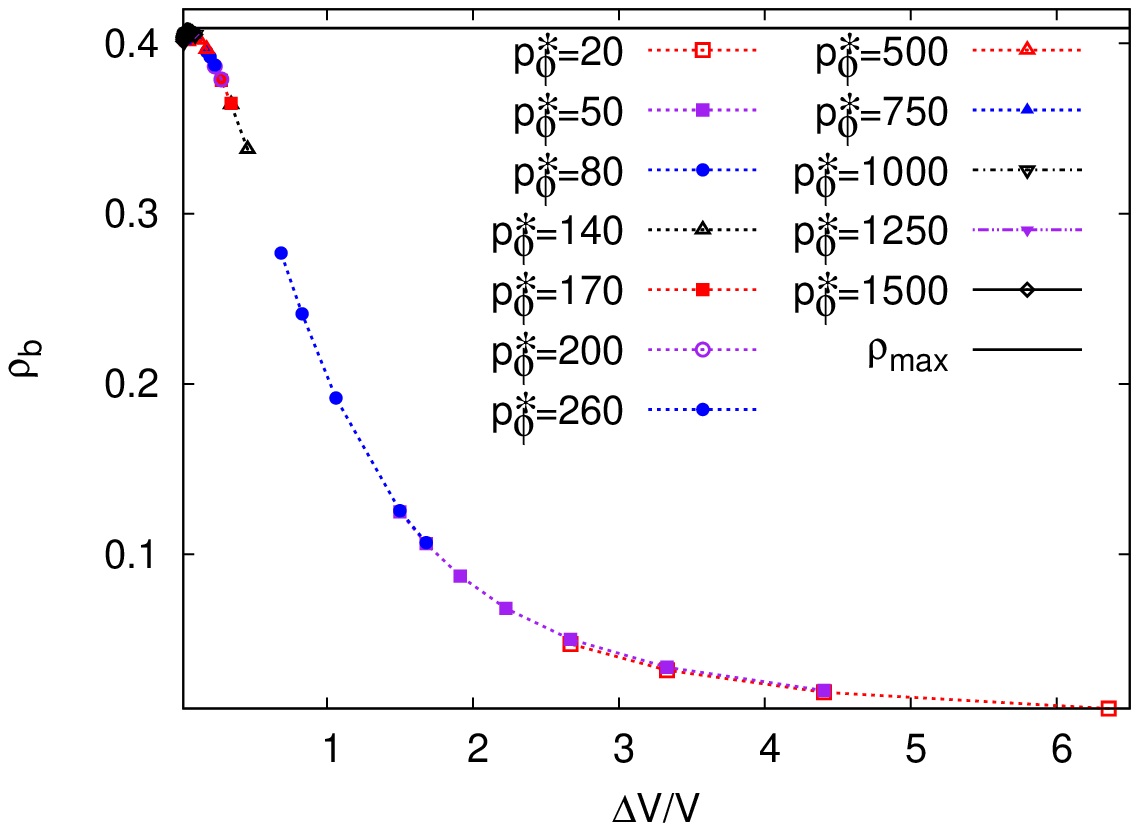}
\label{fig:rhovm2}
}
\subfigure[]
{
\includegraphics[width=0.8\textwidth]{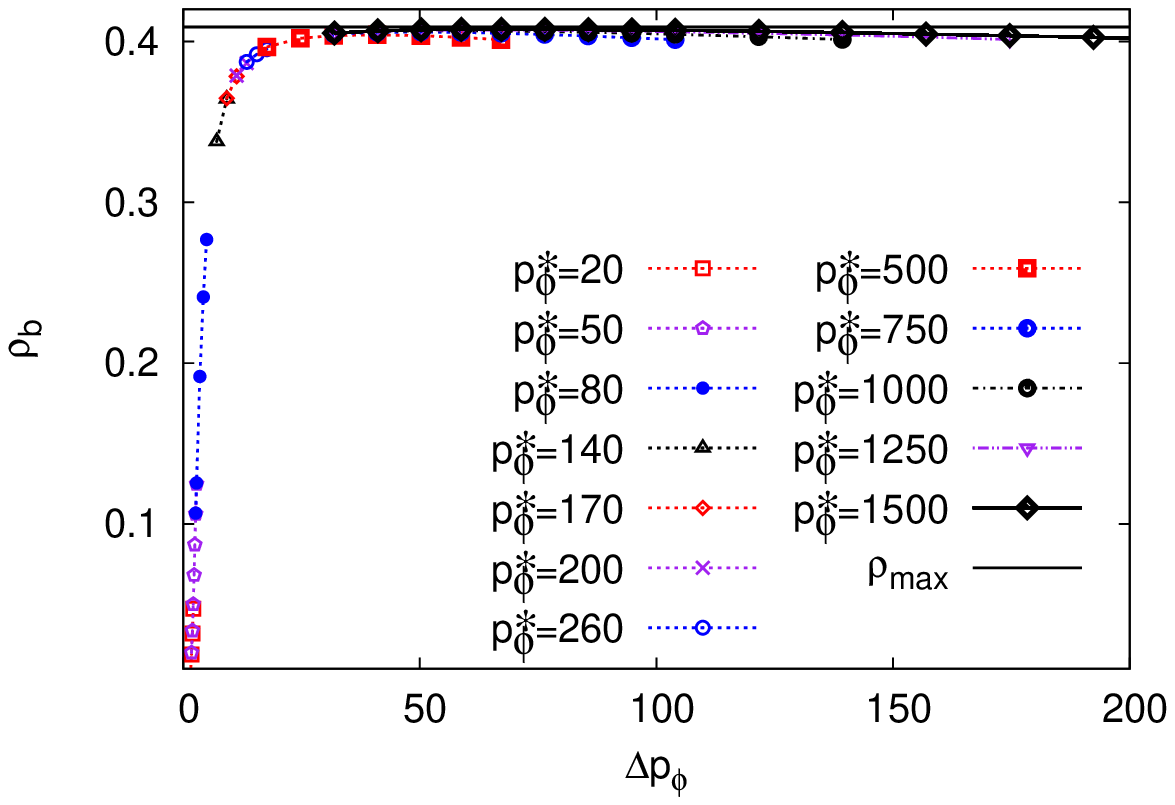}
\label{fig:rhopphim2}
}
\caption{Panel (a) and (b) respectively show the variation of the energy 
  density at the bounce $\rho_{\rm b}$ with respect to the initial relative 
  volume dispersion $\Delta V/V$ and dispersion in field momentum 
  $\Delta p_\phi$ for method-2 initial
  data. In the large $\Delta V/V$ (small $\Delta p_\phi$) regime the energy density is significantly
  smaller than $\rho_{\rm max}$, whereas for small $\Delta V/V$ (large 
  $\Delta p_\phi$ corresponding to sharply peaked states) the energy density
  is close to but still smaller than $\rho_{\rm max}$.}
\label{fig:rhom2}
\end{figure}

\end{enumerate}

\subsection{Method-3: Rotated Wheeler-DeWitt initial state}

\begin{enumerate}
\item \fref{fig:reldispa} shows the behavior of the relative dispersion in
  volume $\Delta V/V$ with respect to $\Delta p_\phi$ in the regime far from
  the bounce in the expanding branch. It is evident that in the large
  $p_\phi$ regime, the relative volume dispersion has a minimum at a certain
  value of $\Delta p_\phi$. This minimum occurs at $\Delta p_\phi =
  \widetilde{\Delta p_\phi} = (3\pi {p_\phi^*}2 G\hbar^2)^{1/4}$.
  As discussed in Sec. \ref{sec:initialdata} this
  is exactly the location at which $\Delta V/V$ as a function of
  $\Delta p_\phi$ has a minimum for method-3 initial data. This is clarified in
  \fref{fig:reldispb} that shows $\f{\Delta p_\phi}{\sqrt{p_\phi^*}}$ plotted
  against $\Delta V/V$.
\begin{figure}[tbh!]
  \subfigure[]
  {
     \includegraphics[width=0.8\textwidth]{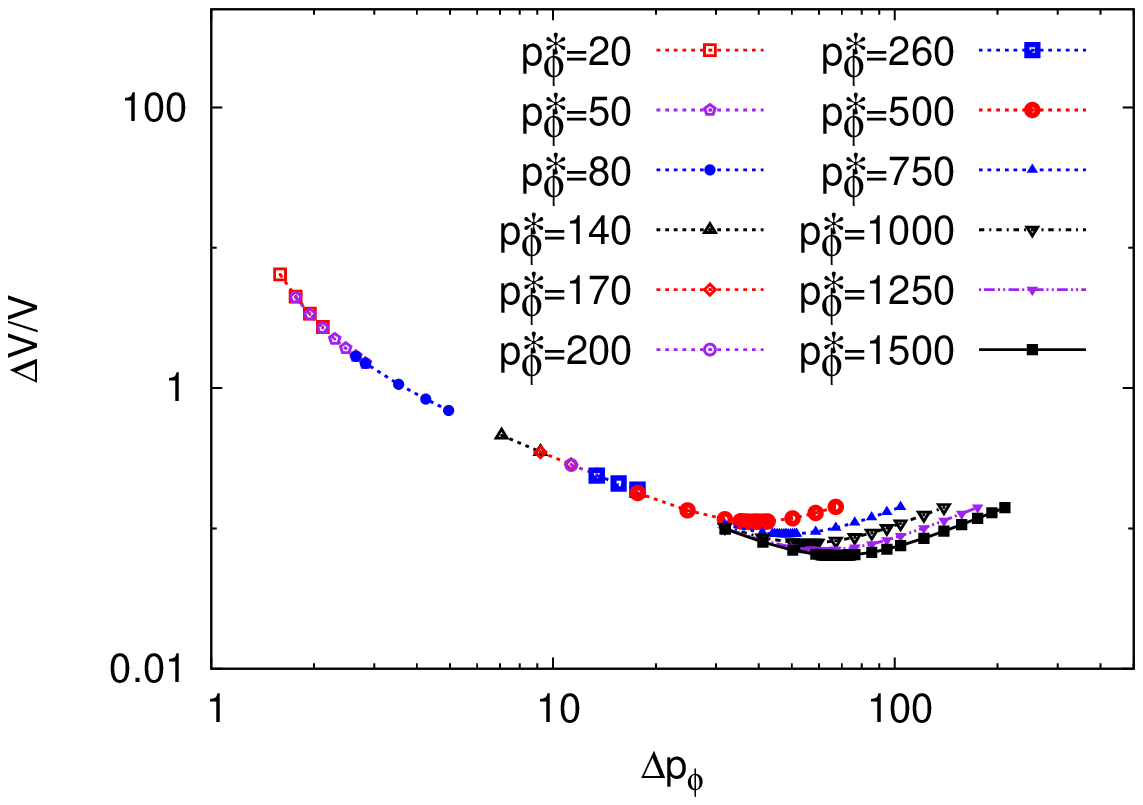}
     \label{fig:reldispa}
  }
  \subfigure[]
  {
	\includegraphics[width=0.8\textwidth]{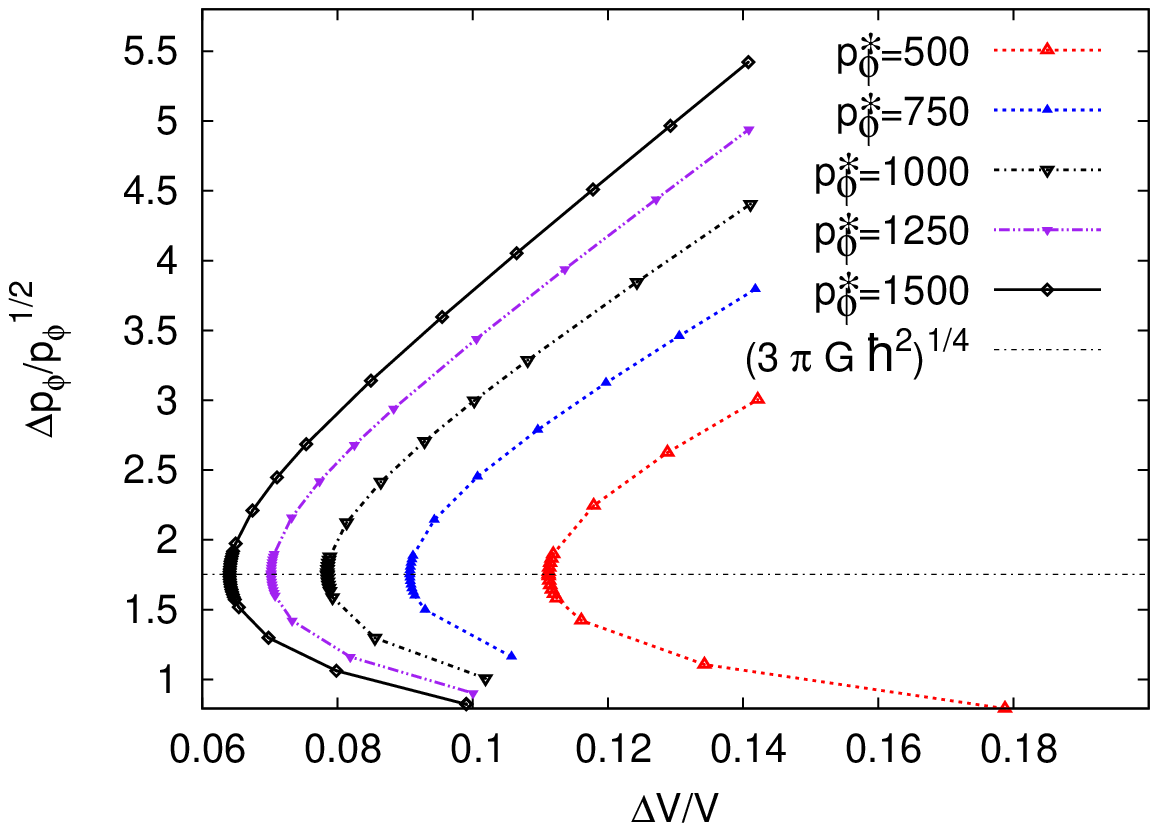}
     \label{fig:reldispb}
  }
  \caption{Panel (a) shows the variation in the value of $\Delta V/V$ with
    $\Delta p_\phi$ for method-3 initial data in the regime far from the bounce
    in the expanding branch. It is evident that $\Delta V/V$ varies
    non-monotonically with increasing $\Delta p_\phi$. Panel (b) shows
    $\Delta p_\phi/\sqrt{p_\phi^*}$ plotted against $\Delta V/V$. We find that $\Delta p_\phi/\sqrt{p_\phi^*}=(3\pi G\hbar^2)^{1/4}$ when
    $\Delta V/V$ is minimal, as discussed in Sec. \ref{sec:initialdata}.}
\label{fig:reldisp}
\end{figure}

\item \fref{fig:vbouncem3delpphi} shows the variation of $\delta$ with respect
  to $\Delta p_\phi$ for several values of $p_\phi^*$.  It is
  interesting to note that $\delta$ varies monotonically with $\Delta p_\phi$.
  As can be seen, the curve is linear in the large $\Delta p_\phi$
  regime, where the initial state is relatively sharply peaked.
  This indicates a power law relation between $\delta$ and $\Delta p_\phi$
  \be
    \delta \propto \left(\Delta p_\phi\right)^\mu,
    \label{eq:powerlawdelta}
  \ee
  where $\mu$ is a constant exponent. A numerical fit yields the value to be
  $\mu=-2.018\pm0.001$ (in the large $p_\phi$ regime).  As we will discuss later in this
  subsection, a similar power-law relation is also present in the behavior
  of the difference in bounce time between the LQC and effective trajectories
  as function of $\Delta p_\phi$. 
  In the small
  $\Delta p_\phi$ region the curve tilts slightly upward compared to straight
  line behavior. 
  \begin{figure}[tbh!]
  \includegraphics[width=0.75\textwidth]{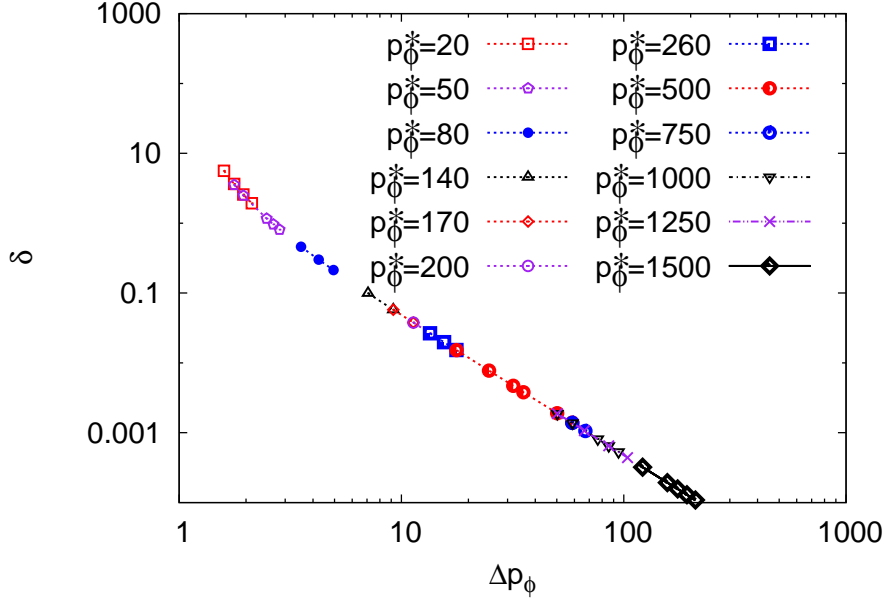}
  \caption{The variation of $\delta$ with respect to the dispersion in the
    field momentum $\Delta p_\phi$ for method-3 initial data is shown.
    In the large $\Delta p_\phi$ regime we observe a linear behavior and
    the slope of the curve is determined to be $\mu=-2.018\pm0.001$.}
\label{fig:vbouncem3delpphi}
\end{figure}
\begin{figure}[tbh!]
  \subfigure[]
  {
    \includegraphics[width=0.8\textwidth]{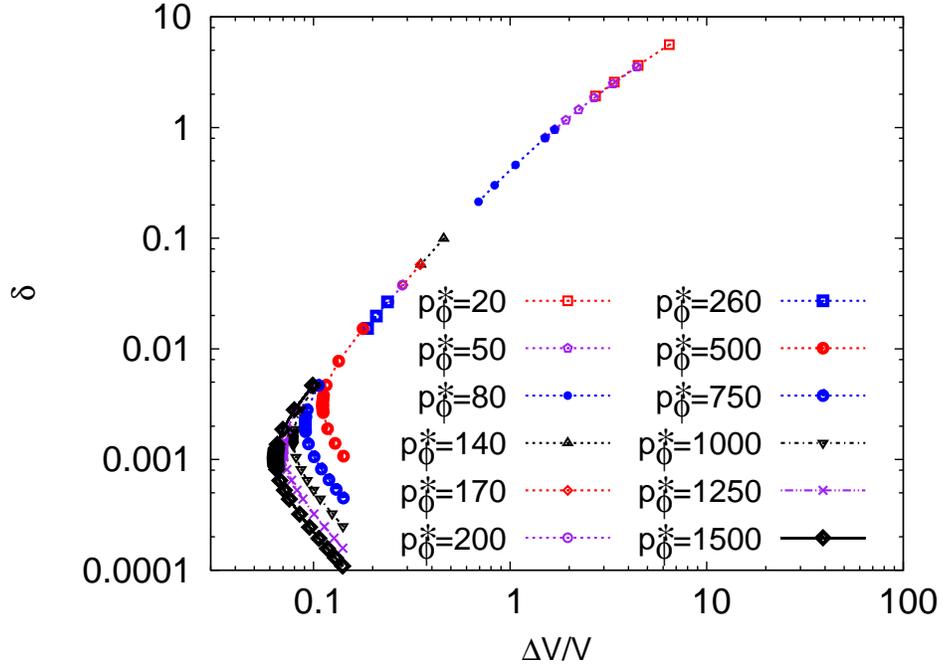}
    \label{fig:vbouncem3a}
  }
  \subfigure[]
  {
    \includegraphics[width=0.8\textwidth]{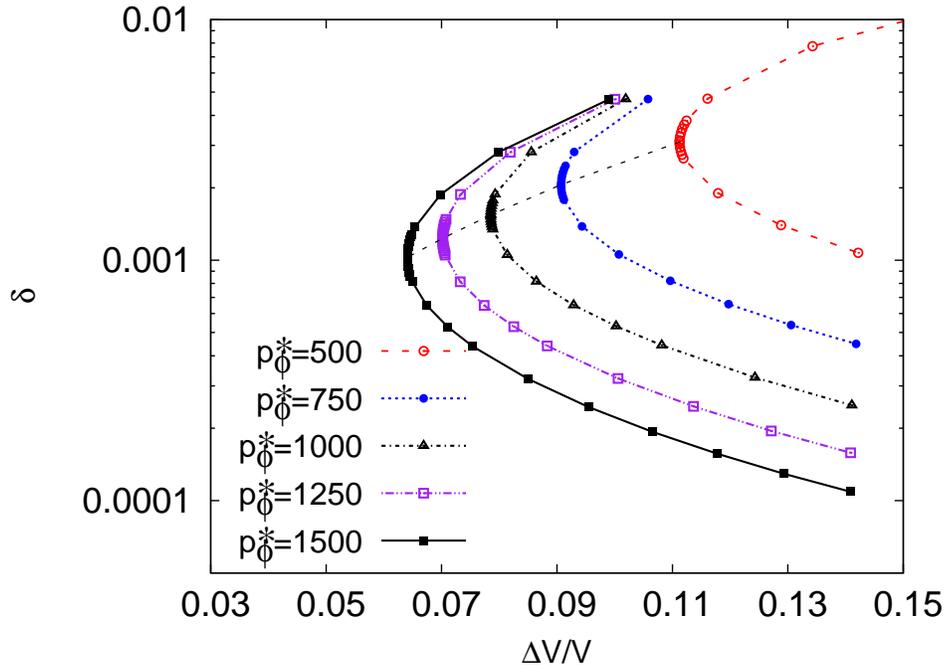}
    \label{fig:vbouncem3b}
  }
\caption{This figure shows the variation of $\delta$ with respect to the relative dispersion in
  volume $\Delta V/V$ for method-3 initial data. Panel (a) shows the full range
  of $\Delta V/V$, while panel (b) is a zoom into the small $\Delta V/V$
  regime. It is clear from panel (a) that $\delta$ increases with increasing
  $\Delta V/V$ in the large $\Delta V/V$ regime. In panel (b) the dashed curve, connecting the turn
  around points of the individual curves for different $p_\phi$ values,
  corresponds to
  $\Delta p_\phi=\widetilde{\Delta p_\phi} = (3\pi {p_\phi^*}^2 G\hbar^2)^{1/4}$.
}
\label{fig:vbouncem3}
\end{figure}

\begin{figure}
\includegraphics[width=0.75\textwidth]{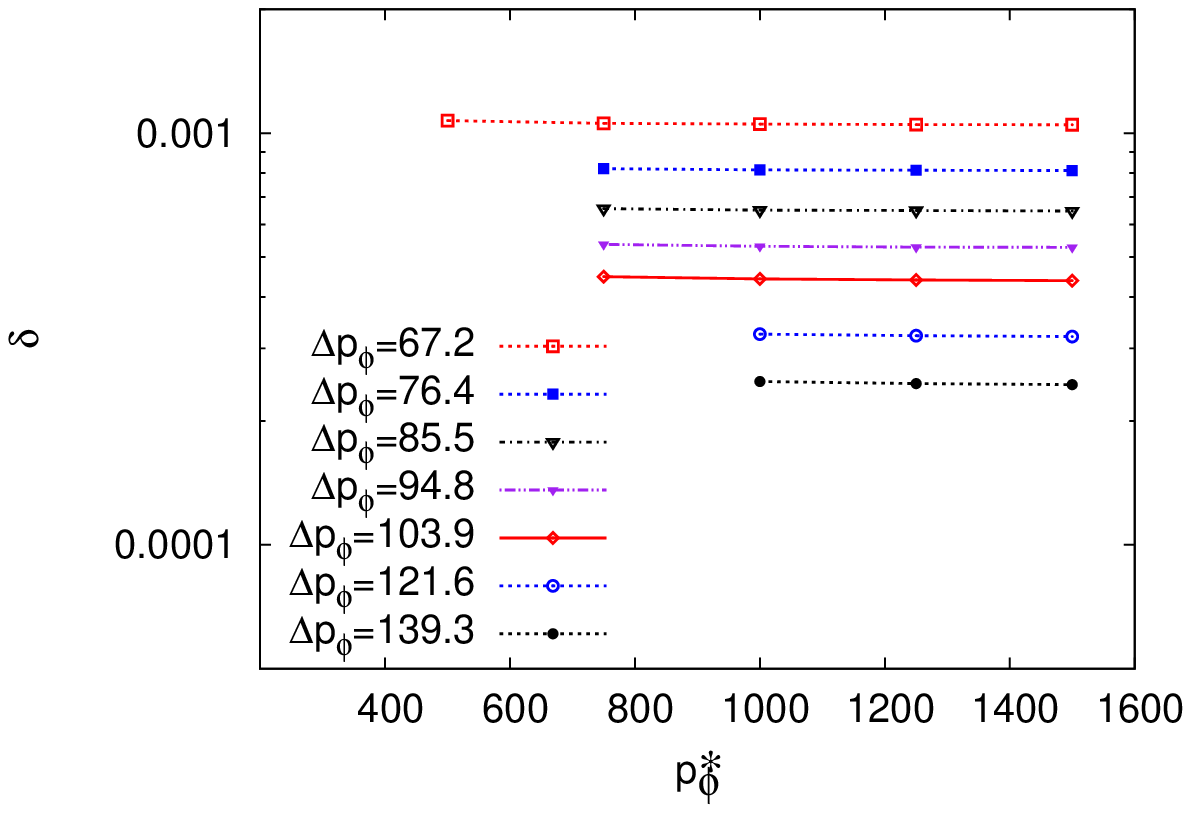}
\caption{This plot shows $\delta$ as function of $p_\phi$ for different values
of $\Delta p_\phi$ for method-3 initial data. For a given $\Delta p_\phi$,
$\delta$ is practically independent of $p_\phi^*$.}
\label{fig:deltam3pphi}
\end{figure}

\item
\fref{fig:vbouncem3a} shows the variation of $\delta$ with respect
 to the relative dispersion in volume $\Delta V/V$. The value of $\delta$ 
 behaves almost linearly in the small $p_\phi^*$
regime where $\Delta V/V$ is large. As the relative dispersion in $V$ decreases,
i.e.\ in the large $p_\phi^*$ regime, and the state becomes more sharply peaked,
the quantity $\delta$ decreases 
indicating that the effective theory serves as a better approximation to the full LQC evolution
in this case with decreasing $\Delta V/V$.
\fref{fig:vbouncem3b} shows the behavior of $\delta$ with the relative volume
dispersion $\Delta V/V$, in the regime  corresponding to small $\Delta V/V$ and
large $p_\phi^*$. Different curves in \fref{fig:vbouncem3b} show
the variation of $\delta$ for different $p_\phi^*$, and the data points on each
curve represent different $\Delta p_\phi$. Since $\Delta V/V$ varies non-monotonically 
with $\Delta p_\phi$ in this regime, the behavior of $\delta$ with varying 
$\Delta V/V$ is also non-monotonic. Recall that the input parameter for method-3 initial 
data are $p_\phi^*$ and $\Delta p_\phi$, hence $\Delta V/V$ is a derived 
quantity.
The dashed curve connecting the turn around points of
different curves corresponds to $\Delta p_\phi=\widetilde{\Delta p_\phi} = (3\pi p_\phi^2 G\hbar^2)^{1/4}$
which is where $\Delta V/V$ is minimum and $\delta$ changes behavior. Below this curve
($\Delta p_\phi>\widetilde{\Delta p_\phi}$) $\delta$ decreases with increasing $\Delta V/V$, while
above the curve ($\Delta p_\phi<\widetilde{\Delta p_\phi}$) $\delta$ increases.
This is summarized in the row for method-3 and the column for fixed $p_\phi^*$ in
Table \ref{tab:summary}.
The figure also shows that for a fixed value of $\Delta V/V$, the quantity 
$\delta$ increases with increasing $p_\phi^*$ above the dashed curve 
($\Delta p_\phi<\widetilde{\Delta p_\phi}$) while  
$\delta$ decreases with increasing $p_\phi^*$ below the dashed curve
($\Delta p_\phi>\widetilde{\Delta p_\phi}$). This is summarized 
in the row for method-3 and the column for fixed $\Delta V/V$ in 
Table \ref{tab:summary}.
\fref{fig:deltam3pphi} shows the dependence of $\delta$ with varying $p_\phi^*$ 
for different values of $\Delta p_\phi$. It can be seen that $\delta$ 
is practically independent of $p_\phi^*$ for a given value of $\Delta p_\phi$
as summarized in the row for method-3 and the column for fixed $\Delta p_\phi$
in Table \ref{tab:summary}.
  \begin{figure}[tbh!]
  \includegraphics[width=0.75\textwidth]{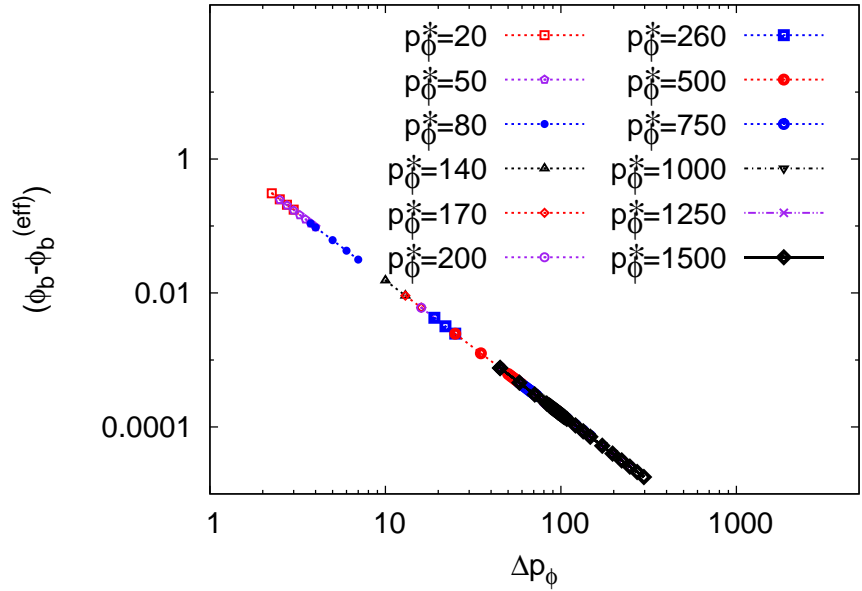}
  \caption{The behavior of the difference between the time
     of bounce in the LQC and the effective dynamical evolution for method-3
     initial data. The curve is linear over the whole range, implying a
     power law relation between
     $\left ({\phi_{\rm b}}-{\phi_{\rm b}}^{(\rm eff)}\right )$ and $\Delta p_\phi$.
     The slope of the curve is determined to be $\nu=-2.015\pm0.003$.}
\label{fig:phibounce}
\end{figure}

\item \fref{fig:phibounce} shows the variation of the difference in the values
  of emergent time `$\phi_{\rm b}$' at the bounce in the LQC and effective
  trajectories $\left ({\phi_{\rm b}}-{\phi_{\rm b}}^{\rm (eff)}\right )$ with
  respect to $\Delta p_\phi$. There is a  power law
  relation between these quantities over the complete range of values of $\Delta p_\phi$ considered:
  \be
    \left ({\phi_{\rm b}}-{\phi_{\rm b}}^{\rm (eff)}\right ) \propto
    \left(\Delta p_\phi\right)^\nu,
    \label{eq:powerlawphi}
  \ee
  with $\nu$ being the power law exponent, which turns out to be $-2.015\pm0.003$. 
 This power law 
  is very similar to the power law relating $\delta$ to $\Delta p_\phi$
  as given in \eref{eq:powerlawdelta} and even has the same exponent
  $\mu=\nu=-2.01$ to three significant figures. Also, $\nu$ 
  is surprisingly the same (within the fitting errorbars) as the exponent in 
  the power law behavior of 
  $\left ({\phi_{\rm b}}-{\phi_{\rm b}}^{\rm (eff)}\right )$ for method-2, which is 
  $-2.017\pm0.008$.

\item \fref{fig:rhom3} shows the variation of the energy density at the bounce
  $\rho_{\rm b}$ with respect to $\Delta V/V$ (upper panel) and $\Delta p_\phi$
  (lower panel). As the relative dispersion in volume increases,
  the energy density at the bounce decreases. In the small $\Delta V/V$ regime,
  where the state is sharply peaked in volume, the energy density is close to
  but smaller than $\rho_{\rm max}$.
  \begin{figure}[tbh!]
    \includegraphics[width=0.75\textwidth]{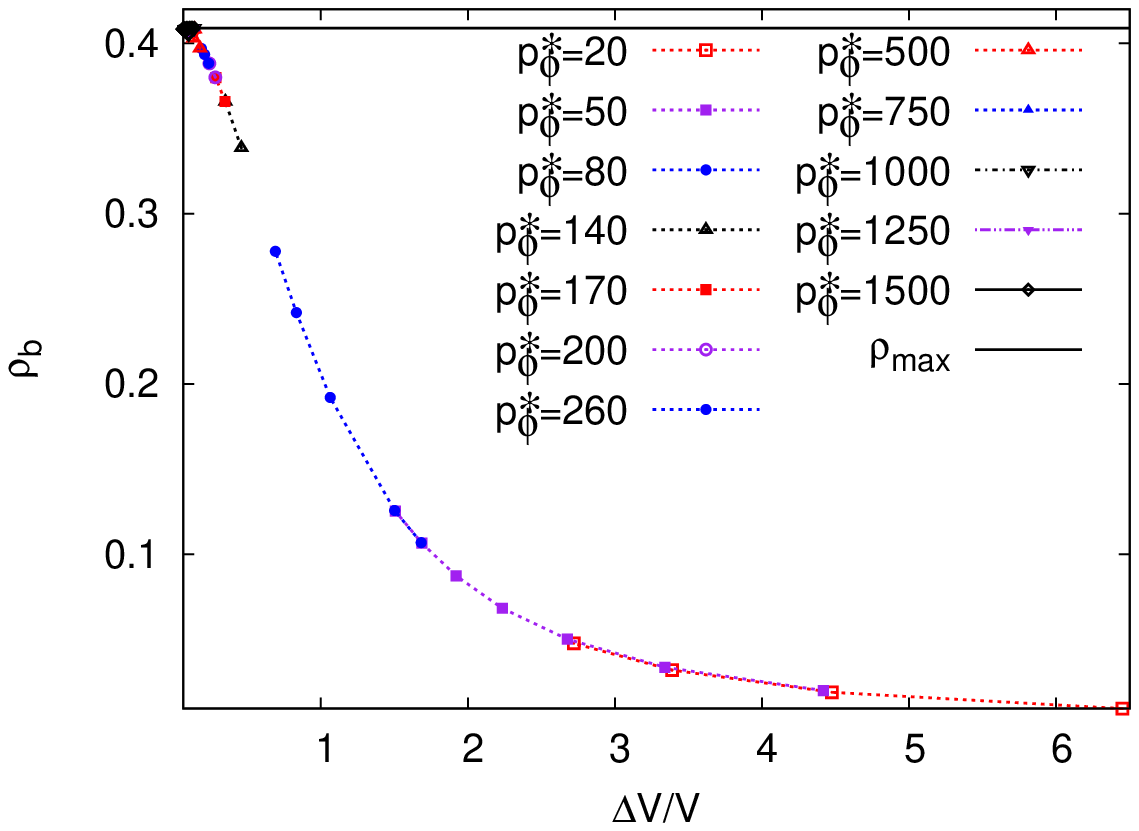}
    \includegraphics[width=0.75\textwidth]{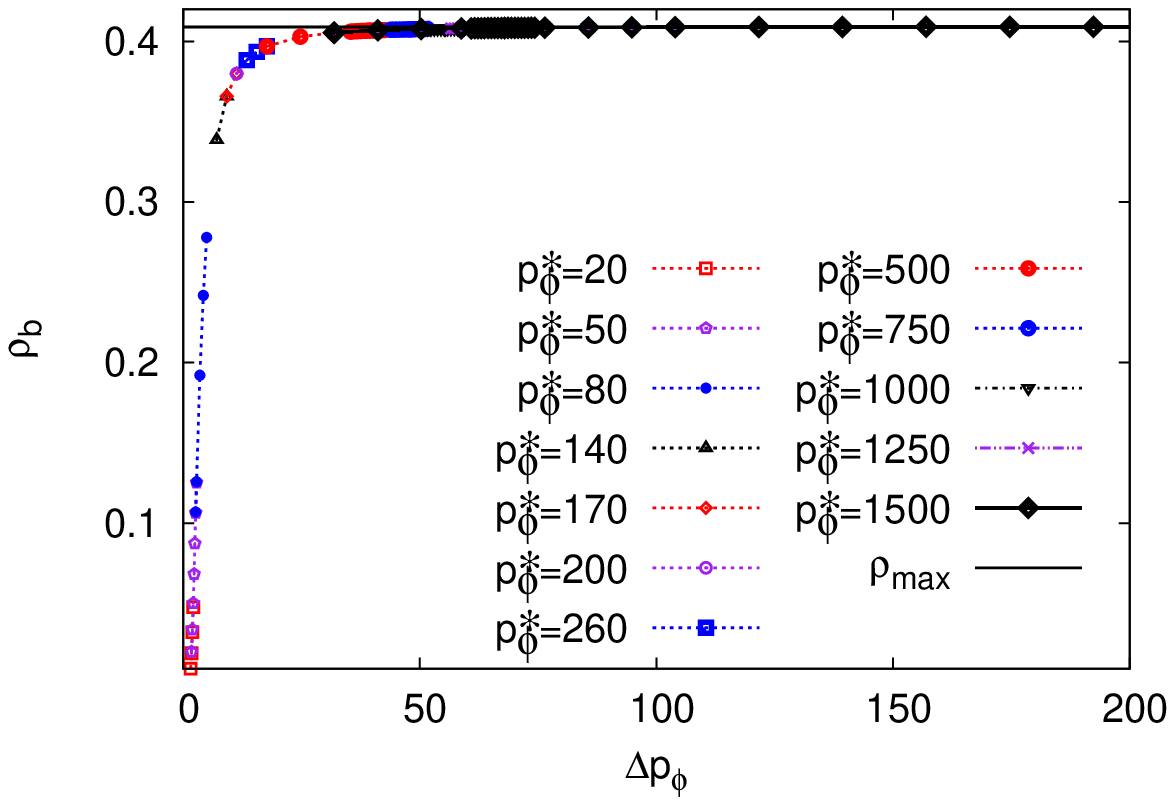}
    \caption{The energy density at the bounce $\rho_{\rm b}$ plotted with
      respect to the initial relative volume dispersion $\Delta V/V$ (upper
      panel) and with respect to the dispersion in the field momentum
      $\Delta p_\phi$ (lower panel) for method-3 initial data. In the large
      $\Delta V/V$ regime (or equivalently for small values of
      $\Delta p_\phi$), $\rho_{\rm  b}$ is much smaller than
      $\rho_{\rm max}$. For sharply peaked states, $\rho_{\rm  b}$ is
      close to, but always smaller than $\rho_{\rm crit}$.}
\label{fig:rhom3}
\end{figure}

\end{enumerate}

\section{Discussion}\label{sec:disc}
In this paper we have presented a rigorous numerical study of the evolution of states 
in loop quantum cosmology for a flat FRW universe in the presence of a massless scalar field, 
including states which are widely spread and bounce close to the Planck volume. Due to the underlying discrete quantum geometry, 
the evolution of states is non-singular and the big bang singularity is replaced by a quantum bounce. Previous numerical studies in LQC demonstrated the 
existence of a bounce for sharply peaked states corresponding to large values of the field momentum $p_\phi$. Such states lead to a bounce volume 
much greater than the Planck volume. The evolution of such states turns out to be in excellent agreement with the effective spacetime description 
derived from an effective Hamiltonian \cite{josh,vt,psvt}. In this manuscript, we have extended the results on the quantum bounce to states which are very widely spread and which bounce much closer to the Planck volume. Investigations along these lines provide a very rigorous test for the validity of the effective dynamics.   Numerical investigation of such states, on the other hand, present severe computational 
challenges which limit the types of states and the values of parameters for which a stable 
numerical evolution could be performed with previously used techniques. In order to overcome the computational limitations and study 
the numerical evolution of more general classes of states, we recently proposed a numerical 
scheme, Chimera, which proves to be a robust and efficient way to perform the simulations (see Ref. \cite{dgs1} for details). 
For example, as discussed in this paper, numerical 
simulations of very widely spread states, which would have taken extremely long time using conventional methods, can be performed in less than an hour on a workstation 
using the Chimera scheme.

In the present paper, we have used the Chimera scheme to 
study the LQC evolution of three different kinds of states and investigated the quantitative 
difference between the effective and LQC trajectories. We presented the results of a large 
number of simulations with various values of the relative volume dispersion $\Delta V/V$, 
spread in the field momentum $\Delta p_\phi$ and the field momentum $p_\phi$ at which the 
states are initially peaked. We found that the occurrence of the quantum bounce is a generic 
feature of all the different kinds of initial states we explored. This is in
agreement with the results in Ref. \cite{acs}, obtained via an exactly solvable model albeit 
for a slightly different quantum 
Hamiltonian constraint of the spatially flat isotropic model.
We also found surprising quantitative results concerning the difference between 
the LQC and the corresponding effective trajectories for different types of
initial data. As a generic feature, we found that the effective theory always overestimates 
the energy density at the bounce and underestimates the bounce volume, and that 
the effective theory remains a good approximation unless the states are widely spread. The energy 
density at the bounce always turns out to be bounded above by $\rho_{\rm{max}}$ predicted by the 
exactly solvable model \cite{acs} and the effective theory (\eref{rhomax_eff}).
Our analysis also shows that the nature of departure of the effective theory from LQC 
depends on the type of initial data, as well as the initial
data parameters.
To quantify these deviations we defined a quantity $\delta$ as the fractional difference in the 
bounce volume in the two descriptions. It turns out that $\delta$ can be large 
for small $p_\phi$, for which the bounce happens closer to the Planck volume. 
However, $\delta$ depends on the choice of initial state in a subtle way.
For a fixed value of $\Delta V/V$, $\delta$ always increases with decreasing $p_\phi$ 
for method-1 and method-2 initial states, while for method-3 the behavior depends on whether 
$\Delta p_\phi$ is greater or smaller than a certain value. 
In the large $p_\phi$ regime, where $\delta$ is small,  
we found interesting non-monotonic variations of $\delta$ with $\Delta V/V$ and 
$\Delta p_\phi$. This behavior is similar to the non-monotonic dependence of 
$\Delta V/V$ on $\Delta p_\phi$ for the initial state. This was also noted in Ref. 
\cite{montoya_corichi2} 
for a different construction of initial state and a different quantum 
constraint. For all states and all initial conditions considered in this 
paper, our results are also
in accordance with the study of the fluctuations of states across the bounce, and 
obey the corresponding triangle inequality \cite{kp} which places restrictions on the 
variation of the fluctuations across the bounce. 
Thus, in addition to strengthening some of the previous results on the quantum bounce, the 
results shown in this paper bring out some surprising features as well. For example, conventionally 
one would have expected that $\delta$ would continue to increase with increasing 
volume dispersion of the state. However, our analysis shows that this is not true in general. 

It can be concluded from the numerical results obtained in this paper that the effective theory 
is in good agreement with the LQC evolution for states which are peaked
at a classical macroscopic universe at late times. 
Notable differences between the effective and the LQC trajectories only occur
for states with large relative fluctuations. In the derivation of the modified Friedmann equation in Ref. \cite{vt}, such states lead to further corrections. 
Incorporating these corrections in the effective dynamics would lead to a better agreement with the numerical simulations presented in this analysis.
 However, a more rigorous understanding is needed in the effective descriptions to gain insights on the way departures depend non-monotonically on 
fluctuations, and the way those departures are affected by the choice of the method of construction of the initial states. In addition, our results show that the role of fluctuations in the 
matter variable is as equally important as of the the volume variable in understanding reliability of the effective 
dynamics.\footnote{It will be interesting to investigate in detail the effects of matter fluctuations in the arguments presented in Ref. \cite{rovelli_wilson}. Study of these fluctuations is expected to yield more insights on the role of the infinite limit of the fiducial cell on the reliability of 
the effective dynamics. }
The matter and geometry fluctuations have 
non-trivial interdependence which can be different for different constructions of the initial sates. Our results show that deciphering the reliability of effective descriptions 
in LQC is a subtle problem and an extra care is needed in generalizing conclusions from the effective descriptions.

Moreover, although the differences between the effective and the LQC 
trajectories for sharply peaked states corresponding to large field momenta 
are small, these differences depend on the parameters of the initial data in 
an unexpected fashion. Whether the effective theory becomes more or less 
reliable as one varies the parameters of the initial state, depends in a subtle 
way on the initial data parameters and the way the initial state is constructed. 
Note that in this paper, we have considered the effective description which is 
based on the embedding approach \cite{josh, vt}. The fate of the effective 
description based on the truncation approach \cite{mb_rev_eff} needs to be 
rigorously tested in a similar way, to make any reliable comparison between 
the two approaches to the effective theory and the loop quantum evolution.

Most of the phenomenological studies of various cosmological models and 
observational predictions of LQC are based on the effective description of
LQC. The rigorous numerical simulations and the study of the differences between 
the effective and LQC trajectories, presented in this paper, open new avenues to 
understand the implications of the small departures between the
effective theory and full quantum evolution on the observational signatures for widely spread states, 
e.g.\ along the lines of those computed in Refs. \cite{aan1,aan3}. 
For a better understanding of these modifications, one needs to extend the numerical 
studies to more 
general classes of models. This will be reported in upcoming works where we have used the 
Chimera scheme to study the numerical evolution of squeezed states and states with non-gaussian
waveforms for a massless scalar field \cite{dgms} as well as more general matter models such as 
with an Ekpyrotic like potential \cite{cyclic}.  
These upcoming works along with the results presented in 
this paper provide a benchmark to numerically test the effective theory, and  complement 
insights from the analytical derivation of the corrections to the
effective equations for more general classes of states and matter models 
\cite{psvt}. Availability of these corrections promises to provide a more rigorous understanding of the new physics resulting from the quantum bounce 
and its effects on cosmological perturbations.

\acknowledgements
We would like to thank Abhay Ashtekar, Steve Brandt, Frank L\"offler and Miguel Megevand for useful discussions,  
and Martin Bojowald and Edward Wilson-Ewing for comments. 
This work is supported by a grant from John Templeton Foundation and by NSF grant 
PHYS1068743. The opinions expressed in this publication are those of authors and do not 
necessarily reflect the views of John Templeton Foundation. BG would also like to acknowledge 
support from the Coates Scholar Research Award and the Dissertation Year Fellowship of the 
Louisiana State University. This material is based upon work supported by HPC@LSU computing resources.

\section*{References}

\end{document}